%% file: main.tex
\documentclass{aa} 
\usepackage{makecell}
\usepackage{graphicx}
\usepackage{txfonts}
\usepackage{natbib}
\usepackage{amsmath}
\usepackage{ amssymb }
\usepackage{color}
\usepackage{array}
\usepackage{multirow}
\usepackage{subfig}
\usepackage{xspace, siunitx}
\usepackage{xcolor}
\usepackage{url}
\usepackage{rotating}
\usepackage{pdflscape}

\newcommand{\equ}[1]{Eq.~\ref{eq:#1}}

\newcommand{\fig}[1]{Fig.~\ref{fig:#1}}

\newcommand{\tab}[1]{Table~\ref{tab:#1}}

\newcommand{\sect}[1]{Sect.~\ref{sec:#1}}
\newcommand{\app}[1]{Appendix~\ref{app:#1}}

\newcommand{\lcdm}[0]{$\Lambda$CDM\xspace}
\newcommand{\atlas}{ATLAS$^{3\mathrm{D}}$\xspace}
\newcommand{\rhoio}{\ensuremath{\rho_{10}}\xspace}
\newcommand{\mjam}{\ensuremath{M_\mathrm{JAM}}\xspace}

\newcommand{\reff}{\ensuremath{R_\mathrm{e}}\xspace}
\newcommand{\bilmat}{\citet{bil20}\xspace}
\newcommand{\msun}{\ensuremath{\,M_\sun}\xspace}
\newcommand{\rotsup}{\ensuremath{\lambda_{R_\mathrm{e}}^\mathrm{N}}\xspace}

\begin{document}

\title{{Origin of the differences in rotational support among early-type galaxies: The case of galaxies outside clusters}} 

\titlerunning{Kinematic transformation of ETGs in MATLAS}

\author{M. B\'{i}lek\inst{1,2,3}
\and
P.-A. Duc\inst{3}
\and
E. Sola\inst{3}
}
\institute{LERMA, Observatoire de Paris, CNRS, PSL Univ., Sorbonne Univ., 75014 Paris, France\\
    \email{michal.bilek@obspm.fr}
        \and
        Coll\`ege de France, 11 place Marcelin Berthelot, 75005 Paris, France
        \and
        Universit\'e de Strasbourg, CNRS, Observatoire astronomique de Strasbourg (ObAS), UMR 7550, 67000 Strasbourg, France\\
}

\date{Received ...; accepted ...}
%

\abstract
{Early-type galaxies (ETGs) are divided into slow and fast rotators (FRs and SRs) according to the degree of ordered rotation of their stellar populations. Cosmological hydrodynamical simulations indicate that galaxies form as FRs before their rotational support  decreases, usually because of mergers.
}
{We aimed to investigate this process observationally for galaxies outside of clusters.
}
{ We made use of the fact that different merger types leave different traces that have different lifetimes. We statistically analyzed multiple characteristics of galaxies that are expected to be influenced by mergers{, such as  tidal features, kinematically distinct cores, and stellar ages.} They were taken from the MATLAS and \atlas databases. Through multilinear regression we identified the quantities that, at a fixed mass and environmental density of the galaxy, significantly correlate with a measure of the ordered rotation of the galaxy, \rotsup. 
}
{We found a negative correlation of the rotational support with the occurrence of tidal disturbances and kinematic substructures, and a positive correlation with metallicity and metallicity gradients. For massive galaxies, the rotational support correlates negatively with the abundance of alpha elements,  and for the galaxies in low-density environments, it correlates negatively with the central photometric cuspiness.  These and additional literature observational constraints are explained the easiest if the mergers that decreased the rotational support of ETGs were typically  minor, wet, and happening at $z\approx 2$. {They did not form the currently observed tidal features.} The observed frequency of tidal features implies a merging rate of 0.07-0.2 per Gyr. This is insufficient  to explain the observed growth of the radii of ETGs with redshift by mergers.}
{}


\maketitle

\section{Introduction} \label{sec:intro}

The formation of early-type galaxies (ETGs) seems to have proceeded in two phases  \citep{oser10,naab14}. The first phase, a ``wet clumpy collapse'' \citep{thomas99}, is a period of intensive in situ star formation. Galaxies were assembled through numerous gas-rich mergers or a smooth accretion of gas { from} cosmic filaments. This formation stage is witnessed by the chemical composition of ETGs, which  is best explained  by the monolithic-collapse models \citep{thomas99,kobayashi04,pipino10,pipino13}. {These models} reproduce the observed values of metallicity and abundance of $\alpha$-elements, and the correlations of these quantities with the masses of the galaxies.  The end of the first phase of formation of ETGs is characterized by the termination of star formation in the galaxies. It happens approximately at a redshift of two, but this limit is not universal. Observations suggest that some ETGs, particularly the most massive ones or those with the highest stellar surface densities, were already quenched as soon as at $z=7-9$, while others continued forming stars much longer, up to $z=0.5-2$ \citep{mcdermid15,gonzalez17, forrest19,estrada20,carnall20,costantin21,tacchella22}.

{There are many ways for a galaxy to become quenched, the relative importance of which has not yet been fully clarified.} For the most massive galaxies, the mass quenching mechanism seems plausible \citep{keres05,dekel06}: when the falling interagalactic gas reaches the halos of galaxies that are massive enough, it is shock-heated so much that the cooling time exceeds the age of the Universe. The infalling clouds of cold gas are not able to reach the galaxy and get dissolved in the hot gas halo of the galaxy \citep{afruni19}.  The filaments of cold interagalactic gas are able to penetrate the hot circumgalactic gaseous halos only before  $z=1.5-2$. Therefore, the massive galaxies are currently mostly passive. A number of mechanisms have been identified for the lighter galaxies.  They include, for galaxies falling to galaxy clusters, the mechanisms of starvation \citep{larson80} or gas shocks \citep{bitsakis16,ardila18,bitsakis19}. Other spirals get quenched in centers of clusters by ram-pressure striping. Theoretical arguments suggest that even if a galaxy has a substantial gas content, the bare presence of a spheroidal component can  postpone or prevent star formation; this is the so called ``morphological quenching'' \citep{martig09,martig13}. The role of the activity of galactic nuclei in the quenching of galaxies is still debated (see \citealp{harrison17} for a  review). In any case, the chemical composition of ETGs indicates that the dominant quenching mechanism has to cut the ETGs progenitors off from the inflow of fresh intergalactic  gas \citep{peng15,trussler20}, which is called ``strangulation''. The same data also show signs of a removal of the cold gas from the galaxies, particularly from the less massive ones, but this is an effect of the secondary importance \citep{trussler20}.

The monolithic collapse phase and quenching are still not sufficient to explain the observational data. The passive galaxies at redshift around two are more compact  \citep{daddi05, trujillo06, vandokkum09} and disky \citep{toft07,hill19}  than the local ETGs.  The simplest explanation of galaxy expansion involves a bare mass loss of the galaxy by stellar evolution and gaseous outflows caused by active galactic nuclei \citep{fan08,damjanov09,fan10}. However, cosmological simulations and some observational evidence \citep{naab09,trujillo11} rather suggest that primarily minor mergers are responsible for the expansion.  This is supposed to happen from about  $z=2$ until today, and it is called the second phase of ETG formation. This would explain several other observations, such as the existence of tidal features (e.g., \citealp{MC83,atkinson13,duc15,bil20}), the properties of globular cluster systems \citep{cote98},  { and} the flattening of metallicity gradients at the outskirts of ETGs \citep{zibetti19,oyarzun19}.  Finally, it has been observed that star formation can be renewed in ETGs after they have been quenched \citep{thomas10,serra14,gavazzi18,mancini19,yildiz20}.  The new stars  then form in a disk.

Some ETGs were likely formed outside of the two-phase scenario by binary mergers of spirals \citep{toomre77}. While this mechanism was popular in the past, it turns out that ETGs, particularly the massive ones, {are formed only rarely through this mechanism. \citep{thomas99,naab09b,harrissaasfee,krajnovic11}.}

The unsolved problems of ETG formation in {$\Lambda$-cold-dark-matter (\lcdm) cosmological} models are generally related to a too gradual formation of ETGs: the main signs are the frequent observations of very massive quenched galaxies at very large redshifts \citep{hill17,schreiber18,merlin19,faisst19,forrest19,stevans21,carnall22},  and the problems with reproducing the values and correlations of the abundance of the $\alpha$-elements \citep{thomas99,thomas02,thomas03,nagashima05,pipino09,delucia17,okamoto17,vincenzo18}.

Early-type galaxies seem to form bimodal statistical distributions in the space   of their properties. This is why they are divided into slow and fast rotators  (SRs and FRs, see  \citealp{cappellari16} for a review). Fast rotators show a regular rotational pattern in the kinematic maps of their inner  stellar populations \citep{emsellem11}. Their kinematic axes are aligned well with the minor photometric axes of the galaxies. Fast rotators turn out to be intrinsically oblate axisymmetric ellipsoids \citep{weijmans14,foster17}.  Kinematic maps of SRs instead either do not show any rotation, or show complex features, such as kinematically distinct components \citep{emsellem07,emsellem11}. Slow rotators are rounder than FRs and are weakly triaxial \citep{weijmans14,foster17,li18b}, and their kinematic and photometric axes do not align well \citep{krajnovic11,ene18}.  Slow rotators are also typically more massive than FRs \citep{emsellem11}. The vast majority of ETGs are FRs, but SRs prevail among the most massive ETGs \citep{emsellem11}. This is because the degree of rotation is probably primarily a function of the mass of the galaxy \citep{veale17,brough17,greene17} and the most massive galaxies are located in the centers of their groups or clusters, even though some works suggest that the rotator type is also influenced  independently by the density of the environment of the galaxy \citep{graham19,graham19b,vandesande21b}. While the primary motivation for introducing the FRs and SRs was the distinct appearance of their kinematic maps, many works rely on quantitative definitions, which are supposed to be roughly equivalent to the morphological definition. Here we build on the  widely used parametric separation criterion by \citet{emsellem11}: ETGs having their $\lambda/\sqrt{\epsilon}$ below the value of 0.31 are classified as SRs while the rest are classified as FRs. Here $\epsilon$ stands for the apparent ellipticity of the galaxy within the half-light radius, and the parameter $\lambda$ quantifies the relative importance of the ordered rotation and velocity dispersion in supporting the galaxy.  

{Simulations provide us with insights into the formation of FRs and SRs. The}  44 high-resolution zoom-in simulations by \citet{naab14} showed that there are many ways to form an SR or an FR in terms of the number of mergers, their mass ratios, and gas fractions. The slowest rotators, however, were formed by many minor mergers.  The early work by \citet{naab14} had the disadvantage that it could not  indicate how important the individual formation channels of SRs and FRs are. Ideally, this would be shown by cosmological hydrodynamical simulations.  Such a route was followed by \citet{penoyre17}, who inspected the Illustris simulation. They found that at very high redshifts, all galaxies are FRs, with a high rotational support. The progenitors of the current SRs and FRs are still nearly indistinguishable at $z=1$ in terms of the distributions of stellar masses and of the $\lambda$ parameter. The rotation support was found to decrease after major mergers but very massive galaxies decreased it even in periods without mergers.
\citet{lagos18} undertook a similar approach with the EAGLE \citep{schaye15} and HYDRANGEA \citep{bahe17} simulations. They again found a link between mergers and a decrease in the degree of rotational support. They found that galaxies that experienced dry mergers in the simulations, either minor or major, usually ended up with a lower degree of rotation than galaxies that underwent wet or no mergers. Some mergers can increase the rotational support, but such mergers are in the minority.
A small fraction of SRs did not experience any mergers and such galaxies inhabited low-spin halos. Nevertheless, in most cases, the transformation from FRs to SRs happens through mergers. The formation of SRs from FRs in the EAGLE simulation was confirmed by \citet{lagos22} and in the MAGNETICUM simulation by \citet{schulze18}. 

Cosmological hydrodynamic simulations still have relatively low resolution. This might be the reason why the stellar kinematics of the simulated galaxies do not fully reproduce the reality. For example, \citet{ebrova20} noted that the kinematically distinct components in Illustris  are too large,   \citet{lagos22} pointed out the nonrealistic radial profiles of velocity dispersion of the  galaxies in EAGLE, and \citet{schulze18} reported a population of overly flattened SRs in MAGNETICUM.

In this paper we investigate the details of the decrease in the rotational support of ETGs  observationally for galaxies outside of galaxy clusters.
In order to quantify how much the transformation has progressed in a given galaxy, we heuristically exploit the parameter $\rotsup = \lambda/\sqrt{\epsilon}$ \citep{emsellem11}, which we call the rotational support.   We note that more recent works use more elaborate criteria than $\rotsup <$ or $> 0.31$ to classify galaxies into SRs and FRs. The newer criteria are supposed to capture the bimodality of ETGs better. For example, \citet{cappellari16} define SRs as satisfying the conditions $\lambda < 0.08 + \epsilon/4$ and $\epsilon<0.4$. Here we build on the older SR and FR separation criterion that uses the \rotsup parameter because it seems obvious how to use it to continuously quantify the stage of the kinematical transformation. It is possible that future studies will find a more suitable quantification of the transformation stage than the \rotsup parameter.

The main idea of this paper is the following. We assume, inspired by the simulations, that all ETGs form initially with a  relatively high and a relatively universal value of rotational support, and then the rotational support is decreased by mergers. Next, we assume that if the mergers have been very important in the evolution of the galaxy until now, the galaxy would, in most cases, be observed to have a low value of the rotational support at the current cosmic epoch. The mergers change various parameters of the galaxy. The traces of the mergers depend on whether the merger is gas rich or gas poor, and on the mass ratio of the merging galaxies. In addition, the different merger signs have different lifetimes. We make use of that and investigate the correlations of various merger-sensitive indicators with the rotational support in order to study the mergers that are responsible for the decrease in the rotational support. In order to reduce confounding effects, {we studied the correlations} at a fixed mass and environmental density through mulitilinear regression.

The paper is organized as follows. {Our data sources are  presented in \sect{data}.} In \sect{indicators}, we list the merger-sensitive parameters that we exploit to derive the characteristics of the mergers that caused the decrease in the rotational support of ETGs. For each parameter, we explain how it is expected to change after different types of mergers. \sect{survival} is devoted to the estimation of how long after a merger a galaxy appears morphologically disturbed. The main method of the paper is described in \sect{corr}, where we explain how we determined whether and how different merger sensitive parameters correlate with the \rotsup parameter for galaxies at a fixed mass and environmental density, and present the results. These results and additional literature findings are then used in \sect{interp} to deduce what types of mergers were typically decreasing the rotational support of galaxies and when they happened. Given that we a get different time of the kinematic transformation than predicted by cosmological simulations, we verify our conclusion by independent methods in \sect{late}. As a by-product, we estimate in \sect{mergingfreq} the current merging rate of our galaxies from the incidence of tidal disturbances. In \sect{theor}, we show that many observational findings about the occurrence of FRs and SRs can be explained as simple consequences of the fact that galaxy mergers usually decrease the rotational support of galaxies. We synthesize our findings in \sect{picture}, where we propose how typical FRs, SRs, and spiral galaxies form. The paper is summarized in \sect{sum}. In \app{simplecorr} we provide the correlations of the various merger-sensitive parameters with galaxy mass, environment density, and rotational support. We also compare the merger-sensitive parameters in FRs and SRs.

For the conversion between look-back time and redshift, we used Ned Wright's cosmology calculator\footnote{\url{http://www.astro.ucla.edu/~wright/CosmoCalc.html}} \citep{wright06} with the cosmological parameters $H_0 = 69.6$\,km\,s$^{-1}$\,Mpc$^{-1}$, $\Omega_\mathrm{M}=0.286$ and $\Omega_\mathrm{vac}=0.714$ \citep{bennett14}.

\section{Data}
\label{sec:data}
Our work is based on the public data provided by the \atlas \citep{cappellari11a}\footnote{\url{http://www-astro.physics.ox.ac.uk/atlas3d/}} and MATLAS \citep{duc15,bil20}\footnote{\url{http://obas-matlas.u-strasbg.fr}} surveys. The \atlas survey   targets nearby ($<42$\,Mpc) massive ($M_K<-21.5$\,mag) ETGs (galaxies lacking spiral arms) and is volume complete. The survey strives to collect all possible information about its targets. All of the data we used in this paper was taken from \atlas, except for those on the photometric irregularities (tidal features, irregular outer isophotes) and dust lanes; that information was taken from the catalog of \bilmat extracted from the MATLAS survey. The MATLAS survey took very deep ($\sim 28.5$\,mag\,arcsec$^{-2}$) wide-field ($1\times 1\degr$) optical (the $u^*$, $g^\prime$, $r^\prime$, and $i^\prime$ bands) images of all \atlas ETGs,  except for those in the Virgo Cluster, with the 3.6\,m Canada-France-Hawaii Telescope, employing observing and data processing strategies optimized for detecting large-scale low-surface-brightness structures.{ We used the MATLAS sample \citep{bil20} where we excluded the two galaxies without mass measurement (PGC\,056772 and PGC\,061468). In total, the sample considered here contains 175 objects. The list is available as online material.} Importantly, the MATLAS survey avoided the cluster environments, and therefore our results pertain only to galaxies in low- to medium-density environments such as  galaxy groups. 

In this paper, we characterize the mass of a galaxy by \mjam, adopted from \citet{cappellari13a}. It is  defined as $\mjam = (M/L)_eL$, where $L$ is the total luminosity of the galaxy and $(M/L)_e$ is the effective dynamical mass-to-light ratio of the galaxy within the galactocentric distance of one $\reff$  derived using Jeans Anisotropic Modeling \citep{cappellari08}. \citet{cappellari13a} showed that the median fraction of dark matter mass within a sphere of radius \reff is 13\%; thus the stellar mass of the galaxy can be estimated as 0.87\,\mjam, or, in the logarithmic scale, the stellar mass is  lower by 0.06\,dex.  The 15th, 50th, and 85th percentiles of the distribution of $\log(\mjam/M_\sun)$ in our sample are 10.2, 10.6, and 11.0, respectively. 

The environmental density of a galaxy in this paper is quantified by the \rhoio parameter of \citet{cappellari11b}, defined as the mean density of galaxies inside a sphere that is centered on the target galaxy and  contains the ten nearest neighbors. The 15th, 50th, and 85th, percentiles of the distribution of $\log(\rhoio/\mathrm{Mpc}^{-3})$ in our sample are -2.2, -1.6, and -1.0, respectively. The highest \rhoio reaches $1.66\,$Mpc$^{-3}$ for NGC\,4623.

In our work, we quantify the rotational support through the parameter 
\begin{equation}
    \rotsup = \frac{\lambda}{\sqrt{\epsilon}},
\end{equation}
introduced by \cite{emsellem11}. Here $\lambda$ quantifies the degree of ordered rotation of the galaxy, as it is observed projected on the sky, and $\epsilon$ the apparent ellipticity, see \citet{emsellem11} for details. Both of the quantities are measured within one \reff from the center of the galaxy. The values were taken  from \citet{emsellem11}.  The 15th, 50th, and 85th percentiles of the distribution of \rotsup in our sample are 0.32, 0.77, and 1.0, respectively. The distribution of our sample in the space of galaxy mass - environmental density - rotational support is shown in \fig{mrs}. 
{The used values of \mjam, \rhoio and \rotsup are available in the online material.}

\section{Traces of past mergers}
\label{sec:indicators}
\begin{figure*}[ht!]
        \centering
        \includegraphics[width=17cm]{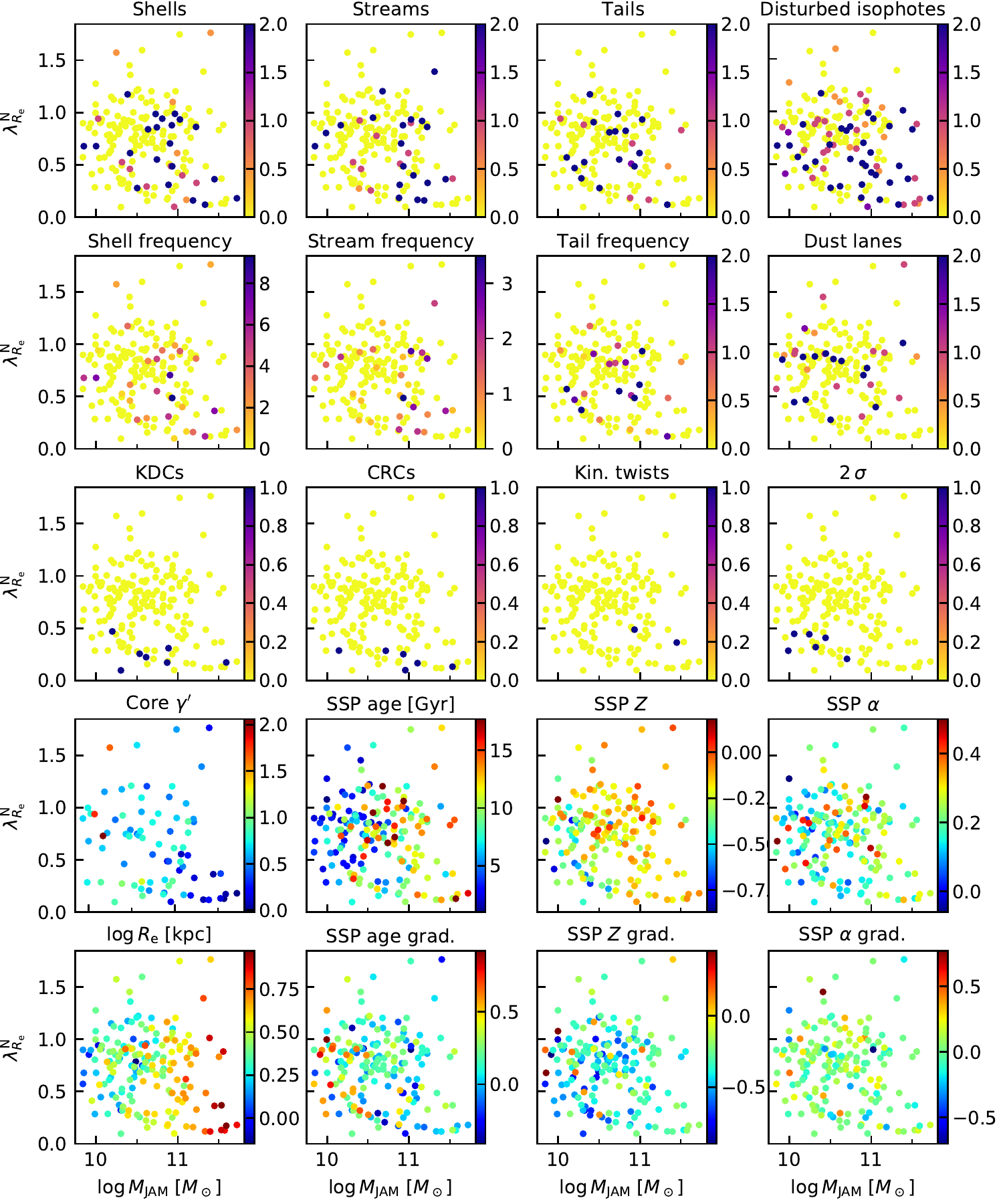}
        \caption{Investigated merger-sensitive parameters as functions of the rotational support (vertical axis in each tile) and the \mjam mass of the galaxy (horizontal axis in each tile). Each point represents one galaxy. The colors of the points indicate the values of the quantities stated in the tiles of the figures, coded according  to the  color bars given to the right of each tile. For shells, streams, tails, disturbed isophotes, and dust lanes, the values of zero, one, or two indicate that the feature is not present, is likely present, or is certainly present, respectively. The frequency of shells, streams, and tails indicates the number of these features in the galaxy. For KDCs, CRCs, KTs, and the $2\,\sigma$ features, the values of zero or one mean that the kinematic substructure is not or is present in the galaxy, respectively. } 
        \label{fig:2d-allMI}
\end{figure*}

{In this section, we describe the merger-sensitive parameters that we exploit later in the paper. It shall be kept in mind, as we discuss in \sect{interp}, that none of them are totally reliable -- each can be influenced by other processes than  mergers.} Figure~\ref{fig:2d-allMI} shows the distribution of {these merger indicators  in our galaxy sample}  in the space of rotational support versus mass of the galaxy. Figure~\ref{fig:2d-allrho} shows the same but the mass is replaced by the environmental density. In \app{simplecorr}, we inspect the correlations of mass, environmental density and the rotational support with the different merger-sensitive parameters. In that appendix we also compare the values of these parameters between FRs and SRs. The used values of the merger-sensitive parameters listed in the following subsections are available in the online material. It summarizes also the corresponding data sources.

\subsection{Tidal disturbances}
Galaxy mergers leave morphological irregularities in the images of the galaxies. They can remain observable for several gigayears before the system relaxes or before the existing tidal features are destroyed by another merger. The survivability depends, for example, on the mass ratio of the merged galaxies, their orbital configuration, their morphological types, and the density of the environment. {An attempt to quantify the lifetimes of tidal features was made by \cite{mancillas19}, and we discuss the issue further in \sect{survival}.} A catalog of various types of tidal disturbances was presented in \bilmat. It was constructed by a visual inspection of deep optical images of the galaxies in the MATLAS survey. We consider the following types of tidal disturbances.

\textit{Shells.} These are arc-like sharp photometric jumps whose center of curvature most often coincides with the core of the galaxy. The conditions of their formation are still under debate but generally radial mergers are preferred \citep{hendel15,amorisco15}, but see, e.g., the simulation by \citealp{ebrova20}). While in older works they were attributed to minor mergers (see the review in \citealp{bildiz}), recent studies have suggested that shells often form also in intermediate  or even major mergers \citep{pop18,kadofong18}. 

\textit{Streams.} They appear as thin, long structures, sometimes  going through the center of the host galaxy, but most of time wrapping around it. Generally, they can be attributed to non-radial minor mergers \citep{hendel15,amorisco15} while even major mergers involving disks can produce similar features \citep{wang12}. In that case, typically there are several streams, or they are accompanied by other tidal disturbances. 

\textit{Tails.} Tails are morphologically similar to streams but they have a higher  thickness \citep{sola22}, up to the size of the host galaxy, and they are always attached to it.
{The presence of a tail or tails in a galaxy indicates that the  galaxy is just in the process of disruption by a massive neighbor, or that the galaxy is a remnant a of past major merger.}

\textit{Disturbed outer isophotes.} In a relaxed ETG, outer isophotes are axially symmetric about the minor and major photometric axes. An interaction breaks this symmetry, making the isophotes lopsided or irregular. In many mergers, the irregularities would take the form of some of the tidal features above. However, once the tidal features become too old, they lose their distinct form and they appear just as irregular isophotes. Disturbed outer isophotes can also signify an ongoing or distant interaction.  

Shells, streams, and tails are collectively called the ``tidal features''. In this paper we call the union of tidal features and disturbed outer isophotes  ``tidal disturbances''. To quantify whether a given tidal disturbance is present in the galaxy, we adopted the rating from \bilmat. They list, for each of the tidal disturbances, its rating expressing the visual prominence of the given feature in the galaxy. A value of zero indicates that the feature is not present, a value of one that it is possibly present, and a value of two that it is certainly present in the galaxy. We also took the frequency of each type of tidal feature in a galaxy, which indicates how many tidal features of the given type are present in the galaxy, from \bilmat. Several types of tidal disturbances can be present in a galaxy at the same time.  Whenever we wanted to quantify whether any tidal disturbance is present in the galaxy, we took the maximum of the ratings of the individual types of tidal disturbances.

\subsection{Dust lanes}
Dust is common in spiral galaxies but less so in ETGs. This suggests that dust gets into ETGs by the accretion of spirals. Indeed, ETGs with tidal disturbances often show prominent dust lanes. We took the information about the occurrence of  dust lanes from the catalog by \bilmat, who rate the presence of dust lanes similarly to tidal disturbances: ranging from zero for no dust lanes to two for prominent dust lanes.

\subsection{Kinematic substructures}
The majority of ETGs shows regular ordered rotation around the photometric minor axis \citep{krajnovic11,emsellem11}. The are, however, exceptions\footnote{We follow here the classification by \citet{krajnovic11}.}:
``counter-rotating cores'' (CRCs) rotate around the same axis as the rest of the galaxy below $\sim1\,\reff$ but in the opposite sense; ``Kinematically distinct cores'' (KDCs) do not share the rotation axis with the rest of the galaxy; and finally ``kinematic twists'' (KTs) are characterized by a gradual change of the kinematic position angle through the volume of the galaxy below $\sim 1\reff$. The so-called double $\sigma$ (or $2\sigma$) galaxies show, in their maps of velocity dispersion,  two peaks that lie on the major  photometric axis of the galaxy and their separation is greater than half of the effective radius of the galaxy. The peaks likely arise because there are two counter-rotating stellar disks in these galaxies \citep{krajnovic11}. 
Together, we call KDCs, CRCs, KTs, and the $2\sigma$ features ``kinematic substructures''. We quantified the presence of a given type of  kinematic substructures by a value of one, and its absence by zero. {Kinematic substructures can form in mergers, even if alternative formation channels  exist  \citep{ebrova20b,young20}.}


\subsection{Effective radius}\label{sec:re}
At the redshifts over about 1-2, passive galaxies of any fixed mass are around five times smaller  than the local ETGs  \citep{daddi05, trujillo06, vandokkum09}. The growth of the effective radius is usually attributed to galaxy mergers: the initial potential energy of the two separated galaxies is transferred into the internal energy of the merger remnant. It was shown that a larger expansion is achieved if a given mass is accreted through minor mergers than through major mergers \citep{naab09}. The number of mergers that caused the expansion is expected to be relatively low.  For example, \citet{trujillo11} calculated that the growth of radius since $z\approx0.8$ can be achieved by $3\pm1$ mergers of the mass ratio of 1:3 or by $8\pm2$ mergers of the mass ratio of 1:10. Some local ETGs might even have coincidentally avoided any substantial mergers since $z=1-2$, as their properties suggest \citep{martin18,beasley18,yildirim17}. Therefore, if SRs and FRs had different merger histories, we expect a difference in their effective radii at a fixed stellar mass. For the effective radii, we took the $R_\mathrm{e}^\mathrm{maj}$ parameters from \citet{cappellari13a}. They were obtained as the major axes of multi-Gaussian fits of the galaxies.  They were published in the angular scale. We converted them into kiloparsecs assuming the distances of the galaxies given in  \citet{cappellari11a}.

\subsection{Inner photometric profile}
The photometric profiles of ETGs are generally described well by a S\'ersic profile. However, high-resolution images, mainly those taken by the Hubble Space Telescope, reveal that there are deviations from it in the centers of ETGs (\citealp{lauer95}, see \citealp{graham13} for a review). Some inner photometric profiles  follow a power law, creating an excess of light with respect to the S\'ersic profile fitted to a wider radial range. These are called the cusps. On the contrary, the derivative of the photometric profile can nearly reach zero in galaxy centers, such that there is deficit of light with respect to a S\'ersic profile. Such features are called the cores. Cored profiles are found typically in bright galaxies but, interestingly, stellar densities are higher in the low-luminosity galaxies with cuspy profiles \citep{faber97}.  The formation of cores is usually explained in the following way. After a merger of two galaxies, there are two supermassive black holes in the merger remnant galaxy. They experience dynamical friction and sink toward the center of the galaxy. Their potential energy transforms into the kinetic energy of stars, which are subsequently ejected from the center of the galaxy, so that a central core forms \citep[e.g.,][]{faber97,miloslavljevic02}. However, gas-rich mergers tend to form new stars in the center, so that a remnant of a gas-rich merger can eventually become more cuspy than the progenitor galaxies \citep[e.g.,][]{hopkins09b,hahn13}. We use here the parameter $\gamma^\prime$ tabulated by \citet{krajnovic20} to quantify the shape of the  inner photometric profile. It comes from fits by the Nuker law. Higher values of $\gamma^\prime$ indicate more cuspy profiles.

\subsection{Chemical composition}

The stellar metallicity of galaxies, $[Z/H]$, is known to increase with the mass of the galaxy and the density of its environment \citep{peng10,maiolino19}. In addition, star-forming galaxies tend to have lower metallicities than passive galaxies of the same mass (e.g., \citealp{thomas10,peng15,maiolino19}). Similarly, the abundance of $\alpha$-elements, $[\alpha/\mathrm{Fe}]$, is generally higher for more massive ETGs \citep{thomas10,mcdermid15} while the $\alpha$-abundance is lower in spirals \citep{proctor02,peletier07, ganda07,scott17,parikh21}. The chemical composition of galaxies can thus be expected to depend on their merger history. A high value of  $\alpha$-enhancement signifies rapid star formation (e.g., \citealp{thomas99}). The metallicity of ETGs grows toward their centers \citep[e.g., ][]{spolaor09, kuntschner10, scott13,li18}, while the $\alpha$-elements show nearly no gradient \citep{rowlands18}.

Dry major mergers tend to flatten the preexisting metallicity gradients simply because of the mixing of stellar populations \citep{dimatteo09}. On the other hand, minor mergers tend to steepen metallicity gradients, since minor mergers tend to deposit the low-metallicity material of the small galaxies at high radii  \citep{amorisco17,karademir19}. Simulations indicate that gas-rich major mergers decrease both the central metallicity and the metallicity gradient, since tidal torques bring the metal-poor gas in galaxy centers {(\citealp{kobayashi04,montuori10,perez11,navarrogonzalez13,hirschmann15,taylor17}, but see \citealp{hopkins09})}, as also indicated observationally \citep{rupke10,maiolino19}. This means that the central starburst is not powerful enough to increase the central metallicity.  On the other hand, the central starburst is able to increase the central enrichment by alpha elements, which is hinted at both by simulations \citep{perez11} and observations \citep{pierce05}. The values of  metallicity and $\alpha$-abundance we used here are the measurements by \citet{mcdermid15} within one effective radius, based on single-stellar-population (SSP) models. We checked that our results did not change if we used the star-formation-history ages published in the same work, which are based on full spectral fitting.  Stellar population gradients were taken from \citet{krajnovic20} and again are derived from SSP models. A higher value of the metallicity gradient means a less negative, and therefore a flatter, gradient.
Chemical properties derived from SSP models are known to preferentially trace the properties of the old stellar populations present in the galaxy \citep{serra07}.

Early-type galaxies can also form stars in situ during the second phase of assembly. There are indeed observational indications that star formation can be renewed in ETGs after some period of passivity \citep{thomas10,bil20,yildiz20}.  Simulations show that the stars formed in situ can constitute a substantial fraction of the stars of the resulting ETG too \citep[e.g.,][]{penoyre17,lagos18}. One can argue that these in situ stars are formed by freshly accreted gas from the intergalactic medium, and therefore the new stars decrease the total metallicity of the galaxy. However, here we rely on the hypothesis that the stars formed in the second phase of ETG assembly do not cause a decrease in the metallicity of the galaxy. The first reason for this is the fact that the rejuvenated ETGs in the sample of \citet{thomas10}, which are about 2\,Gyr old, show the same or rather somewhat higher metallicities than the standard, old ETGs of the same velocity dispersion. Second, according to the calculations of \citet{trussler20}, a typical star-forming galaxy transforms into a typical passive galaxy with the typical metallicity relatively quickly, on the characteristic time scale of around 2\,Gyr. An ETG that experienced a temporal burst of star formation would likely turn back to the standard mass-metallicity relation even faster.

\subsection{Stellar ages}
Ages of galaxies generally grow toward higher masses, earlier morphologies, and denser environments \citep[e.g.,][]{thomas10,mcdermid15,gonzalez17}.  Stellar populations inside of galaxies generally become older toward the centers of galaxies \citep{spolaor09,kuntschner10,li18}. Dry mergers lead to population mixing, while wet mergers can give rise to completely new stars in the merger remnant. Both dry and wet major mergers tend to flatten age gradients, while minor mergers, which deposit material far from the center, can increase them.

Here we consider the stellar ages, expressed in gigayears, measured by the SSP method from line indices. The measurements were taken from \citet{mcdermid15} and pertain to the stellar population in one effective radius. The ages are defined as the time passed since half of the stellar population was formed. The SSP age is know to be biased toward the younger stellar populations in the galaxy \citep{serra07}. Uncertainties in age measurements grow quickly with the age. The age gradients we use here were taken from \citet{krajnovic20}.

\begin{table*}
	\caption{Predictability of the investigated merger-sensitive parameters from the rotational support, \rotsup. }
	\label{tab:rottarall}
	\centering
	\begin{tabular}{l|ll|ll|ll|ll|ll}
		\hline\hline
 & \multicolumn{2}{c|}{All}& \multicolumn{2}{c|}{$\log \mjam<11$} & \multicolumn{2}{c}{$\log \mjam\geq11$} & \multicolumn{2}{c|}{$\log \rhoio<-2$} & \multicolumn{2}{c}{$\log \rhoio\geq-2$}\\
Parameter & sign & $p$ [\%] & sign & $p$ [\%] & sign & $p$ [\%] & sign & $p$ [\%] & sign & $p$ [\%] \\
\hline
\input{tables_maintext/Ftest_rotmes_tar-all.txt}
\hline
	\end{tabular}
	\tablefoot{The sign sub-column means the sign of the correlation, while the $p$ subcolumn shows the probability that there is actually no correlation.}
\end{table*}

\section{Survivability of tidal disturbances}
\label{sec:survival}
Estimating the distribution of the lifetimes of tidal disturbances is a difficult task. {Once a tidal disturbance is induced by a merger, it will have a morphology of a tidal feature, that is shells, streams, tails, or their combination.  \citet{mancillas19} did the first step toward the estimation of the lifetimes of different types of tidal features by inspecting one zoom-in hydrodynamic simulation of a galaxy in a \lcdm Universe. With this approach, they included the destruction of tidal features by subsequent galaxy interaction. Of specific interest for this work,  they investigated whether the features would be detected by the MATLAS survey. The tails were found to have the shortest lifetimes, namely below 1\,Gyr. The longest lifetimes were for the shells, which lived for up to 4\,Gyr. As the tidal disturbance evolves, it will eventually lose its characteristic morphology of a shell, tail, or a stream. After that it will be observable as disturbed outer isophotes.
\citet{mancillas19}  did not discuss the lifetimes of disturbed outer isophotes.  In our paper, we assume that it will take at least 4\,Gyr before any tidal disturbance disappears completely, regardless of its initial morphology. }

We make an analytic upper estimate of the survivability of tidal disturbances in our galaxies.   We ignore the fact that tidal features can be destroyed by new galaxy interactions \citep{mancillas19}. Then tidal features disappear because of the phase mixing mechanism (e.g., \citealp{mo10}), and therefore 
the characteristic time scale of the dissolving of the tidal disturbances is the orbital period of a star at the position of the tidal disturbance.

We measured the sizes of the irregularities in the galaxies with disturbed isophotes in MATLAS images and found that they are usually ten times larger than the effective radii. Therefore, we approximated the radii of the outermost isophotes as $R_\mathrm{OI} = 10\reff$.  We estimated the orbital period at  $R_\mathrm{OI}$ as
\begin{equation}
    T_\mathrm{orb} = \frac{2\pi\,R_\mathrm{OI}}{\sqrt{-a\,R_\mathrm{OI}}},
\end{equation}
where $a$ stands for the gravitational acceleration at $R_\mathrm{OI}$. That was determined from the empirical radial acceleration relation {(\citealp{lelli17}, see also \citealp{milg83})} and the total stellar mass of the galaxy, which is  $0.87\mjam$. The results of \citet{bil19} indicate that the gravitational acceleration in ETGs might be even stronger than expected from the radial acceleration relation, in line with our effort to give an upper estimate for the survivability time. We assumed that tidal disturbances live at most for $10T_\mathrm{orb}$ after the merger. We then obtained that mean maximal survival time for our galaxy sample is $9\pm3$\,Gyr, nearly independently of galaxy mass. Therefore, we adopted 9\,Gyr as the maximum lifetime of tidal disturbances. This suggests that most of the tidal disturbances must have been formed before $z=1.4$ {, which means during the second phase of ETG assembly.}

\section{Correlations of merger-sensitive parameters with \rotsup at a fixed mass and environmental density}
\label{sec:corr}
Here we describe our  methodology. As we explained in \sect{intro}, we assume that galaxies gradually {lose} rotational support primarily by mergers.  {At the present cosmic epoch, different galaxies are observed in different stages of the transformation.} Some of the galaxies would still be classified as FRs according to their value of \rotsup, while others would already be classified as SRs.

Later in this paper, we deduce when these mergers were primarily happening and what type they typically were (i.e., wet or dry, and minor or major). This is found by determining how the merger-sensitive parameters change with the rotational support. However, it is necessary to take into account that the merger-sensitive parameters would correlate with the rotational support even if mergers did not influence the rotational support at all, for the reasons described below.

The typical number of mergers that a galaxy experiences is expected to be an increasing function of the current mass of the galaxy, because more massive galaxies have stronger gravitational fields and are more extended, and therefore dynamical friction can be effective up to larger distances. Therefore, their effective cross-section for mergers is expected to be larger than for galaxies of a lower mass. Next, the typical number of mergers that a galaxy experiences is expected to increase with the density of its environment because there are more galaxies in the vicinity available for merging\footnote{Actually, in galaxy clusters mergers can be prevented because of high relative velocities of the galaxies \citep{ghigna98,mihos03}, but we do not have cluster galaxies in our sample.}. The typical amount of stellar mass formed in situ is expected to be an increasing function of the current galaxy mass, again because a more massive galaxy is able to attract the accreted gas from a larger distance than a less massive galaxy.  The typical amount of stellar mass formed in situ is expected to decrease with the density of environment because in these environments the effects of starvation, strangulation, ram pressure striping, and shock heating take place.
As a result, the correlations of the rotational support with mass and environmental density would induce correlations of the rotational support with merger-sensitive parameters, even if the mergers did not influence the rotational support.  In statistical literature, this is called the confounding effect.   In \app{simplecorr}, we check that the rotational support correlates significantly with galaxy mass and partly also with environmental density, and thus the necessary condition for the confounding effect to happen is satisfied. The way to eliminate the confounding effect is to inspect the correlations of the quantities of interest at fixed values of the confounding quantities. In our case, this means inspecting the correlations  of  merger-sensitive  parameters with the rotational support for galaxies of a fixed mass and environmental density.

The main method of this paper is based on the assumption that if we consider galaxies of a fixed mass and environmental density, then mergers typically have a larger importance for galaxies with a lower rotational support than for those with a higher rotational support. The 
word ``typical'' is important here because, for example, some mergers with a specific orbital configuration and gas content can spin up the galaxy \citep{dimatteo09b,qu10,naab14,penoyre17}, or because the rotational support depends also on the orientation of the galaxy with respect to the direction to the observer.  The trends of the importance of mergers and rotational support are thus expected to be valid only in the statistical sense. 

Ideally, one should sort the galaxy sample into narrow bins of mass and environmental density, and inspect in them the correlations between the rotational support and the merger-sensitive parameters. This is not possible for our sample because it is too small. We thus used another method that is able to eliminate or mitigate the confounding effect, namely the multilinear regression.  
 
In particular, we made multilinear fits of each merger-sensitive parameter as a function of galaxy mass, environmental density, and rotational support:
\begin{equation}
    [\textrm{parameter}] = b+a_M\log\mjam +a_\rho \log\rhoio +a_\mathrm{KS}\rotsup.
    \label{eq:fit}
\end{equation}
The fitted coefficients are listed in \app{fits}. The coefficient at the rotational support in \equ{fit}, $a_\mathrm{KS}$, can be used for assessing the correlation between the rotational support and the investigated merger-sensitive parameter without the confounding effects of mass and environmental density, provided that the relation between the quantities is well described by \equ{fit}. In this method, one essentially substitutes the actual value of the merger-sensitive parameter of a given galaxy by the value  predicted by the formula in \equ{fit} on the basis of the \mjam, \rhoio, and \rotsup of the galaxy.   It then remains to be ascertained whether the correlation of the merger-sensitive parameter and the rotational support at a fixed mass and environmental density is statistically significant. To this end, we made another multilinear fit for the merger-sensitive parameter, but only as a function of the galaxy mass and environmental density: 
\begin{equation}
    [\textrm{parameter}] = \tilde{b}+\tilde{a}_M\log\mjam +\tilde{a}_\rho \log\rhoio.
    \label{eq:fit2}
\end{equation}
The statistical significance of the correlation of the given merger-sensitive parameter with the rotational support at a fixed mass and environmental density was evaluated through an $F$-test applied to the residuals of the fitting by \equ{fit} and \equ{fit2}. The $F$-test indicated whether the addition of the rotational support among the independent quantities improved the quality of the fit significantly. In other words, we found in this way whether the rotational support provides any information on the given merger-sensitive parameter if the mass and environmental density of the galaxy are already known. We accepted the significance level of 5\% for the $F$-test. This means that the probability that the test will indicate a significant correlation between the quantities, if the quantities actually do not correlate, is 5\%.

The results are presented in \tab{rottarall}. The second main column represents the results for our whole galaxy sample. The third and fourth main columns represent the results of our test when applied only to galaxies that have an  $\mjam$ mass lower or a higher than $10^{11}\msun$, respectively. This limit is motivated by the theoretical expectations described in  \sect{intro}.  Similarly, in the last two main columns of \tab{rottarall}, we divided our galaxy sample by the environmental density at $\rhoio=-2$. We are not aware of any past works that would show that galaxy properties abruptly change at a particular value of environmental density (e.g., separating filaments from galaxy groups).  We chose our  separating value on the basis of \fig{2d-allrho} because, at this value, the incidences of tidal disturbances, dust lanes, and KDCs seem to change abruptly. We explored whether the results changed if we used other quantifications of the environmental density instead of $\log \rhoio$, namely the parameters $\Sigma_3$ and $\nu_{10}$ of \citet{cappellari11b}, and we found no substantial difference.

In   \tab{rottarall}, the sign subcolumns  specify whether the given merger indicator typically increases or decreases toward a higher value of the rotational support for galaxies at a fixed mass and environmental density. This is  the sign of the fitted parameter $a_{KS}$ in \equ{fit}. The $p$ subcolumns  of the table give the $p$-value of the $F$-test -- that is the probability that the inclusion of the rotational support in the multilinear fit of the given merger-sensitive parameter actually does not improve the fit substantially, even if the $F$-test indicates so. Thus, the lower the $p$-value, the more significant the correlation of the rotational support with the given parameter.  The emphasized values signalize that  including the \rotsup term in the multilinear model improves the fit of the merger indicators at the  significance level of 5\% (bold font) or 1\% (bold and larger font).

The table tells us that the rotational support indeed helps us to predict some of the  merger-sensitive parameters, even if we already know the galaxy mass and environmental density. Namely, in the sample as whole, we find that for galaxies at a fixed mass and environment density, the rotational support correlates negatively with the incidence of disturbed isophotes, as well as with the incidences of KDCs, CRCs, and the $2\,\sigma$ features. On the other hand, the correlation is positive with a regular kinematic appearance. Next, we find that for galaxies at a fixed mass and environment density, the rotational support correlates positively with the values of metallicity and metallicity gradients (i.e., the galaxies of a higher rotational support usually have flatter metallicity gradients, since the gradients are usually negative). The most significant correlations of the rotational support are those with metallicity and with the presence of a KDC. Again, we point out that not detecting a significant correlation  does not imply that the correlation does not exist. It can just be too weak to be detected in the current data.

Moving to the group of galaxies with $\log\mjam\geq11$, for galaxies at a fixed mass and environmental density, we detected only one significant correlation: galaxies of low rotational support are more likely to  have a relatively high abundance  of $\alpha$-elements.  On the contrary, for the less massive part of our sample, we basically obtained  the same  correlations as for the whole sample. The exceptions are the correlations of the rotational support with disturbed isophotes and metallicity gradients,  which slightly miss our significance threshold.

In the low-density subsample ($\log\,\rhoio<-2$), we noted two substantial changes with respect to the full sample. First of all, at a fixed mass and environmental density, the central photometric cuspiness significantly decreases toward higher values of  rotational support. In the high-density part of the sample, the trend is opposite, yet {insignificant:  the cuspiness tends to increase with increasing rotational support.} The second substantial difference is the loss of a significant correlation between the rotational support and the incidence of disturbed isophotes with respect to the complete galaxy sample.  In the high-density subsample, we detect most of the significant correlations detected in the full  sample. The first exception is the one between the rotational support and regular rotation, whose significance misses somewhat our limit. The second change is the lack of correlation of the rotational support with the metallicity gradient.

It is known that metallicity correlates strongly with the stellar velocity dispersion of galaxies \citep{thomas10,mcdermid15}. We thus tested if the strong correlation of the rotational support and metallicity disappeared if we fixed not only the mass and environmental density, but also the stellar velocity dispersion taken from \citet{cappellari13a}, namely the velocity dispersion within the isophote counting half of the galaxy luminosity. We made a multilinear fitting and an $F$-test as before. The sense  of the correlation of the rotational support and metallicity remained as in  \tab{rottarall} and the significance did not change much.

It is useful to define the cleaned merger-sensitive parameters as:
\begin{equation}
    [\textrm{cleaned parameter}] = [\textrm{parameter}] - (b+a_M\log\mjam +a_\rho \log\rhoio),
\end{equation}
where the values of $b$, $a_M$, and $a_\rho$ were obtained by fitting \equ{fit} to the data. The cleaning allowed us to compare the values of the merger-sensitive parameters of galaxies without being affected by the confounding effects of mass and environmental density. {The values of the cleaned parameters are provided in the online material.} Figure~\ref{fig:cleaned} shows our galaxy sample in the space of the cleaned parameters versus the rotational support for all investigated merger-sensitive parameters.

\begin{table*}
\caption{Summary of the constraints on the typical mergers that {caused the decrease} in the ordered rotation of ETGs.}
\label{tab:constraints}
\begin{tabular}{l|cccc}
\hline\hline
Observation  & \makecell[c]{Time since \\ transformation  {[Gyr]}} &   \makecell[c]{Indicates\\ minor mergers?} &   \makecell[c]{Indicates\\ wet mergers?} &   \makecell[c]{From our\\ sample?} \\ \hline
Tidal disturbances & >4 & n & n & Y \\
KDCs & <12? & n & n & Y\\
KDCs + tidal disturbances & >4  & n & n & Y\\
\makecell[l]{Stellar age + $\alpha$-elements \\ \hspace{1em} (massive ETGs)} & >10 & n & Y & Y \\
Galaxy counts at high $z$ & >8-9  & n & n & n \\
\makecell[l]{Ellipticities of massive quenched \\ \hspace{1em} galaxies  at high $z$} & >10  & n & n & n \\
\makecell[l]{S\'ersic indices of massive \\ \hspace{1em} quenched  galaxies  at high $z$} & >12  & n & n & n \\
\makecell[l]{Luminosities, surface brightnesses \\\hspace{1em} and effective radii of  brightest\\ \hspace{1em} cluster galaxies  at high $z$} & >10  & n & n & n \\
Inner phot. profiles
\\\hspace{1em}(low environmental density) & >10 & n & Y  & Y\\
Dust lanes & >1  & n & n & Y\\
Effective radii & >12  & n & Y & Y\\ 
Metallicity gradients & & Y & n & Y\\ \hline
\end{tabular}
\tablefoot{Description of the columns. {\bf Observation}: The observable  providing constraints on the properties of mergers that caused the kinematic transformation. The order is as discussed in the text. {\bf Time since transformation}: The typical time of the kinematic transformation implied by the given observable. A question mark indicates a speculative estimate. {\bf Indicates minor mergers?} and {\bf Indicates wet mergers?}: The columns show whether the given observable implies that the mergers causing the kinematic transformation were minor or wet, respectively.  The symbols ``Y'' and ``n'' indicate yes and no, respectively. None of the observing facts considered in this study indicate that the mergers were major or dry.  {\bf From our sample?}: Indicates whether the given observational data are those analyzed in this paper or come from literature, with the meaning of symbols as before. The data from literature sources might be less telling for the galaxy sample investigated in this paper. }
\end{table*}

\section{Deducing the mechanism of decreasing the rotational support of ETGs from observations}
\label{sec:interp}
Cosmological simulations of  galaxy formation predict that ETGs decrease the level of their rotational support with time and that mergers play a substantial role in that (\sect{intro}), which we take as a basic assumption in this paper. 
In this section, we exploit the correlations of the rotational support with merger-sensitive parameters at a fixed galaxy mass and environment density, as found in \sect{corr},   to learn more about this transformation process from observations. We make use the fact that the different merger-sensitive parameters have different lifetimes and are sensitive to different types of mergers.
We add further constraints based on findings from the literature. Relying on a constraint provided by a single  parameter can be misleading since the merger-sensitive parameters can be influenced by other mechanisms than by mergers. It turned out to be advantageous that we are working here with many merger-sensitive parameters, because this allowed us to confirm some of our conclusions by several independent pieces of evidence. This allowed us to reduce the probability of a mistake caused by measurement errors or by misinterpreting the data. The individual pieces of evidence we found are summarized in \tab{constraints}.  In the following paragraphs, we aim to find, for every constraint, not only an interpretation based on mergers, but also on other mechanisms. It turns out that every observation requires another alternative mechanism, which is often speculative or not underpinned by quantitative models. On the other hand, mergers appear as a more solid and universal explanation of all observations.

In \sect{corr} we found that, at a fixed mass and environmental density, the metallicity of a galaxy generally decreases with the decreasing rotational support. This is the most significant correlation we found. {A similar result was found by \citet{bernardi19}. }This observational finding agrees with our assumption that the rotational support decreases because of mergers of smaller galaxies, because metallicity generally decreases with galaxy mass.  Nevertheless, as with the other merger indicators, the metallicity of a galaxy is not influenced only by mergers, but also by other mechanisms or factors. These include the balance between the energy of gas outflow and the depth of the potential well of the galaxy \citep{pipino10}, the star formation history of the galaxy and its initial mass function \citep[e.g.,][]{matteucci14}, a removal of gas cold gas from the galaxy, or an interruption of the inflows of intergalactic gas to the galaxy \citep{peng15,trussler20}. One can speculate that during the monolithic collapse phase, the formation of galaxies with a low rotational support is faster, and thus some of the alternative mechanisms are more effective than when the galaxy has a lower rotational support.

We detected that, at a fixed mass and environmental density, galaxies with a lower rotational support have a significantly higher incidence of disturbed outer isophotes. With a decreasing rotational support, the incidences of the different types of tidal features increase too, even if these correlations are not statistically significant.  All types of tidal disturbances become particularly rare above the rotational support of 1.0, as Figs.~\ref{fig:2d-allMI} and \ref{fig:cleaned} show. This indicates that even at a fixed mass and environmental density, the galaxies with a lower rotational support experience more mergers. The tidal disturbances had to be {younger than the maximum lifetime} of tidal disturbances, which we estimated in \sect{survival} as 9\,Gyr. On the other hand, it is important to note that even among the galaxies with the lowest rotational support, most galaxies do not show tidal disturbances. {Indeed, out} of  the 25 SRs in our sample (i.e., their $\rotsup<0.31$), only eight have their cleaned indicator of disturbed isophotes higher than zero. Thus, if we ask for the typical {cosmic epoch} when mergers were decreasing the rotational support, it had to be before the minimum lifetime of {disturbed isophotes}, that is, before 4\,Gyr (\sect{survival}). There is no guarantee that the observed tidal disturbances were caused by the mergers that contributed most to the decreasing the rotational support. The findings suggest the possibility that galaxies with a lower rotational support live in environments where galaxy interactions are more common compared to those that have the same mass and environment density but a higher rotational support. As for the alternative explanations that do not rely on mergers, tidal disturbances could have been caused by non-merging galaxy flybys, which are expected to be common by alternative theories of gravity \citep{bil18,bil19d}. If the rotational support of ETGs was set by internal processes, then we have to speculate that galaxies with a lower rotational support experience more non-merging flybys than galaxies with a higher rotational support. Tidal disturbances might also  dissolve faster if the gravitational potential of the galaxy is flatter because the stars would not continue orbiting in the plane of collision, as in a spherical potential. It has indeed been found that FRs are oblate ellipsoids while SRs are closer to being spherical \citep{cappellari07,weijmans14,foster17, li18b}.

Next, we detected that, at a fixed mass and environmental density, galaxies with a lower rotational support possess KDCs, CRCs, and the $2\,\sigma$ features more often. This suggests that these kinematic substructures formed in the same mergers that were responsible for decreasing the rotational support of the galaxies. This would likely mean that the decrease in rotational support did not happen during the era when galaxies were mostly gaseous, that is, say before the redshift of three (more than 12\,Gyr ago), because a streaming of gas flows in opposing directions in the same system is difficult. This maximum age estimate agrees well with the result of \citet{ebrova20b}, who found that the oldest KDC in the cosmological hydrodynamical simulation Illustris is 11.4\,Gyr old at the current epoch.  This observational  constraint is, however, not very strong, since only relatively few galaxies have a KDC or CRC -- the typical epoch of decreasing the rotational support might have been earlier.  On the other hand, one should keep in mind that kinematic substructures can have a different origin than mergers{, such as non-merging galaxy flybys \citep{hau94,young20}, projection effects \citep{statler91}, and sequential accretion of gas from different cosmic filaments, \citep{algorry14}, see also \citep{ebrova20b}.} {Also, we can speculate that a kinematic substructure might be more difficult to observe if its host galaxy has a strong net rotation, because the substructure would have a lower contrast in the kinematic map}. Yet another speculative possibility is that during the monolithic collapse phase, the angular momentum of the inflowing gas  changed its direction at some point.

An interesting constraint on the typical time of the kinematic transformation appears once we note that, at a fixed galaxy mass and environmental density,  the incidence of KDCs, CRCs, and the $2\,\sigma$ features does not correlate significantly with the presence of tidal disturbances. We found this by the combination of the multilinear fitting and an $F$-test, just as we did when looking for correlations of our merger-sensitive parameters with \rotsup. Taken in another way, out of the 19 galaxies with KDCs, CRCs, or the $2\,\sigma$ features, only five have a positive cleaned parameter of tidal disturbances (that is more prominent tidal disturbances than typical for galaxies of the given mass and environment density), which is $26\pm12\%$ (Poisson error assumed). If we count only the galaxies with KDCs or CRCs, only five of 13  ($38\pm17\%$) have a positive cleaned parameter of tidal disturbances. This is not an excess compared to the whole investigated galaxy sample, where the fraction  of galaxies with a positive cleaned parameter of tidal disturbances is $36\pm5\%$.  This suggests that the transformation of the rotational support typically occurred before the minimum lifetime of the tidal disturbances, which is before 4\,Gyr, {otherwise most galaxies with kinematic substructures would show a positive cleaned parameter of tidal disturbances}.

We did not detect any significant correlation between the rotational support and the stellar age for galaxies at a fixed mass and environmental density. This indicates that the decrease in the rotational support did not typically occur by the means of recent wet mergers that would form a lot of new stars. For the massive part of our galaxy sample ($\log \mjam>11$), we can combine this with additional constraints to get a more complete picture. For them, we detected that at a fixed mass and environmental density, a lower  rotational support implies a higher  abundance of $\alpha$-elements. This is a signature of wet mergers, not dry. For these massive galaxies, we do not expect substantial recent in situ star formation that would be responsible for the high $\alpha$-abundance. This is suggested both by the reconstructed observed star formation histories \citep{mcdermid15}  and cosmological simulations \citep{penoyre17}. To explain these constraints simultaneously, we propose that the mergers that  lowered the rotational support of the massive galaxies had to be wet, but occurring at high redshifts, so that the difference in stellar age with respect to the galaxies with a higher rotational support cannot be measured. The mergers preferably had to happen when ETGs were still forming stars, which is at about $z>2$, (about 10\,Gyr ago). This view is supported by the observations of quiescent galaxies at $z=1.6$ (9.6\,Gyr ago)  by \citet{onodera15}. Their galaxies mostly had stellar masses over $10^{11}\,M_\sun$, and most of these galaxies are SRs \citep{emsellem11}.  \citet{onodera15} found that if their galaxies evolved passively since the redshift of  the observation, their metallicity and $\alpha$-abundance  would agree excellently with that of the ETGs in the nearby Universe. One could perhaps explain  the negative correlation of the  $\alpha$-abundance with rotational support without mergers. For example if the progenitor gas cloud had a higher angular momentum,  it settled more slowly, such that the star formation was less bursty, and that resulted in a stellar population that is poor in $\alpha$-elements.

Another interesting constraint on the time of the lowering of the rotational support by mergers comes from  the evolution of the cosmic spatial density of quiescent galaxies with redshift. Recent results suggest that the cosmic number density of quiescent galaxies with stellar masses over $10^{11}\,M_\sun$ (i.e.those that are mostly SRs in the nearby Universe), have not evolved at least since $z=1.4$ {(9.1\,Gyr ago, \citealp{kawinwanichakij20})}.  This suggests that such galaxies do not experience any substantial mass growth by mergers.  The lighter galaxies continued growing to a later time: the cosmic number density for the quiescent galaxies with stellar masses over $10^{10}\,M_\sun$ has not evolved since $z=1$ (7.8\,Gyr ago). We do not have lighter galaxies in our sample. The mergers that would cause the decrease in the rotational support  bring some material and increase the stellar mass of the galaxies. {Altogether, the observations show that galaxies of the masses investigated in this paper stopped evolving 8-9\,Gyr ago. During the earlier epochs, the galaxies could change their masses either by mergers or in situ star formation. Thus, we can conclude that the mergers happened more than 8-9\,Gyr ago.} One might argue against this constraint: in the late cosmic times, galaxies grow more by minor mergers than by major mergers and the minor  mergers tend to deposit their material at the outskirts of galaxies \citep{amorisco17,karademir19}. This material is difficult to detect because of its low surface brightness, but it can comprise a non-negligible fraction of the stellar mass of the galaxy \citep{huang18}. However, it is then questionable if mergers that deposit material at the outskirts of galaxies can decrease the rotational support of the galaxy that is measured within one effective radius of the galaxy. In addition, the sample of \citet{huang18} consisted of extremely massive galaxies (logarithmic stellar masses over 11.4); for the MATLAS sample, \citet{duc15} found the halos of our galaxies to contain, on average, about 5\% of the total luminosities of the galaxies (confirmed by another method in Sola et al. in prep).

Observations of massive galaxies at high redshifts provide yet further clues as to the time of formation of  massive SRs. It turns out that quenched galaxies at higher redshifts are generally flatter than the local ETGs. The exceptions are the galaxies with logarithmic stellar masses over 11.3 that appear always round in projection, at least to $z=2$ \citep{chang13}. If we assume that galaxies form first as rotating disks and then they transform to pressure-supported spheroids by mergers, then the most massive galaxies, which are usually are SRs, had to be formed before that redshift. 

In addition, the flat quiescent galaxies at high redshifts also have lower S\'esic indices, resembling the spiral galaxies in the local Universe in this regard. The increase in the typical S\'ersic index of quiescent galaxies in time can be attributed to mergers \citep{schweizer82,hilz13}. \citet{lustig21} nevertheless found that the  quiescent galaxies with stellar masses around $10^{11}\,M_\sun$ already had high S\'esic indices around 12\,Gyr ago ($z = 3$), suggesting that the mergers happened before that time. The flat objects with low S\'ersic indices,  which prevail among quiescent galaxies at high redshifts, can be the progenitors of the local FRs, which prevail also in the population of the local ETGs. In addition, it is observed that effective radii, surface brightnesses, and luminosities of brightest cluster galaxies have not evolved at least  for 10\,Gyr ($z=1.8$) \citep{chu21,chu22}.

Another constraint on the time of the transformation of the rotational support is provided by  the inner photometric profiles. For our sample as a whole, we did not detect any trend of the $\gamma^\prime$ parameter with the rotational support at a fixed mass and environmental density. We only found  that the galaxies belonging to the low-density subsample ($\log \rhoio<-2$) with a lower rotational support have more cuspy profiles. This disfavors dry mergers as the cause of the decrease in the rotational support, because such mergers should make the profiles more cored (this was already suggested for low-mass SRs by \citealp{krajnovic20}). This suggests that mergers that decreased the rotational support were gas-rich, such that the increase in cores by the merging of the central black holes was balanced by the formation of new cusps, as explained in \sect{indicators}. This fits in our picture where the decrease in the rotational support happens primarily at high redshifts, when ETGs were still forming stars. Combining the lack of correlation between the rotational support and central photometric profile with the old ages of stellar populations, we expect that the mergers happened typically before $z=2$ (10\,Gyr ago). The higher cuspiness  of galaxies with a low rotational support in the low-density environments  suggests that in these environments the mergers were particularly gas-rich, in agreement with the environment-morphology relation. This was already  suggested by \citet{krajnovic20}. We note that cores in ETGs can also be induced during the stage of the  monolithic collapse \citep{nipoti06}, or as a result of repeated removals of gas from centers of galaxies by activity of galactic nuclei \citep{vandervlugt19}. In order to explain the correlation found without mergers, these two alternative core-forming processes would have to be stronger for galaxies with a higher rotational support, which is again a speculative possibility. Our results could be somewhat biased by the fact that we had the information about the central photometric profile for a much lower number of galaxies than for the other investigated merger-sensitive parameters; namely, the slope of the inner photometric profile is available for 80 out of our 175 galaxies.

We did not detect any statistically significant correlation between the incidence of dust lanes and the rotational support for galaxies at a fixed mass and environmental density. Given that the lifetime of dust is expected to be under 1\,Gyr \citep{patil07}, the mergers that decreased the rotational support likely typically happened before that time. The same result  would be expected if the mergers were dry, which is, however, disfavored by  some of the previous pieces of evidence.

We also did not detect a statistically significant correlation of the effective radius with the rotational support at fixed galaxy mass and environmental density.  Here we assume that the decrease in the rotational support happens primarily by mergers, which implies that at a fixed galaxy mass and environment density, the galaxies with a lower rotational support should have larger effective radii if the mergers were dry, as explained in \sect{indicators}. This suggests that the decrease in the rotational support happened primarily  when the galaxies were still mostly gaseous, say at $z>3$, because the expansion by mergers works only for dissipationless systems.  Alternatively, one might argue that the growth of the radius of the galaxies with a low rotational support by mergers was approximately the same as the growth of radius of the galaxies with a higher rotational support by in situ star formation.

Finally, we found that at a fixed galaxy mass and environmental density, the galaxies with a lower rotational support have steeper metallicity gradients\footnote{We noted the apparent contradiction with the results of \citet{krajnovic20} who reported less steep metallicity gradients in SRs compared to FRs. There are several reasons for this: 1)  \citet{krajnovic20} did not compare the $\gamma^\prime$ parameter for FRs and SRs of the same mass and environmental density; 2) while they analyzed the \atlas sample just as we did, our sample contains only its MATLAS subset, which avoids the Virgo cluster; 3) \citet{krajnovic20} applied a binary  separation of ETGs to FRs and SRs. }. As explained in \sect{indicators}, this signifies that mergers that decreased the rotational support were preferably  minor.  The gradients we used, being derived by the SSP method, are biased toward the old stellar populations. This further supports the hypothesis that the transformation happened a long time ago. The correlation is stronger in the low-density environment subsample. This can be explained by the accreted galaxies being more metal poor, as can be expected due to the fact that the metallicity of galaxies decreases toward low-density environments at a fixed galaxy mass. {The steeper metallicity gradient of the galaxies with a low \rotsup with respect to the galaxies with a high \rotsup could also be explained by the fact that the latter routinely experience major mergers}. This, however, goes against our assumptions and several  observational findings stated above.

Another option that remains to be discussed is that the galaxies that currently have a low rotational support first experienced the ancient wet mergers at $z>2$, which changed the values of the merger-sensitive parameters,  but these mergers were not those that decreased the rotational support. The rotational support was instead changed at a lower redshift by mergers that did not influence the merger-sensitive parameters. We dismiss this option because it seems too fine-tuned and opposes Occam's razor. In addition, it contradicts the abovementioned observed constancy of volume density of quenched galaxies since $z=1-1.4$, and because in \sect{late} we find that mergers in the current Universe are very rare. 

To summarize this section, the observations can be explained easiest if the mergers that decreased the rotational support  were wet and minor. {They had to happen when the ETG progenitors were not completely { gaseous}, but still contained a substantial fraction of gas. This is at around $z=2$ or even before, depending probably on the mass and surface density of the galaxy (see \sect{intro}).} 
The individual constraints on the time and way of the transformation described above are summarized in \tab{constraints}. This way of galaxies forming with a low rotational support resembles the pictures of the monolith collapse from the simulations of \citet{kobayashi04}, where the galaxy is rapidly assembled through many wet minor mergers. A wet clumpy collapse is what was also assumed in the successful chemical model of \citet{thomas99}. A similar scenario was claimed to explain the chemical properties of KCDs \citep{bender92}.  Also, the simulations of \citet{naab14} showed that the galaxies with the lowest $\lambda$  had to be formed by multiple minor mergers. The higher incidence of tidal disturbances observed in the galaxies with a low rotational support suggests that such galaxies live in environments where galaxy interactions are more common, such as the intersections of cosmic filaments. The late interactions, however, do not seem to be determinant for the properties of the galaxies, which we check again below in \sect{late}.

The observational evidence does not seem to agree with the prediction of cosmological simulations that the transformation of the kinematic structure of ETGs happened after the redshift of one  (8\,Gyr ago)  \citep{penoyre17,lagos18}. This seems to be  another manifestation of the problem of the too gradual formation of ETGs in cosmological simulations (see \sect{intro}). The ideal solution to determine  when the transformation occurred would be to observe spatially resolved stellar kinematics at high redshifts. {Fist observations of this type start appearing nowadays }\citep{newman18,cole20} but we have to wait until data become available for a statistically meaningful sample of galaxies.

It is interesting to put the typical time of the transformation that we determined, the redshift of about two, in the context of the other events happening in the Universe at the same epoch. The redshift of two is the time when the global star formation in the Universe, dominated by massive galaxies, started to drop \citep{madau14,liu18,wilkins19}. At the same time, the frequency of galaxy interactions dropped too \citep{huertascompany15,ventou17,ventou19}. Before $z=2$, the interstellar medium of star-forming galaxies was moving randomly, under the effect of turbulence, and the gas was forming a lot of giant gas clumps, which could hold a substantial fraction of the baryonic mass of the galaxies. At $z\sim 2$, ordered motions in star-forming galaxies started to prevail \citep{simons17}. At the same time, galaxies attained the standard Hubble morphologies known from the local Universe, instead of showing prominent giant gas clumps \citep{mortlock13, lee13}.  The redshift of about two thus seems to be a plausible point at which the kinematic morphology {settles}. Indeed, it has been proposed that the kinematic morphology is more fundamental than the Hubble photometric morphology \citep{cappellari11b}, and therefore settling both morphologies at the same time appears logical.

\begin{figure*}
        \centering
        \includegraphics[width=17cm]{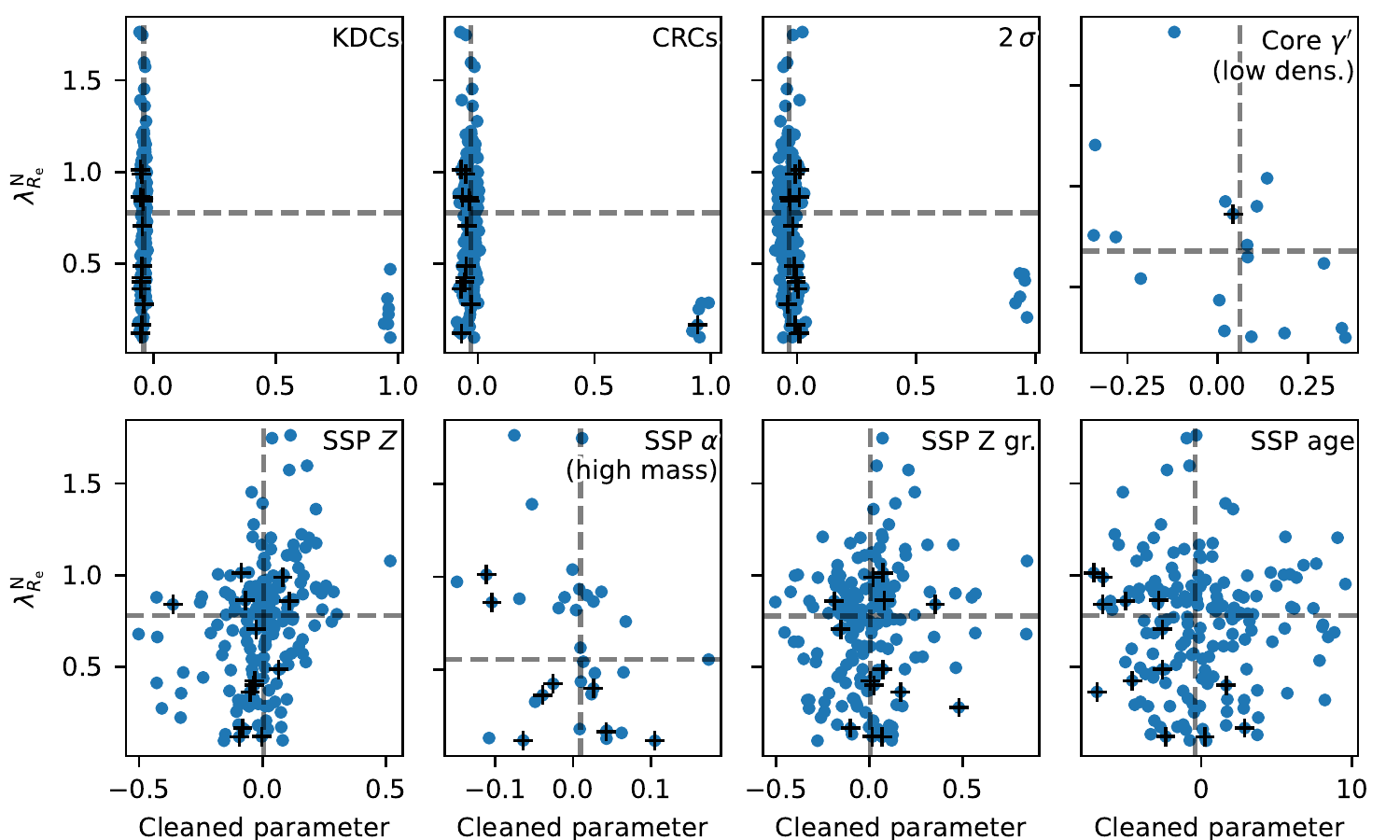}
        \caption{Demonstration that the significant correlations between the rotational support and the merger-sensitive parameters detected above were not caused by recent major mergers. 
        All galaxies from the sample are shown as  points, while the candidates for the recent major merger remnants are highlighted by crosses. The horizontal coordinate of each panel shows the cleaned value of the given parameter. The vertical axis shows the rotational support. {The dashed lines indicate the median values calculated for all available data.} } 
        \label{fig:majmerg}
\end{figure*}

\section{Sanity check: The role of the late galactic interactions in the decrease in the rotational support}
\label{sec:late}
In the previous section, we found multiple independent pieces of evidence that  the rotational support is decreased by mergers typically before the redshift of two.  It might be concerning that we also found that  galaxies with a lower rotational support  possess tidal disturbances more often at a fixed mass and environmental density. This observation alone could also  be interpreted to mean that the rotational support is decreased before the maximum lifetime of tidal features, which is 9\,Gyr (while this contradicts the other findings from the previous section). In this section we therefore inspected in more detail the role of the late interactions, whose signatures can still be observed as tidal features, in decreasing the rotational support.  {It seems that the late interactions do not significantly influence galaxies.}

\subsection{Recent major mergers do not contribute to the correlations with rotational support}
\label{sec:majmerg}
In \sect{corr}, we detected correlations between the rotational support and some of the merger-sensitive galaxy parameters. We used the correlations to deduce what mergers caused {the decrease} in the rotational support and found that they happened at high redshifts. In this section, we perform a sanity check on the early drop in the rotational support.  
According to the Illustris simulation, at $z=1$ (8\,Gyr ago) the progenitors of the current-day FRs and SRs had indistinguishable distributions of the $\lambda$ parameter \citep{penoyre17}.
{Galaxy merging was relatively rare in the Universe in these late epochs. We expect every galaxy to have had relatively few of these recent mergers \citep[e.g.,][and see \sect{mergingfreq}]{rodriguez15} since $z=1$ (7.8\,Gyr ago), typically about one or five, depending on the mass ratio and whether we trust rather observations or simulations. These few mergers had to set the current $\lambda$ of galaxies. There would be many galaxies that did  not receive any substantial merger, and such objects would not show any correlation between the merger sensitive parameters and $\rotsup$. The correlations that we detected would be caused only by the galaxies that experienced substantial merging.} 
{Galaxies showing large morphological disturbances (i.e., probable remnants of recent major mergers), would then contribute a large portion of the points that induce the detected correlations.}
We test this in this section and find that that it does not seem to be the case.

We identified the following 14 galaxies as the most serious candidates for major merger remnants, because of their morphology in the deep MATLAS images\footnote{{Images of all MATLAS galaxies are available at \url{http://obas-matlas.u-strasbg.fr}.}}: NGC\,0474, NGC\,1222, NGC\,2764, NGC\,3414, NGC\,3610, NGC\,3619, NGC\,3640, NGC\,4382, NGC\,4636, NGC\,4753, NGC\,5485, NGC\,5493, NGC\,5557, and NGC\,5866. {These candidates are identified in the galaxy list in the online material}. They show either strong morphological disturbances, even close to the centers of the galaxies, or their stellar halos are strongly offset from the central parts of the galaxies. {Their indicator of disturbed isophotes is two (13 cases) or one (1 case). The galaxies} often show complex tidal features and unrelaxed dust patches. Such galaxies underwent relatively strong interactions recently (before the lifetime of tidal features, see \sect{survival}). We avoided the galaxies that seem to be involved in an ongoing interaction with their neighbors, or the galaxies whose morphology is difficult assess, for example because of pollution by the light scattered from a nearby star, or because the galaxies overlap with neighbors in projection. Our selection is likely biased in favor of mergers involving spiral or lenticular galaxies. Such mergers typically produce more distinct photometric irregularities than major mergers of pressure-supported galaxies. Our sample of major merger candidates spans both mass bins and both the density bins defined in \sect{corr}.

We compared the properties of the major merger candidates to the less-disturbed galaxies in our sample in \fig{majmerg}. It shows the galaxies in the space of the cleaned merger-sensitive parameters versus the rotational support. The figure shows only the parameters that were identified in \sect{corr} to correlate statistically significantly with the rotational support at a fixed mass and environment density of the galaxy.  If the statistically significant correlation was detected only for one of the considered galaxy subsamples (that is the low- and high-mass subsamples, and the low and high environmental density subsamples), then the figure shows only the galaxies belonging to that subsample. In addition, to stress our arguments (see below), we also show the galaxies in the plane of the cleaned SSP age versus the rotational support, where the galaxies do not form a significant correlation. Every galaxy in \fig{majmerg} is represented by a point, and the major merger candidates are highlighted by crosses. The median values of the cleaned parameters and of the rotational support for all the depicted galaxies are marked by the dashed lines. We did not include among the plots in this figure the disturbed outer isophotes because their presence was the main criterion to select the major merger candidates -- the major merger candidates would be offset from the other galaxies in the sample by definition. In agreement with what was found in \sect{corr}, \fig{majmerg} shows that the major merger candidates have relatively low rotational support, compared to other galaxies.

{The figure shows that the major merger candidates do not seem to contribute to the correlations}. For example, it is striking that almost none of the candidates possess a KDC or other kinematic substructures. For the central photometric slope and the abundance of $\alpha$-elements, there does not seem to be any cloud of points corresponding to the galaxies that did not experience any substantial {recent} merging. The data are rather explained better, such that another process defines the distribution of points in the {space of} cleaned parameters versus \rotsup, and that the major merger candidates are just randomly selected points at the given value of \rotsup. {This  other process could be, in our interpretation, the numerous wet mergers at high redshift.}  In contrast, the distribution of the major merger candidates in the space of the cleaned SSP age versus \rotsup is clearly biased. This is how we expect the major merger candidates to be distributed in the other plots, if the late mergers were responsible for forming the correlations.

{ We also note that the gradients of metallicity for the major merger candidates seem to be more positive (i.e. flatter)  than for the other galaxies.  At the same time, the major merger candidates have a low \rotsup compared to the others. Together, this} means that the major merger candidates actually make the detected correlation weaker. {One interpretation is that these recent major mergers erase a correlation that was induced by  the ancient numerous minor mergers.}

The interactions that created the recent major merger candidates thus do not seem to be important for shaping the correlations that we used {to infer how the rotational support of galaxies is lowered.} This suggests that the mergers that lowered the rotational support had to be more major or numerous in order to induce the observed trends.

\subsection{Events forming observable tidal disturbances are rare}
\label{sec:mergingfreq}
Here we provide an estimate of how many interactions the galaxies in our sample experience per gigayear, on the basis of the observed incidence of tidal disturbances. The interactions appear to be relatively rare. 

We assume that every tidal disturbance has the same life time, $T_L = 4-9\,$Gyr (see \sect{survival}), that the frequency of the disturbance-forming events is the same  for all the galaxies in the sample under consideration, and that the frequency is constant in time.We count only the independent interactions. This means, for example, that the interaction of a given galaxy with another galaxy and its satellite is counted as one interaction. Then the number of interactions that created tidal disturbances in a given galaxy, $n$, follows a Poisson probability distribution:
\begin{equation}
    P(n~\textrm{interactions}) = \frac{\mu^ne^{-\mu}}{n!}.
\end{equation}
One can estimate the parameter of the distribution, $\mu$, from the fraction of galaxies that are observed to have a tidal disturbance, $f$, by making use of the fact that 
\begin{equation}
    f = P(n>0) = 1-P(n=0) = 1-e^{-\mu}.
\end{equation}
This gives $\mu = -\ln(1-f)$. The mean of a Poisson distribution is the parameter $\mu$ itself, and therefore the mean number of events that formed tidal disturbances in a galaxy in the considered sample can be estimated as 
\begin{equation}
\langle n \rangle   = -\ln(1-f).
\label{equ:nest}
\end{equation}
If galaxies typically experience a large number of events forming tidal disturbances in $T_L$, then we expect tidal disturbances to be present in a large fraction of galaxies.

In the census of the tidal disturbances in the MATLAS sample, \bilmat found $f$ for the total sample to be 41\%, if we count both  the certain and likely detections. The likely detections include the galaxies where the tidal disturbances were too faint to be detected with certainty, or the galaxies whose images had an inferior quality. This value of $f$ is thus rather the upper limit of the true value. With this fraction of galaxies with tidal disturbances, we found that every galaxy typically experienced only 0.51 independent interactions in the last $T_L$. For massive galaxies with $\log \mjam >11$, \bilmat found that $f$ increases to 60\%, meaning that such galaxies experienced on average 0.91 disturbing events in $T_L$. Some tidal disturbances might had been undetected because of unfavorable projection effects. \citet{mancillas19} investigated this issue in simulations and found that streams and tails are not sensitive to projection effects, but shells are. Nevertheless, because several shells are usually present in a galaxy, it is possible to detect at least one of them from any line of sight. Moreover, that work did not consider galaxies with disturbed isophotes. Doing so would decrease the sensitivity to the projection effects. Even the most massive galaxies, therefore, have experienced just about one interaction in the last 4-9\,Gyr. This also includes the minor mergers forming stellar streams. For the high-mass galaxies, we obtained the rate of interactions forming tidal disturbances of 0.23 per Gyr for the minimum $T_L$, or 0.1 for the maximum one. This is in reasonable agreement with the merging rates  estimated from the spatial density of close galactic pairs \citep{man16}. Out of the galaxies with $\mjam<10^{11}\,M_\sun$ in MATLAS, 36\% have likely or certain detections of tidal disturbances. Equation~\ref{equ:nest} then implies that only 0.44 interactions happened in these lighter galaxies in the time $T_L$, meaning that the merging rate is at most 0.11 per Gyr.

We were able to estimate the frequency of major mergers from the number of major merger candidates from the last section. The 14 galaxies constitute 8\% of the whole MATLAS sample. From here we obtained $\langle n \rangle = 0.08$, or at most 0.02 major mergers per Gyr, for the minimum lifetime of tidal disturbances. This agrees with the major merger rates found by the counts of galaxy pairs \citep{mundy17}. This shows that major mergers are extremely rare in the current Universe. 

Thus, we again find evidence against the decrease in the rotational support within the lifetime of tidal disturbances. The increased incidence of tidal disturbances in galaxies with a low rotational support thus rather reflects that galaxies with a low rotational support have continued to receive more mergers even until today, but the late mergers do not usually change the properties of the galaxies substantially.

The above estimates of merging frequencies can be used to test the hypothesis that quenched galaxies increase their radius with redshift because of mergers. \citet{trujillo11} estimated the number of mergers necessary for explaining the observed evolution of radii of quenched galaxies with redshift. The masses of their galaxies were similar to ours. For the redshift of 0.4, which corresponds to the look-back time of about the minimum lifetime of tidal disturbances, they give $1.4\pm0.3$ mergers for the mass ratio of 1:3 or $3.0\pm0.7$ mergers for the mass ratio 1:10. The value for the major mergers is somewhat higher than the number of interactions that we estimated for the massive galaxies from the frequency of tidal disturbances, but still in the $2\sigma$ uncertainty limit. Nevertheless, we have to remember that many of the disturbances, especially the tidal streams, likely come from minor mergers and others are false detections. For the low-mass galaxies, the discrepancy is even higher. If we instead adopt the maximum lifetime of tidal disturbances, which corresponds to $z=0.8$, the estimates of \citet{trujillo11} predict even a 2.5 times higher number of mergers than since the redshift of 0.4. This all suggests that the expansion of ETGs with redshift must be partly attributed to  other mechanisms than mergers \citep{fan08,damjanov09,fan10,ishibashi13,vandervlugt19}. \citet{man16} and \citet{newman12} arrived to the same conclusion from the counts of ongoing merger candidates. The need for  alternative mechanisms is particularly strong for the low-mass galaxies. Indeed, the chemical models by \citet{trussler20} predict that gas outflows become more and more important toward ETGs with lower masses.

Further, it is worth noting that in the simulation inspected by \citet{mancillas19}, at least one tidal stream in the reach of the MATLAS survey was detected for any time and for any line of sight. This contrasts with the large fraction of observed galaxies without any tidal disturbances; only 16\% of galaxies in MATLAS have certain or likely detection of streams. This again points to a too extended formation of ETGs in cosmological simulations.
 
\begin{figure*}
        \centering
        \includegraphics[width=17cm]{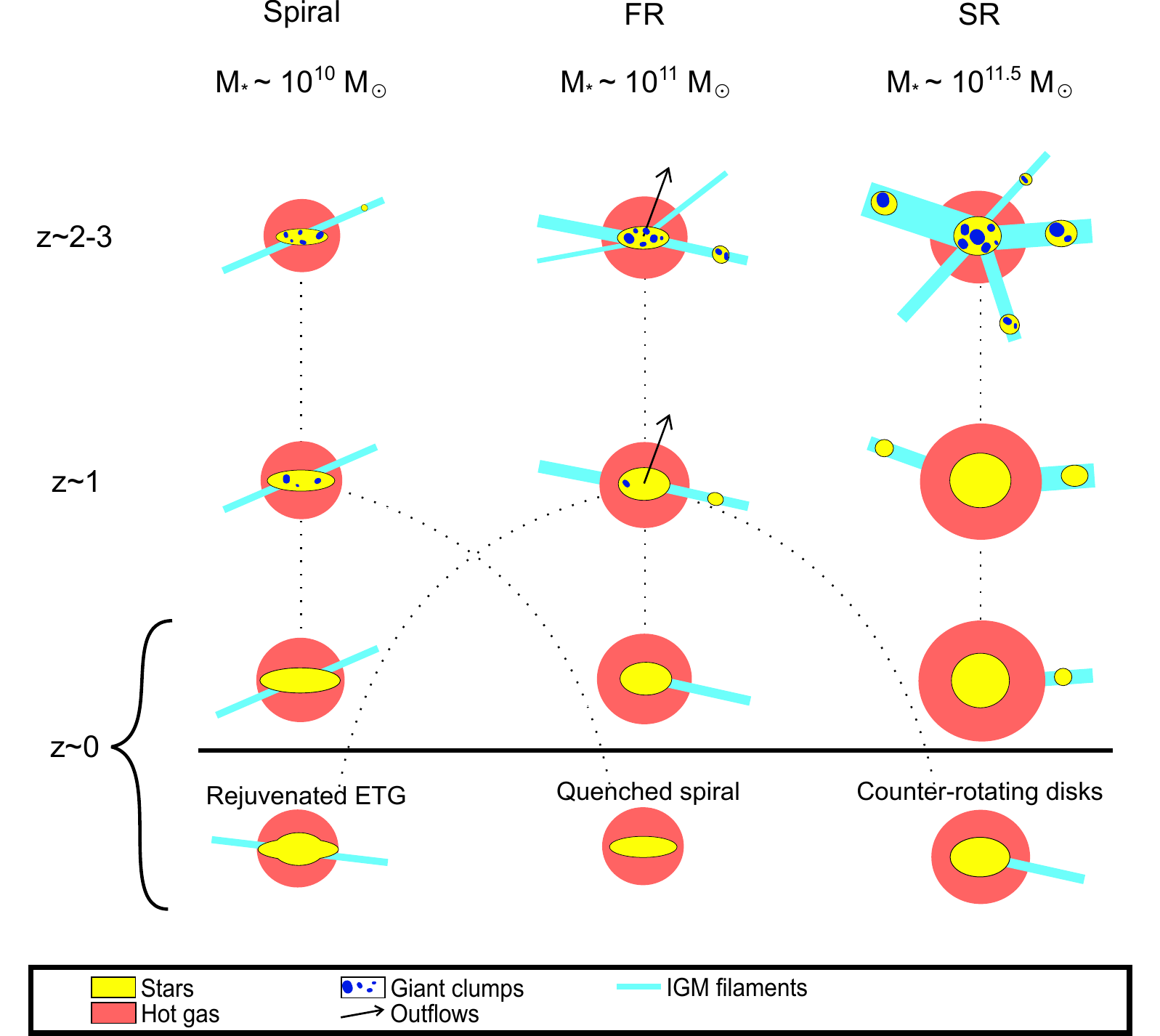}
        \caption{Proposed scenario for the formation of massive galaxies {outside of galaxy clusters}.  The columns from left to right  correspond to the formation of a typical spiral, FR, and SR. At the top of each column, we state the typical current stellar  mass of the respective galaxy type. The last row of the figure {below the thick black line shows some 
        nontypical ways of forming the different galaxy types.} {The dotted lines mark the evolutionary tracks of the different galaxy types.} } 
        \label{fig:galform}
\end{figure*}

\begin{figure}[h!]
 \resizebox{\hsize}{!}{\includegraphics{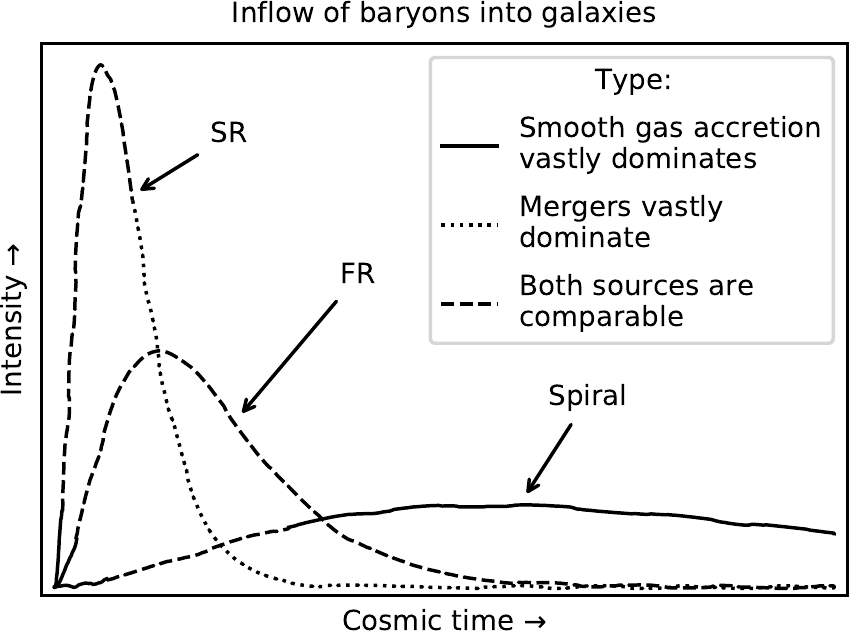}}
        \caption{{Proposed  assembly of different types of massive non-cluster galaxies. The curves schematically show the intensity and type of inflow of baryonic material into the galaxies over time.}}   \label{fig:inflow}
\end{figure}

\section{Corollaries of the assumption of the decrease in the rotational support by mergers}
\label{sec:theor}
{This paper is based on the assumption that galaxies are formed with a high rotational support, which is then decreased primarily in mergers. In this section we point out that this assumption alone  can explain several observational facts.}

If a galaxy gained a substantial fraction of mass through mergers, it will probably end with a low rotational support. Many massive galaxies and the central galaxies of clusters and galaxy groups are expected to have formed in this way \citep{dubinski98,mihos03,delucia07,penoyre17}. It follows from here that these galaxies can be expected to  preferably be SRs. Indeed, the fraction of SRs increases with galaxy mass and the central galaxies are usually SRs \citep{emsellem11,houghton13,deugenio13,brough17,veale17,vandesande21b}.

Rotational support cannot be lower than zero. If galaxy mergers usually decrease the rotational support, then the galaxies that experienced many mergers will have their \rotsup parameters clustered toward zero. This explains the finding that the distribution of galaxies in the $\lambda$ plane versus stellar mass is bimodal, with one of the peaks near $\lambda=0$ \citep{vandesande21} (even if similar studies can be affected by the sample selection bias, as demonstrated in \citealp{graham18}). Next, it is observed that for galaxies of a fixed stellar mass, the prominence of the peak near $\lambda=0$ increases with the stellar mass, which can be explained by the greater fraction of material gained by accretion by the massive galaxies.

A galaxy that has gained a large fraction of mass through many mergers from different directions can be expected to lose any sign of the initial rotation. They would acquire a new kinematic morphology. \citet{cappellari16} divided galaxies with a low value of rotational support into misclassified and genuine SRs. The two classes have different kinematic structures. The orbits of stars in the genuine SRs are randomly oriented, while the misclassified SRs seem to consist of two counterrotating disks. {The} genuine SRs probably correspond to the galaxies that experienced a large number of mergers. The misclassified SRs might have formed from FRs that experienced one merger that coincidentally had a suitable orbital configuration. {Indeed, the massive galaxies that experience many mergers, are typically observed to be genuine SRs}\citep{cappellari16}.

It is worth noting that the galaxy sample investigated here is expected to contain only a few genuine SRs. Such galaxies have a $\rotsup<0.31$ and $\mjam>2\times10^{11}\,M_\sun$ \citep{cappellari16}. Even if our sample does not contain many genuine SRs, its members still experienced mergers. The final stage of this process is the formation of genuine SRs. {Our approach here resembles the studies of the formation of quenched galaxies that make use of the green valley galaxies that have not quenched completely yet.}\citep[e.g.,][]{trussler20,carleton20,noirot22}.

\section{Final picture of the formation of SRs, FRs, and massive spirals}
\label{sec:picture}
In this section, we address the typical formation paths undertaken by spiral galaxies, and fast and slow rotator ETGs. The proposed scenarios are outlined  in \fig{galform}. Our picture for the formation of the different galaxy types is based on the assumption that the {most predictive} factor of the evolution of a galaxy is its final mass. The primary dependence on mass agrees, for example, with the findings from  \app{simplecorr} and with  the fact that the morphology of galaxies correlates strongly with their mass:  spirals are typically less massive than  ETGs \citep{read05,blanton09,kelvin14} and  FRs are less massive than SRs (Figure~\ref{fig:mrs}, \citealp{emsellem11,cappellari16,veale17,brough17,greene17,graham18}). The first three rows of  \fig{galform} depict the typical time evolution of a standard galaxy. In the last row, we included some of the nontypical ways the discussed galaxy types are formed.

We begin with the proposed formation scenario of a typical SR, which is depicted in the right column of \fig{galform}. It is based mostly on our findings for galaxies outside of galaxy clusters, but we argue later that even the cluster SRs follow a similar formation path. Before the redshift of about two, the galaxy rapidly grows by in situ star formation (\sect{intro}). It experiences numerous wet mergers. This leads to a high enrichment of the stellar population by $\alpha$-elements. The present-day slow rotation  suggests that the accreted material was arriving from multiple directions, a process that is favored if the galaxy was located at the intersection of the multiple cosmic filaments that bring in the galaxy intergalactic medium (IGM) and other, smaller, galaxies. The location of SRs at the intersections of cosmic filaments does not necessarily imply that they live in regions of extremely high concentrations of galaxies. 
Some filaments may indeed host  a low number of galaxies that are sparsely distributed. 
Part of the inflowing IGM is shock-heated, forming a halo of hot gas around the galaxy (\sect{intro}). At the redshifts above two, the cold IGM is still able to penetrate the hot halo of the massive galaxies and feed the star formation in the galaxies. According to the successful chemical models of monolithic collapse by \citet{pipino10}, the galaxy quenches because of supernova winds. The quenching proceeds outside-in.

The mergers experienced by the galaxy contribute to the decrease in the rotation support of the galaxy. Nevertheless, there are some additional mechanisms that might contribute as well. We consider two equal gas clouds that have the same apocentric distance with respect to the center of a spherical galaxy, but one cloud is on a circular orbit and the other is on a radial orbit. The binding energy of the gas cloud on the circular orbit is lower. The supernova winds that are supposed to quench the galaxy will thus preferentially remove the gas cloud from the galaxy on the circular orbit. The $\alpha$-enhancement identified for the galaxies with a low rotational support suggests that SRs experience a stronger feedback than FRs of the same mass. Similarly, the non-detection of a correlation between the effective radius and the  rotational support at a fixed mass and environment density indicates that the original potential energy of the accreted galaxies must be removed from the system, either in the form of radiation or gas outflows, a process that will preferentially remove the gas clouds on originally circular trajectories.

It is relevant to note that the gas-rich galaxies that we identified as the progenitors of SRs contain many giant gas clumps at high redshifts \citep{conselice08,elmegreen09,elmegreen14,cava18}. Their masses can reach up to $10^9\,M_\sun$ and are probably gravitationally self-bound  \citep{bournaud14,tadaki18,fensch20}. Their sizes are often comparable to the sizes of the most distant quiescent galaxies ($\sim 1\,$kpc). Once their original host galaxy is accreted by the future SR and turns into tidal debris, these gas clouds move on randomly oriented orbits and contribute to the turbulent nature of gas around the future SRs.

At the redshift of around one, the formation of the SRs becomes much quieter. By the redshift of around two, the IGM filaments can no longer  reach the center of the galaxy (\sect{intro}). All arriving IGM is shock-heated so that it contributes to the hot halo. The frequency of mergers also decreases. The mergers become drier, because generally all massive galaxies become gas poorer in later cosmic epochs, and partly because of the quenching of the satellites in the hot halo of the main galaxy. Toward the redshift of zero, the frequency of mergers decreases even further. The possibility that SRs reside at the  intersections of several filaments is strengthened by the fact that we found that with {the decrease in} the rotational support of the galaxies, the incidence of disturbed isophotes increases at a fixed mass and environmental density.

Here we come to the question of the formation of the cluster SRs. For the non-cluster SRs, we explained above the observational constraints by the SRs residing on intersections of cosmic filaments. But this is exactly the location of galaxy clusters and massive groups. The reasoning above thus hold true also for the central galaxies of these structures. Our assumption on the transformation of the kinematic structure by mergers alone has an interesting implication: the most massive galaxies are expected to experience the highest number of mergers partly because of the strength of their gravitational field and partly because they tend to be the central galaxies of their environments. { In the end}, the galaxy gains most of its mass by mergers. This object type differs from FRs and spirals by a relatively low fraction of stellar mass formed in situ. As galaxy groups and clusters merge together, the central SRs of the original structures can become noncentral galaxies of the new structures. The central SRs of groups and  clusters are expected to experience, compared to the non-cluster SRs, many dry mergers \citep{dubinski98,mihos03,delucia07}. These produce the observed central photometric cores \citep[e.g.,][]{krajnovic20}.  One has to remember that even the central cluster galaxies were gas rich at early cosmic epochs and their rotational support might have already decreased to the current value by that time.

The formation of spirals seems to be the reverse of the formation of SRs.  Most spiral galaxies do not have classical bulges and many do not contain  detectable stellar halos \citep{kormendy10,peebles10,fisher11,merritt16}. Such features are supposed to form mostly by mergers \citep{naab03,bournaud05,bournaud07}. This suggests that spirals form much more smoothly than ETGs: either completely by smooth accretion of the IGM \citep{sancisi08}, with a small contribution of minor mergers, or possibly by very gas-rich mergers at high redshifts. The low abundance of classical bulges in the observed spirals and the non-detection of the stellar halos are difficult to reproduce in current cosmological simulations of galaxy formation \citep{brooks16,peebles20,merritt20}. Spiral galaxies possess hot gas halos at least down to the stellar mass of $10^9\,M_\sun$ \citep{li14cgm,li17cgm}  but cold gas streams are able to go through the halos \citep{dekel06}.

The formation of FRs seems to be an interpolation between the formation of spirals and SRs. At the redshift of two to three, a typical FR faced fewer mergers than a typical SR and it grew more by in situ star formation. If the FR formed in a node of the cosmic web, then only a few filaments were joining in the node, or some of the filaments conducted much more material into the node than the other filaments. As the galaxy contained a lot of gas, it became gravitationally unstable \citep{tadaki18} and formed massive giant clumps, which helped to dynamically heat the disk, make it thicker \citep{bournaud14,clarke19}, and produce a bulge \citep{noguchi99,elmegreen08,hopkins12}. Mergers contribute to the thickening of the disk and to the formation of the bulge. Another source of thickening of the disk might be the strong bars that are hinted at by observations of submillimeter galaxies \citep{gullberg19,hodge19,dudzeviciute20}. Similarly to SRs, at $z=1$ the merging activity has decreased, just as the inflow of fresh intergalactic gas. Because of the typically lower masses of the FRs, the gas is partly able to reach the galaxies. This is suggested by the presence of molecular gas in many galaxies in the given mass range \citep{young11}, or the non-negligible star formation implied by the reconstructed star formation histories \citep{mcdermid15}. In addition to mass quenching, star formation is further suppressed by morphological quenching and possibly other mechanisms.  Our results from \sect{mergingfreq} suggest that the frequency of mergers is too low to explain the growth of radii of quenched galaxies with redshift and the discrepancy is the strongest  for the galaxies with $\mjam<10^{11}\,M_\sun$. Therefore, an additional mechanism is needed in these galaxies, such as gas outflows (\sect{intro}). The need for outflows is hinted at also by the chemical composition of these galaxies \citep{trussler20}.

{Figure~\ref{fig:inflow}} schematically shows curves of the proposed intensity and source of the accreted material by different types of galaxies as functions of the cosmic time. It is inspired by studies of stellar populations \citep{thomas10,pipino13,mcdermid15,gonzalez17} and the merging histories deduced above. The shapes of the curves reflect the ``galaxy downsizing'' phenomenon \citep[e.g.,][]{neistein06}; in other words, that the high-mass galaxies form first and the less massive galaxies form later.

Not all galaxies follow {the typical paths described above}. For example, a spiral galaxy  can be quenched by starvation and turn into an ETG after entering the hot gas halo of a galaxy cluster. The opposite transition is also possible. There are signatures that formerly passive galaxies acquire star-forming disks and become bulged spirals \citep{delarosa19}. Mergers of special relative orbital configurations might lead to the formation of galaxies with two counter-rotating stellar populations. They would then be classified as SRs according to our accepted parametric criterion.

\section{Summary and conclusions}
\label{sec:sum}
Cosmological hydrodynamical simulations suggest that galaxies are formed  with a high degree of rotational support, which decreases later. Mergers play a substantial role in the transformation. According to these simulations, this transition happened less than about 8\,Gyr ago (redshift of one). In this paper we have investigated this transformation observationally, for ETGs outside of galaxy clusters, initially assuming that mergers are solely responsible for the transformation. To quantify how much the transformation proceeded in a given galaxy until now, we primarily made use of  the parameter \rotsup introduced by \citet{emsellem11}  and called the rotational support here. By definition, SRs have a rotational support lower than 0.31, and FRs have a rotational support higher than this. We  exploited  multilinear regressions on data  from the \atlas and MATLAS surveys in order to study the correlations of various parameters that are expected to be sensitive to mergers with the rotational support for galaxies of a fixed mass and environmental density. The results were then interpreted making use of the facts that every merger indicator has a different lifetime and is sensitive to a different type of merger (minor or major, wet or dry). It was crucial to study the correlations at a fixed mass and environmental density in order to account for the so-called confounding effect, that is, the fact that \rotsup correlates with the galaxy mass and with  the environmental density, and that different merger-sensitive parameters correlate with these quantities as well. We also considered other observational results from the literature, mostly from  observations of the high-redshift Universe. We can summarize our results as follows:

\begin{enumerate}
     \item The observations indeed agree with the hypothesis that the rotational support of galaxies is initially high and then it decreases due to successive  mergers. {At a fixed stellar mass and environmental density, the galaxies with a low rotational support more often contain tidal disturbances and kinematic substructures compared with the galaxies with a higher rotational support.} In addition, the metallicity of the galaxies with a lower \rotsup is typically lower.
    \item We found many pieces of evidence showing that the decrease in the rotational support  happened  preferentially toward the end of the first phase of the ETG assembly, when the galaxies still contained a substantial fraction of gas. This means around $z=2$ {or perhaps before}, depending on the mass or surface density of the galaxy (\sect{intro}). 
    This is supported by the fact that the majority of present-day SRs do not show tidal disturbances, and that, at a fixed mass and environmental density, the galaxies with a lower rotational support do not have substantially different effective radii than galaxies with a higher rotational support. The mergers had to be typically wet, for example because at a fixed mass and environmental density, the galaxies with low rotational support have a higher abundance of $\alpha$-elements, as if the mergers decreasing the kinematic stage were causing starbursts.  The mergers also had to be minor to account for the fact that, at a fixed mass and environment density, galaxies with a low rotational support tend to have steeper metallicity gradients. The evidence for the early time of kinematic transformation is particularly strong for high-mass galaxies ($\mjam>10^{11}\,M_\sun$). For them, observations by other authors  of their probable progenitors at high redshifts  indicate that they already had spherical shapes and their current stellar masses at $z=1-2$.  
    The timing of the establishment of the kinematic morphology coincides with that of the photometric morphology (i.e., the Hubble sequence). At that same epoch, ordered motions of gas prevailed over chaotic motions  in star-forming galaxies.
    
    We found that the kinematic transformation of ETGs happened earlier than simulations predict. This agrees with the independent findings of other works that the formation of ETGs in cosmological simulations is too extended. 
    
    \item Galaxies with a lower rotational support  still experience more galaxy interactions today than the galaxies with a higher rotational support at a fixed mass and environmental density. They seem to live in environments where interactions are more frequent. Such recent interactions, however, are rare and do not greatly influence the properties of the galaxies.
    \item From the measured frequency and estimated survival time of the tidal disturbances observed in deep images, we estimated a merger rate of at most 0.23\,Gyr$^{-1}$ for galaxies with $\mjam>10^{11}\,M_\sun$,  or 0.11\,Gyr$^{-1}$ for galaxies with $\mjam<10^{11}\,M_\sun$. We inferred   a frequency of major mergers of  0.02\,Gyr$^{-1}$. These numbers agree with what has been estimated from the frequency of close galaxy pairs in the literature. The frequency of mergers seems to be insufficient to explain the growth of the effective radius of ETGs with time, and therefore additional  mechanisms  are necessary to account for it. 
    \item We have proposed a picture of formation of FRs, SRs, and massive spirals that seems to account for all observational constraints considered in this paper (\sect{picture}). The formation of a typical SR is fast and involves numerous mergers. Spirals typically assemble gradually and smoothly, with most material gained through the accretion of the IGM.  The formation of  typical FRs lies between these two extremes.

\end{enumerate}

\begin{acknowledgements}
We thank the anonymous referee for a constructive report. We thank D.~Krajnovi\'c, M.~Cappellari and I.~Ebrov\'a for valuable comments and discussions. MB is grateful for the financial support by {\it Cercle Gutenberg}.
MB acknowledges the support from the Polish National Science Centre under the grant 2017/26/D/ST9/00449.

        
\end{acknowledgements}

\bibliographystyle{aa}
\bibliography{citace}

\begin{appendix}

\section{Correlations between the merger-sensitive parameters,  rotational support, galaxy mass, and environmental density}\label{app:simplecorr}

In this section we explore whether the galaxy mass \mjam, environmental density \rhoio, and the rotational support \rotsup correlate in our sample with each other, and with the investigated merger-sensitive parameters. We find evidence of such correlations, which is a warning that one should be aware of the confounding effect when investigating the influence of mergers on the rotational support through the merger-sensitive parameters. In addition, we also investigate the differences in the merger sensitive parameters for SRs and FRs.

We explored how the mentioned quantities correlate exploiting Pearson's correlation coefficient. The results are listed in \tab{simplecorr}. In the first three main columns, the sub-column ``sign'' signifies the sign of the correlation.  The sub-column ``$p$'' shows the $p$-value of every correlation, (i.e., the probability that there is actually no correlation). All other relevant  numbers pertaining  the correlations  are provided in \app{fits}.

To summarize the results in  \tab{simplecorr}, we found the following statistically significant correlations between the merger-sensitive parameters and the rotational support.  With increasing rotational support, a galaxy has a decreasing probability of possessing disturbed outer isophotes, containing some form of tidal disturbances, hosting a KDC or a CRC, and having a smaller effective radius. With an increasing rotational support, a galaxy has an increasing probability of not containing any kinematical substructures, of having a more cored central photometric profile, and  a higher metallicity.

We compare the merger-sensitive parameters of FRs and SRs in the fifth main column of \tab{simplecorr}. The first sub-column indicates the average value of the given parameter for SRs. The second sub-column shows the analogous values for FRs. The numbers in parenthesis indicate the uncertainty in the last digit of the mean value. The last column is the probability that the statistical distributions of a given parameter are the same between the two groups of galaxies. This was obtained through the two-sample KS test. The SRs show, with respect to  FRs, a higher incidence and frequency of shells, a higher incidence of KDCs and CRCs, a lower incidence of regular kinematic fields, and they have larger effective radii. All of these differences can be expected simply by the fact that SRs are, on average, more massive than FRs.

\begin{table*}
        \caption{Correlations of the investigated merger-sensitive parameters with different properties of the galaxies. }
        \label{tab:simplecorr}
        \centering
        \begin{tabular}{l|cr|cr|cr|lll}
                \hline\hline
 & \multicolumn{2}{c|}{\rotsup} &  \multicolumn{2}{c|}{$\log \mjam$} & \multicolumn{2}{c|}{$\log \rhoio$} &  \multicolumn{3}{c}{SR/FR}\\ 
 Parameter & sign & $p$ [\%] & sign & $p$ [\%] & sign & $p$ [\%] & Mean SR & Mean FR  & $p$ [\%]\\\hline
 \hline
\input{tables_maintext/simplecorr.txt}
                \hline
        \end{tabular}
        \tablefoot{The sign sub-columns show the sign of the correlation, while the $p$ subcolumns show the probability that there is actually no correlation.  The following columns give the mean value of the merger indicator for galaxies sorted into SRs and FRs. The number in parenthesis indicates the uncertainty in the last digit of the mean. The last sub-column gives the probability that the statistical distributions of the given parameter is the same for SRs and FRs.}
\end{table*}

The correlations with the mass of the galaxy  can be read in the third main column of \tab{simplecorr}. The more massive a galaxy is, the more it is probable that it will host tidal features of any type, or that it will possess disturbed outer isophotes. It will also be more probable that it will host a higher number of shells and streams. More massive galaxies also tend to have a larger effective radius, older stellar populations, a higher metallicity, and a higher content of $\alpha$-elements. On the contrary, more massive galaxies tend to have more negative age gradients.

The correlations of merger-sensitive parameters with the density of environment are shown in the fourth main column of \tab{simplecorr}. We remind the reader that we do not have galaxies in clusters in our sample. We found two correlations that are statistically significant. Galaxies in denser environments have a higher probability of hosting at least some type of tidal features or a tidal disturbance.

For our sample, we found a significant correlation of the rotational support with $\log \mjam$ having a Pearson coefficient of -0.16 with a $p$-value of 3.6\%. This agrees with the dedicated studies (see \sect{intro}). For the correlation of $\log \rhoio$ with the rotational support, we found a Pearson coefficient of -0.06 with a $p$-value of 41\%. In agreement with a previous study of the \atlas sample  \citep{cappellari11b}, we thus found a very weak correlation between kinematics and environment within our subsample of mostly group galaxies. A significant trend between kinematics and environment was previously reported for the Virgo cluster galaxies (see Figs. 6-7 of \citealp{cappellari11b}). For our sample, the correlation of $\log \mjam$ and $\log \rhoio$ has a Pearson coefficient of 0.01 and a $p$-value of 90\%.

\section{Supplementary figures and data}
 \begin{figure}[b!]
        \resizebox{\hsize}{!}{\includegraphics{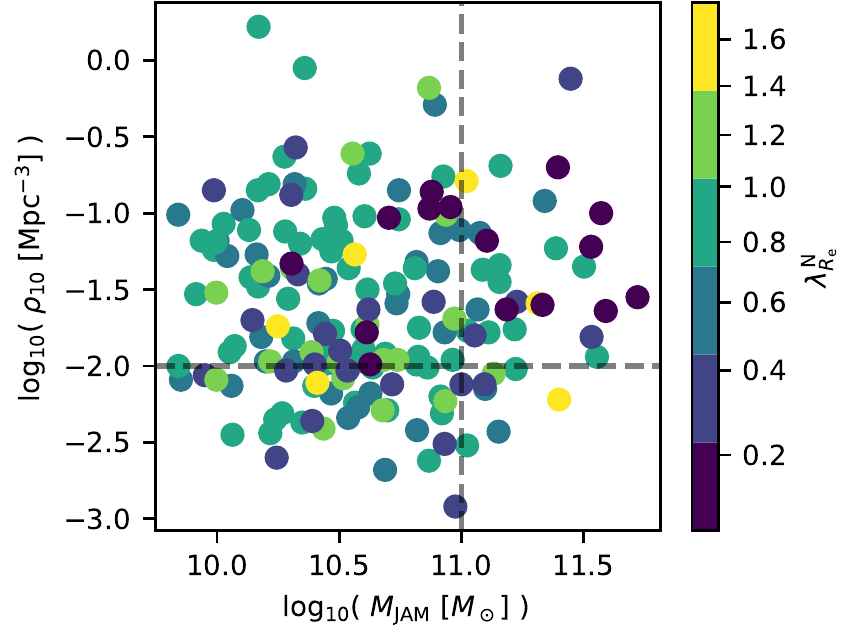}}
        \caption{Rotational support as a function of \mjam and \rhoio. The dashed lines indicate the cuts that we used to define the low-(high-) mass or density subsamples.} 
        \label{fig:mrs}
\end{figure}
\onecolumn
\begin{figure*}
        \centering
        \includegraphics[width=17cm]{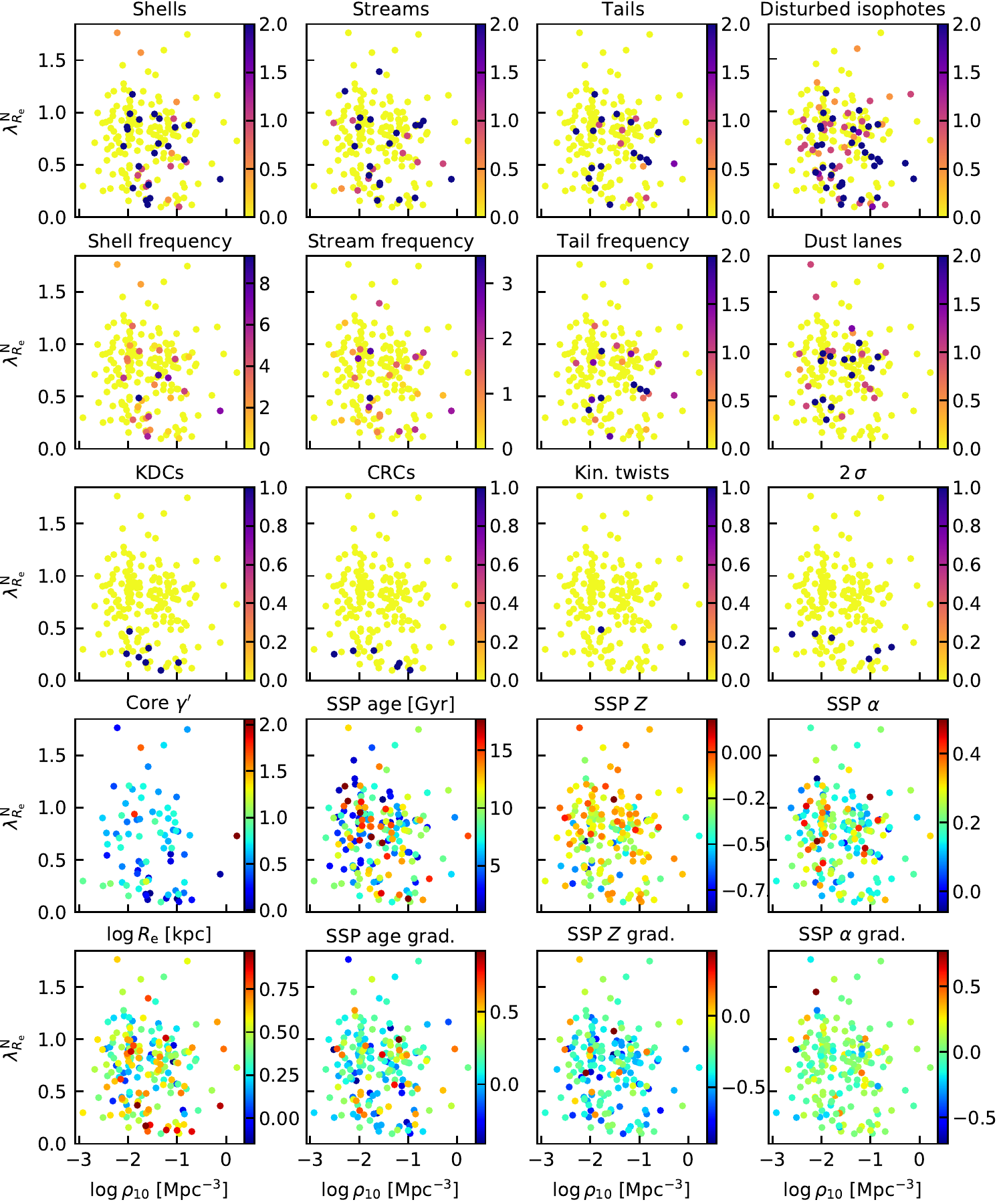}
        \caption{Investigated merger-sensitive parameters as functions of the rotational support (vertical axis in each tile) and the \rhoio environmental density (horizontal axis in each tile). We point out again that our sample does not contain galaxies in clusters. See \fig{2d-allMI} for an explanation of the meaning of the colors of the points.} 
        \label{fig:2d-allrho}
\end{figure*}

\begin{figure*}
        \centering
        \includegraphics[width=17cm]{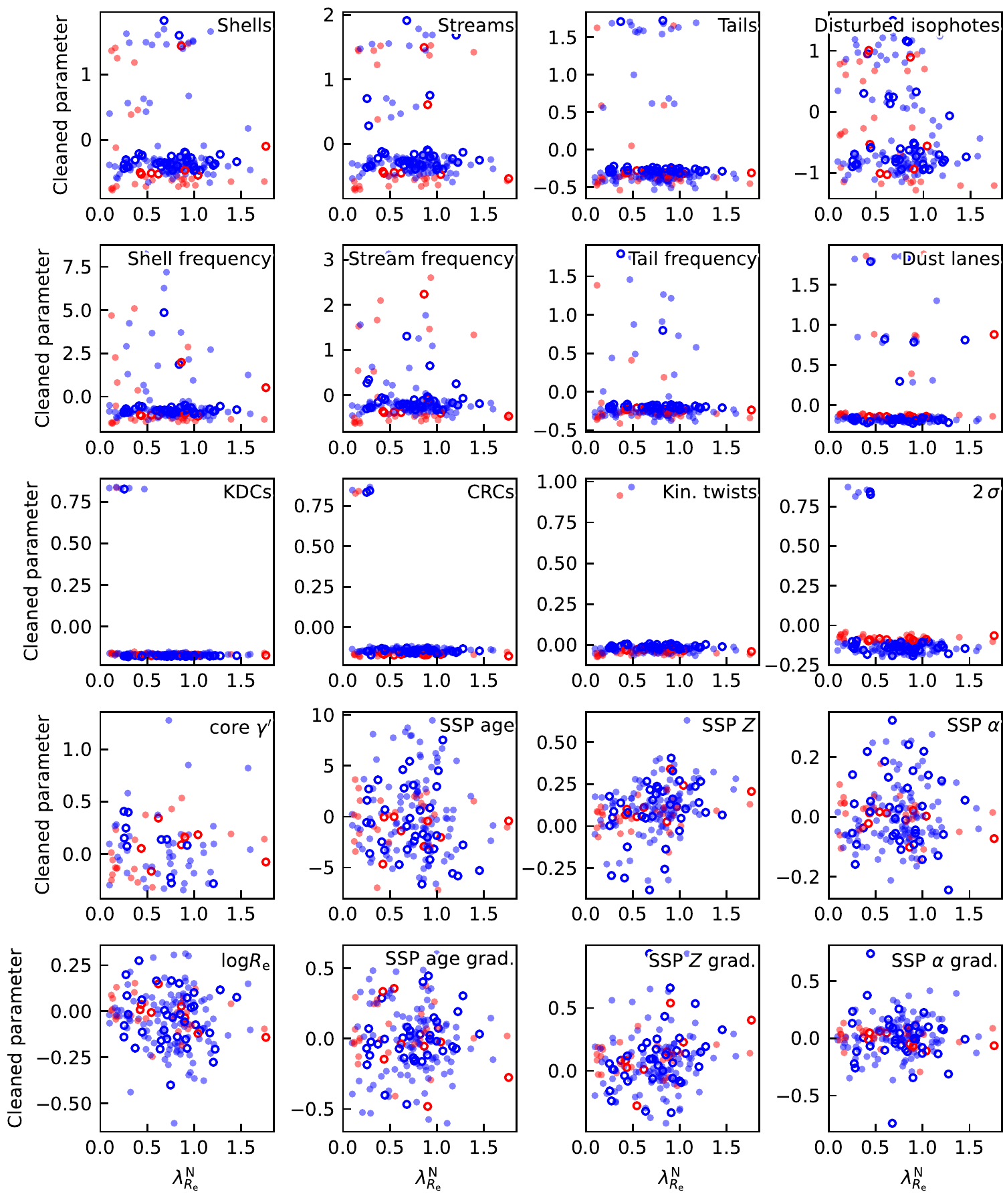}
        \caption{Relation of the cleaned merger-sensitive parameters  (vertical axis in each tile) and the rotational support (horizontal axis in each tile). Red points show galaxies with $\log \mjam\geq11$. Blue points show galaxies with $\log \mjam<11$. Full points show galaxies in environmental densities $\log \rhoio\geq-2$. Empty points show galaxies in environmental densities $\log \rhoio<-2$.} 
        \label{fig:cleaned}
\end{figure*}
\clearpage

\section{Linear fits}
\label{app:fits}

\begin{table*}[!htb]\vspace{-2.5ex}
        \caption{Linear fits of the characteristics of the galaxies in the form \textsl{[parameter]} $= b+ax$, where $x$ is either the rotational support, $\log \mjam$, or $\log \rhoio$.}
        \label{tab:simplefits}
        \centering
        \begin{tabular}{l|llc|llc|llc}
                \hline\hline
 &  \multicolumn{3}{c|}{Rotational support} & \multicolumn{3}{c|}{$\log \mjam$} & \multicolumn{3}{c}{$\log \rhoio$} \\
Parameter  & $b$ & $a$ & $r$ &  $b$ & $a$ & $r$ & $b$ & $a$ & $r$\\
\hline
\input{tables_fits/simplecorrcoeff.txt}
                \hline
        \end{tabular}
        \tablefoot{ The number in parenthesis indicates the uncertainty in the last decimal place of the measured value. The $r$ columns indicate the Pearson correlation coefficient multiplied by one hundred.}
\end{table*}

\begin{table*}\vspace{-60ex}
        \caption{Linear fits of characteristics of the galaxies in the form \textsl{[parameter]} $= b+a_M\log\mjam+a_\rho\log\rhoio+a_\mathrm{KS}\rotsup$ for our sample in its completeness. }\vspace{-2ex}
        \label{tab:rottarfitsall}
        \centering
        \begin{tabular}{llllll}
                \hline\hline
 &  \multicolumn{5}{c}{All}\\
Parameter  & $b$ & $a_M$ &$a_\rho$ & $a_\mathrm{KS}$ & $N_\mathrm{data}$  \\
\hline
\input{"tables_fits/Ftest_rotmes_tar-fits-all.txt"}
                \hline
        \end{tabular}
        \tablefoot{The number in parenthesis indicates the uncertainty in the last decimal place of the measured value.  The column $N_\mathrm{data}$ indicates the number of the available data points.}\vspace{-3ex}
\end{table*}

\begin{table*}
        \caption{Linear fits of characteristics of the galaxies in the form \textsl{[parameter]} $= b+a_M\log\mjam+a_\rho\log\rhoio+a_\mathrm{KS}\rotsup$ for our sample divided into two mass bins. }
        \label{tab:rottarfitsmass}
        \centering
        \begin{tabular}{llllll|lllll}
                \hline\hline
 &  \multicolumn{5}{c}{$\log \mjam<11$} &  \multicolumn{5}{c}{$\log \mjam\geq 11$}\\
Parameter  & $b$ & $a_M$ &$a_\rho$ & $a_\mathrm{KS}$ & $N_\mathrm{data}$ & $b$ & $a_M$ &$a_\rho$ & $a_\mathrm{KS}$ & $N_\mathrm{data}$  \\
\hline
\input{"tables_fits/Ftest_rotmes_tar-fits-masssplit.txt"}
                \hline
        \end{tabular}
        \tablefoot{The number in parenthesis indicates the uncertainty in the last decimal place of the measured value.  The $N_\mathrm{data}$ columns indicate the number of the available data points.}
\end{table*}

\begin{table*}
        \caption{Linear fits of characteristics of the galaxies in the form  \textsl{[parameter]} $= b+a_M\log\mjam+a_\rho\log\rhoio+a_\mathrm{KS}\rotsup$ for our sample divided into two environmental density bins. }
        \label{tab:rottarfitsdens}
        \centering
        \begin{tabular}{llllll|lllll}
                \hline\hline
 &  \multicolumn{5}{c}{$\log \rhoio<-2$} &  \multicolumn{5}{c}{$\log \rhoio\geq -2$}\\
Parameter  & $b$ & $a_M$ &$a_\rho$ & $a_\mathrm{KS}$ & $N_\mathrm{data}$ & $b$ & $a_M$ &$a_\rho$ & $a_\mathrm{KS}$ & $N_\mathrm{data}$  \\
\hline
\input{"tables_fits/Ftest_rotmes_tar-fits-denssplit.txt"}
                \hline
        \end{tabular}
        \tablefoot{The number in parenthesis indicates the uncertainty in the last decimal place of the measured value. The $N_\mathrm{data}$ columns indicate the number of the available data points.}
\end{table*}

\end{appendix}

\end{document}

%% file: tables_maintext/Ftest_rotmes_tar-all.txt
Shells               &  $-$ & $19$ &  $-$ & $53$ &  $-$ & $20$ &  $+$ & $79$ &  $-$ & $13$ \\
Streams              &  $-$ & $34$ &  $-$ & $29$ &  $-$ & $59$ &  $+$ & $88$ &  $-$ & $31$ \\
Tails                &  $-$ & $22$ &  $-$ & $24$ &  $-$ & $46$ &  $-$ & $43$ &  $-$ & $35$ \\
Disturbed isophotes  &  $-$ & {\large $\mathbf{0.64}$} &  $-$ & $6.0$ &  $-$ & $10$ &  $-$ & $29$ &  $-$ & $\mathbf{1.1}$ \\
Any TF               &  $-$ & $22$ &  $-$ & $30$ &  $-$ & $46$ &  $-$ & $95$ &  $-$ & $21$ \\
Any TF or DI         &  $-$ & $13$ &  $-$ & $29$ &  $-$ & $33$ &  $-$ & $36$ &  $-$ & $22$ \\
Shell fr.            &  $-$ & $16$ &  $-$ & $30$ &  $-$ & $39$ &  $+$ & $59$ &  $-$ & $11$ \\
Stream fr.           &  $-$ & $40$ &  $-$ & $14$ &  $-$ & $88$ &  $-$ & $78$ &  $-$ & $46$ \\
Tail fr.             &  $-$ & $20$ &  $-$ & $15$ &  $-$ & $65$ &  $-$ & $29$ &  $-$ & $38$ \\
Dust                 &  $+$ & $49$ &  $+$ & $97$ &  $+$ & $18$ &  $+$ & $33$ &  $+$ & $74$ \\
KDC                  &  $-$ & {\large $\mathbf{0.0072}$} &  $-$ & {\large $\mathbf{0.0037}$} &  $-$ & $47$ &  $-$ & $9.1$ &  $-$ & {\large $\mathbf{0.027}$} \\
CRC                  &  $-$ & {\large $\mathbf{0.014}$} &  $-$ & {\large $\mathbf{0.051}$} &  $-$ & $13$ &  $-$ & $\mathbf{2.7}$ &  $-$ & {\large $\mathbf{0.20}$} \\
KT                   &  $-$ & $36$ &  $-$ & $43$ &  $-$ & $85$& N/A & N/A  &  $-$ & $43$ \\
$2\,\sigma$          &  $-$ & {\large $\mathbf{0.13}$} &  $-$ & {\large $\mathbf{0.058}$}& N/A & N/A  &  $-$ & $26$ &  $-$ & {\large $\mathbf{0.46}$} \\
No kin. feature      &  $+$ & {\large $\mathbf{0.14}$} &  $+$ & {\large $\mathbf{0.0091}$} &  $-$ & $69$ &  $+$ & {\large $\mathbf{0.82}$} &  $+$ & $6.8$ \\
$\log R_e$           &  $-$ & $8.6$ &  $-$ & $30$ &  $-$ & $8.4$ &  $-$ & $21$ &  $-$ & $23$ \\
Core $\gamma^\prime$ &  $+$ & $35$ &  $+$ & $92$ &  $+$ & $19$ &  $-$ & $\mathbf{1.9}$ &  $+$ & $10$ \\
SSP age              &  $-$ & $87$ &  $-$ & $86$ &  $-$ & $76$ &  $-$ & $60$ &  $-$ & $97$ \\
SSP age gr.          &  $-$ & $82$ &  $+$ & $73$ &  $-$ & $70$ &  $+$ & $66$ &  $-$ & $78$ \\
SSP Z                &  $+$ & {\large $\mathbf{0.0025}$} &  $+$ & {\large $\mathbf{0.0075}$} &  $+$ & $15$ &  $+$ & $\mathbf{1.7}$ &  $+$ & {\large $\mathbf{0.089}$} \\
SSP Z gr.            &  $+$ & $\mathbf{3.2}$ &  $+$ & $6.2$ &  $+$ & $13$ &  $+$ & $\mathbf{1.5}$ &  $+$ & $33$ \\
SSP $\alpha$         &  $+$ & $83$ &  $+$ & $32$ &  $-$ & $\mathbf{3.5}$ &  $-$ & $30$ &  $+$ & $26$ \\
SSP $\alpha$ gr.     &  $+$ & $59$ &  $+$ & $31$ &  $-$ & $25$ &  $-$ & $34$ &  $+$ & $11$ \\

%% file: tables_maintext/simplecorr.txt
Shells               & - & 8.7 & + & $\mathbf{1.0}$ & + & 19 & 0.5(2) & 0.26(5) & $\mathbf{\ensuremath{3.1}}$\\
Streams              & - & 16 & + & $\mathbf{0.53}$ & + & 16 & 0.4(1) & 0.25(5) & \ensuremath{\geq25}\\
Tails                & - & 18 & + & 60 & + & 13 & 0.2(1) & 0.22(5) & \ensuremath{\geq25}\\
Disturbed isophotes  & - & {\large $\mathbf{0.090}$} & + & {\large $\mathbf{0.029}$} & + & 7.8 & 0.8(2) & 0.52(6) & \ensuremath{24}\\
Any TF               & - & 10 & + & $\mathbf{1.2}$ & + & $\mathbf{4.6}$ & 0.7(2) & 0.52(7) & \ensuremath{22}\\
Any TF or DI         & - & $\mathbf{2.8}$ & + & {\large $\mathbf{6.3E-3}$} & + & $\mathbf{4.4}$ & 0.9(2) & 0.69(7) & \ensuremath{22}\\
Shell fr.            & - & 8.7 & + & $\mathbf{4.8}$ & + & 29 & 1.0(4) & 0.5(1) & $\mathbf{\ensuremath{1.4}}$\\
Stream fr.           & - & 19 & + & $\mathbf{0.45}$ & + & 20 & 0.3(1) & 0.20(5) & \ensuremath{\geq25}\\
Tail fr.             & - & 15 & + & 50 & + & 14 & 0.11(7) & 0.16(4) & \ensuremath{\geq25}\\
Dust                 & + & 46 & - & 43 & - & 82 & 0.12(9) & 0.26(5) & \ensuremath{15}\\
KDC                  & - & {\large $\mathbf{6.1E-3}$} & + & 60 & - & 92 & 0.24(9) & 0.007(7) & $\mathbf{\ensuremath{\leq0.10}}$\\
CRC                  & - & {\large $\mathbf{6.4E-3}$} & + & 12 & - & 92 & 0.24(9) & 0(0) & $\mathbf{\ensuremath{\leq0.10}}$\\
KT                   & - & 19 & + & 6.1 & + & 11 & 0(0) & 0.013(9) & \ensuremath{\geq25}\\
$2\,\sigma$          & - & $\mathbf{0.44}$ & - & $\mathbf{4.7}$ & + & 67 & 0.08(5) & 0.03(1) & \ensuremath{11}\\
No kin. feature      & + & {\large $\mathbf{0.047}$} & - & 6.0 & - & 17 & 0.32(9) & 0.80(3) & $\mathbf{\ensuremath{\leq0.10}}$\\
$\log R_e $          & - & $\mathbf{1.3}$ & + & {\large $\mathbf{1.5E-27}$} & + & 33 & 0.60(4) & 0.43(2) & $\mathbf{\ensuremath{0.11}}$\\
Core $\gamma^\prime$ & + & $\mathbf{4.1}$ & - & {\large $\mathbf{0.88E-6}$} & - & 59 & 0.59(8) & 0.65(5) & \ensuremath{8.4}\\
SFH age              & - & 14 & + & {\large $\mathbf{0.035E-15}$} & + & 33 & 10.9(4) & 9.9(3) & \ensuremath{13}\\
SSP age              & - & 22 & + & {\large $\mathbf{2.2E-12}$} & + & 42 & 9.6(7) & 8.5(4) & \ensuremath{22}\\
SSP age gr.          & + & 68 & - & {\large $\mathbf{1.1E-3}$} & + & 20 & 0.09(5) & 0.18(2) & \ensuremath{5.1}\\
SSP Z                & + & $\mathbf{0.39}$ & + & {\large $\mathbf{0.25E-6}$} & + & 40 & -0.24(3) & -0.23(1) & \ensuremath{\geq25}\\
SSP Z gr.            & + & 6.9 & + & 14 & - & 30 & -0.36(4) & -0.35(2) & \ensuremath{\geq25}\\
SSP $\alpha$         & - & 87 & + & $\mathbf{1.0}$ & - & 31 & 0.19(2) & 0.214(8) & \ensuremath{\geq25}\\
SSP $\alpha$ gr.     & + & 48 & - & 37 & + & 59 & -0.00(3) & 0.05(1) & \ensuremath{13}\\

%% file: tables_fits/simplecorrcoeff.txt
Shells               & $0.5(1)$ & $-0.3(1)$ & -13 & $-3(1)$ & $0.3(1)$ & 19 & $0.5(1)$ & $0.11(9)$ & 9.9 \\
Streams              & $0.4(1)$ & $-0.2(1)$ & -11 & $-3(1)$ & $0.3(1)$ & 21 & $0.5(1)$ & $0.12(8)$ & 11 \\
Tails                & $0.4(1)$ & $-0.2(1)$ & -10 & $-0(1)$ & $0.1(1)$ & 4.0 & $0.4(1)$ & $0.12(8)$ & 11 \\
Disturbed isophotes  & $1.0(1)$ & $-0.6(2)$ & -25 & $-5(1)$ & $0.5(1)$ & 27 & $0.8(2)$ & $0.2(1)$ & 13 \\
Any TF               & $0.8(2)$ & $-0.3(2)$ & -12 & $-3(2)$ & $0.4(1)$ & 19 & $0.9(2)$ & $0.2(1)$ & 15 \\
Any TF or DI         & $1.0(2)$ & $-0.4(2)$ & -17 & $-6(2)$ & $0.6(1)$ & 30 & $1.1(2)$ & $0.2(1)$ & 15 \\
Shell fr.            & $1.0(3)$ & $-0.6(4)$ & -13 & $-5(3)$ & $0.6(3)$ & 15 & $0.9(4)$ & $0.2(2)$ & 8.0 \\
Stream fr.           & $0.3(1)$ & $-0.2(1)$ & -9.9 & $-3(1)$ & $0.3(1)$ & 21 & $0.4(1)$ & $0.10(8)$ & 9.7 \\
Tail fr.             & $0.27(8)$ & $-0.2(1)$ & -11 & $-0.4(9)$ & $0.06(8)$ & 5.1 & $0.3(1)$ & $0.09(6)$ & 11 \\
Dust                 & $0.2(1)$ & $0.1(1)$ & 5.6 & $1(1)$ & $-0.1(1)$ & -6.0 & $0.2(1)$ & $-0.02(8)$ & -1.8 \\
KDC                  & $0.17(3)$ & $-0.17(4)$ & -30 & $-0.2(4)$ & $0.02(4)$ & 4.0 & $0.04(4)$ & $-0.00(3)$ & -0.78 \\
CRC                  & $0.15(3)$ & $-0.16(4)$ & -30 & $-0.5(3)$ & $0.05(3)$ & 12 & $0.03(4)$ & $-0.00(2)$ & -0.74 \\
KT                   & $0.03(2)$ & $-0.03(2)$ & -9.9 & $-0.4(2)$ & $0.04(2)$ & 14 & $0.05(2)$ & $0.02(1)$ & 12 \\
$2\,\sigma$          & $0.12(3)$ & $-0.12(4)$ & -21 & $0.7(3)$ & $-0.06(3)$ & -15 & $0.05(4)$ & $0.01(2)$ & 3.2 \\
No kin. feature      & $0.48(8)$ & $0.3(1)$ & 26 & $2.3(8)$ & $-0.15(8)$ & -14 & $0.6(1)$ & $-0.08(6)$ & -10 \\
$\log R_e $          & $0.54(4)$ & $-0.12(5)$ & -19 & $-3.4(3)$ & $0.36(3)$ & 72 & $0.50(5)$ & $0.03(3)$ & 7.4 \\
Core $\gamma^\prime$ & $0.50(8)$ & $0.2(1)$ & 23 & $5.9(8)$ & $-0.49(8)$ & -59 & $0.6(1)$ & $-0.04(7)$ & -6.2 \\
SFH age              & $10.8(5)$ & $-1.0(7)$ & -11 & $-36(5)$ & $4.3(4)$ & 61 & $10.7(7)$ & $0.4(4)$ & 7.3 \\
SSP age              & $9.5(8)$ & $-1(1)$ & -9.3 & $-49(7)$ & $5.4(6)$ & 54 & $9(1)$ & $0.5(6)$ & 6.2 \\
SSP age gr.          & $0.15(4)$ & $0.02(6)$ & 3.1 & $2.2(4)$ & $-0.19(4)$ & -33 & $0.24(5)$ & $0.04(3)$ & 9.7 \\
SSP Z                & $-0.31(3)$ & $0.11(4)$ & 22 & $-2.0(3)$ & $0.17(3)$ & 43 & $-0.21(4)$ & $0.02(2)$ & 6.4 \\
SSP Z gr.            & $-0.42(4)$ & $0.09(5)$ & 14 & $-1.0(4)$ & $0.06(4)$ & 11 & $-0.40(5)$ & $-0.03(3)$ & -7.8 \\
SSP $\alpha$         & $0.21(2)$ & $-0.00(2)$ & -1.2 & $-0.3(2)$ & $0.05(2)$ & 19 & $0.19(2)$ & $-0.01(1)$ & -7.6 \\
SSP $\alpha$ gr.     & $0.02(3)$ & $0.03(4)$ & 5.4 & $0.3(3)$ & $-0.03(3)$ & -6.8 & $0.06(4)$ & $0.01(2)$ & 4.1 \\

%% file: tables_fits/Ftest_rotmes_tar-fits-all.txt
Shells               & $-2(1)$ & $0.3(1)$ & $0.10(9)$ & $-0.2(1)$ & 175 \\
Streams              & $-3(1)$ & $0.3(1)$ & $0.11(8)$ & $-0.1(1)$ & 175 \\
Tails                & $0(1)$ & $0.0(1)$ & $0.11(8)$ & $-0.2(1)$ & 175 \\
Disturbed isophotes  & $-3(2)$ & $0.4(1)$ & $0.2(1)$ & $-0.5(2)$ & 174 \\
Any TF               & $-3(2)$ & $0.3(1)$ & $0.2(1)$ & $-0.2(2)$ & 175 \\
Any TF or DI         & $-5(2)$ & $0.6(1)$ & $0.2(1)$ & $-0.3(2)$ & 174 \\
Shell fr.            & $-4(3)$ & $0.5(3)$ & $0.2(2)$ & $-0.5(4)$ & 175 \\
Stream fr.           & $-3(1)$ & $0.3(1)$ & $0.09(8)$ & $-0.1(1)$ & 175 \\
Tail fr.             & $-0.0(9)$ & $0.04(8)$ & $0.08(6)$ & $-0.1(1)$ & 175 \\
Dust                 & $1(1)$ & $-0.1(1)$ & $-0.01(8)$ & $0.1(1)$ & 175 \\
KDC                  & $0.2(4)$ & $-0.00(3)$ & $-0.01(3)$ & $-0.18(4)$ & 175 \\
CRC                  & $-0.2(3)$ & $0.03(3)$ & $-0.01(2)$ & $-0.16(4)$ & 175 \\
KT                   & $-0.3(2)$ & $0.03(2)$ & $0.02(1)$ & $-0.02(2)$ & 175 \\
$2\,\sigma$          & $1.0(3)$ & $-0.08(3)$ & $0.00(2)$ & $-0.13(4)$ & 175 \\
No kin. feature      & $1.6(8)$ & $-0.11(8)$ & $-0.06(6)$ & $0.3(1)$ & 175 \\
$\log R_e$           & $-3.3(3)$ & $0.36(3)$ & $0.01(2)$ & $-0.06(3)$ & 175 \\
Core $\gamma^\prime$ & $5.6(9)$ & $-0.47(8)$ & $-0.05(6)$ & $0.08(8)$ & 79 \\
SSP age              & $-48(7)$ & $5.4(7)$ & $0.4(5)$ & $-0.1(8)$ & 174 \\
SSP age gr.          & $2.3(5)$ & $-0.19(4)$ & $0.04(3)$ & $-0.01(5)$ & 175 \\
SSP Z                & $-2.3(3)$ & $0.18(3)$ & $0.02(2)$ & $0.14(3)$ & 174 \\
SSP Z gr.            & $-1.2(4)$ & $0.07(4)$ & $-0.02(3)$ & $0.11(5)$ & 175 \\
SSP $\alpha$         & $-0.3(2)$ & $0.05(2)$ & $-0.01(1)$ & $0.00(2)$ & 174 \\
SSP $\alpha$ gr.     & $0.3(3)$ & $-0.02(3)$ & $0.01(2)$ & $0.02(4)$ & 175 \\

%% file: tables_fits/Ftest_rotmes_tar-fits-masssplit.txt
Shells               & $-4(2)$ & $0.4(2)$ & $0.12(9)$ & $-0.1(2)$ & 140 & $-1(8)$ & $0.1(7)$ & $0.0(3)$ & $-0.4(3)$ & 35 \\
Streams              & $-2(2)$ & $0.2(2)$ & $0.09(8)$ & $-0.2(2)$ & 140 & $14(9)$ & $-1.1(8)$ & $0.3(3)$ & $-0.2(4)$ & 35 \\
Tails                & $-2(2)$ & $0.2(2)$ & $0.14(9)$ & $-0.2(2)$ & 140 & $3(6)$ & $-0.3(5)$ & $0.0(2)$ & $-0.2(2)$ & 35 \\
Disturbed isophotes  & $-2(2)$ & $0.3(2)$ & $0.1(1)$ & $-0.4(2)$ & 140 & $-7(9)$ & $0.7(8)$ & $0.1(3)$ & $-0.7(4)$ & 34 \\
Any TF               & $-4(2)$ & $0.5(2)$ & $0.2(1)$ & $-0.2(2)$ & 140 & $0(0)$ & $-0.1(9)$ & $0.2(3)$ & $-0.3(4)$ & 35 \\
Any TF or DI         & $-4(2)$ & $0.5(2)$ & $0.2(1)$ & $-0.2(2)$ & 140 & $-5(9)$ & $0.6(8)$ & $0.2(3)$ & $-0.4(4)$ & 34 \\
Shell fr.            & $-6(5)$ & $0.7(4)$ & $0.2(2)$ & $-0.4(4)$ & 140 & $-0(0)$ & $0(2)$ & $0.5(6)$ & $-0.6(7)$ & 35 \\
Stream fr.           & $-1(1)$ & $0.1(1)$ & $0.05(6)$ & $-0.2(1)$ & 140 & $20(0)$ & $-1.2(9)$ & $0.4(3)$ & $-0.1(4)$ & 35 \\
Tail fr.             & $-1(1)$ & $0.2(1)$ & $0.11(7)$ & $-0.2(1)$ & 140 & $2(5)$ & $-0.2(4)$ & $0.0(2)$ & $-0.1(2)$ & 35 \\
Dust                 & $2(2)$ & $-0.2(2)$ & $-0.03(8)$ & $0.0(2)$ & 140 & $-1(6)$ & $0.1(5)$ & $0.1(2)$ & $0.3(2)$ & 35 \\
KDC                  & $-0.1(6)$ & $0.03(5)$ & $-0.01(3)$ & $-0.23(5)$ & 140 & $-3(2)$ & $0.2(2)$ & $-0.04(6)$ & $-0.05(7)$ & 35 \\
CRC                  & $-0.5(5)$ & $0.05(5)$ & $-0.02(2)$ & $-0.16(4)$ & 140 & $1(2)$ & $-0.0(2)$ & $0.06(8)$ & $-0.2(1)$ & 35 \\
KT                   & $-0.3(3)$ & $0.03(2)$ & $-0.00(1)$ & $-0.02(2)$ & 140 & $-0(2)$ & $0.0(1)$ & $0.14(5)$ & $-0.01(6)$ & 35 \\
$2\,\sigma$          & $1.2(6)$ & $-0.10(5)$ & $0.01(3)$ & $-0.19(5)$ & 140& N/A & N/A  & N/A & N/A  & 35 \\
No kin. feature      & $3(1)$ & $-0.2(1)$ & $0.01(6)$ & $0.5(1)$ & 140 & $5(4)$ & $-0.4(4)$ & $-0.5(1)$ & $-0.1(2)$ & 35 \\
$\log R_e$           & $-2.5(5)$ & $0.28(4)$ & $0.01(2)$ & $-0.05(4)$ & 140 & $-5.0(9)$ & $0.51(8)$ & $-0.00(3)$ & $-0.06(3)$ & 35 \\
Core $\gamma^\prime$ & $5(2)$ & $-0.4(2)$ & $-0.03(8)$ & $0.0(1)$ & 53 & $9(3)$ & $-0.8(2)$ & $-0.04(9)$ & $0.1(1)$ & 26 \\
SSP age              & $-60(0)$ & $7(1)$ & $0.6(6)$ & $-0(1)$ & 139 & $-50(0)$ & $6(2)$ & $-0.6(9)$ & $-0(1)$ & 35 \\
SSP age gr.          & $3.2(7)$ & $-0.29(7)$ & $0.02(3)$ & $0.02(6)$ & 140 & $-1(2)$ & $0.1(2)$ & $0.07(7)$ & $-0.03(8)$ & 35 \\
SSP Z                & $-2.8(4)$ & $0.23(4)$ & $0.03(2)$ & $0.17(4)$ & 139 & $-0.6(9)$ & $0.03(8)$ & $-0.01(3)$ & $0.05(4)$ & 35 \\
SSP Z gr.            & $-0.3(7)$ & $-0.02(6)$ & $-0.03(3)$ & $0.12(6)$ & 140 & $-1(2)$ & $0.1(1)$ & $-0.01(5)$ & $0.10(6)$ & 35 \\
SSP $\alpha$         & $-0.4(3)$ & $0.06(3)$ & $-0.02(2)$ & $0.03(3)$ & 139 & $1.0(6)$ & $-0.06(6)$ & $0.01(2)$ & $-0.06(3)$ & 35 \\
SSP $\alpha$ gr.     & $0.1(5)$ & $-0.00(5)$ & $0.01(3)$ & $0.05(5)$ & 140 & $0.3(9)$ & $-0.02(8)$ & $0.02(3)$ & $-0.04(4)$ & 35 \\

%% file: tables_fits/Ftest_rotmes_tar-fits-denssplit.txt
Shells               & $1(2)$ & $0.0(2)$ & $0.8(4)$ & $0.1(2)$ & 43 & $-3(2)$ & $0.3(1)$ & $-0.0(1)$ & $-0.3(2)$ & 131 \\
Streams              & $2(3)$ & $-0.0(2)$ & $0.5(4)$ & $0.0(3)$ & 43 & $-3(1)$ & $0.4(1)$ & $0.1(1)$ & $-0.2(2)$ & 131 \\
Tails                & $1(2)$ & $-0.1(2)$ & $0.3(3)$ & $-0.2(2)$ & 43 & $-0(1)$ & $0.1(1)$ & $0.0(1)$ & $-0.2(2)$ & 131 \\
Disturbed isophotes  & $4(3)$ & $-0.0(3)$ & $1.3(5)$ & $-0.4(3)$ & 42 & $-5(2)$ & $0.5(2)$ & $0.1(1)$ & $-0.5(2)$ & 131 \\
Any TF               & $3(3)$ & $-0.0(3)$ & $1.0(5)$ & $-0.0(3)$ & 43 & $-4(2)$ & $0.4(2)$ & $0.1(2)$ & $-0.3(2)$ & 131 \\
Any TF or DI         & $5(3)$ & $-0.1(3)$ & $1.5(6)$ & $-0.3(4)$ & 42 & $-6(2)$ & $0.7(2)$ & $0.1(2)$ & $-0.3(2)$ & 131 \\
Shell fr.            & $4(5)$ & $-0.1(4)$ & $1.3(8)$ & $0.3(5)$ & 43 & $-5(4)$ & $0.6(4)$ & $-0.1(3)$ & $-0.7(5)$ & 131 \\
Stream fr.           & $1(2)$ & $0.1(2)$ & $0.6(4)$ & $-0.1(2)$ & 43 & $-3(1)$ & $0.3(1)$ & $0.1(1)$ & $-0.1(2)$ & 131 \\
Tail fr.             & $2(2)$ & $-0.1(1)$ & $0.3(3)$ & $-0.2(2)$ & 43 & $-0(1)$ & $0.1(1)$ & $0.03(9)$ & $-0.1(1)$ & 131 \\
Dust                 & $3(2)$ & $-0.2(2)$ & $0.3(4)$ & $0.2(2)$ & 43 & $1(1)$ & $-0.0(1)$ & $-0.1(1)$ & $0.1(2)$ & 131 \\
KDC                  & $0.4(7)$ & $0.00(6)$ & $0.2(1)$ & $-0.12(7)$ & 43 & $0.2(5)$ & $-0.01(4)$ & $-0.05(4)$ & $-0.20(5)$ & 131 \\
CRC                  & $-0.7(9)$ & $0.09(8)$ & $0.0(2)$ & $-0.22(9)$ & 43 & $-0.1(4)$ & $0.02(3)$ & $0.01(3)$ & $-0.14(4)$ & 131 \\
KT                  & N/A & N/A  & N/A & N/A  & 43 & $-0.4(3)$ & $0.04(3)$ & $0.03(2)$ & $-0.03(3)$ & 131 \\
$2\,\sigma$          & $1.5(9)$ & $-0.15(9)$ & $-0.1(2)$ & $-0.1(1)$ & 43 & $0.9(4)$ & $-0.07(3)$ & $0.04(3)$ & $-0.13(4)$ & 131 \\
No kin. feature      & $0(2)$ & $0.1(2)$ & $0.3(3)$ & $0.5(2)$ & 43 & $2(1)$ & $-0.17(9)$ & $-0.12(8)$ & $0.2(1)$ & 131 \\
$\log R_e$           & $-2.9(6)$ & $0.33(6)$ & $0.1(1)$ & $-0.08(7)$ & 43 & $-3.3(3)$ & $0.36(3)$ & $0.03(3)$ & $-0.05(4)$ & 131 \\
Core $\gamma^\prime$ & $3(2)$ & $-0.2(1)$ & $-0.1(2)$ & $-0.3(1)$ & 18 & $6(1)$ & $-0.47(9)$ & $0.01(9)$ & $0.2(1)$ & 61 \\
SSP age              & $-60(0)$ & $7(2)$ & $0(3)$ & $-1(2)$ & 43 & $-44(8)$ & $5.0(8)$ & $-0.3(7)$ & $-0(1)$ & 130 \\
SSP age gr.          & $3.0(9)$ & $-0.28(9)$ & $-0.0(2)$ & $0.0(1)$ & 43 & $2.1(5)$ & $-0.17(5)$ & $0.04(4)$ & $-0.02(6)$ & 131 \\
SSP Z                & $-4.1(7)$ & $0.30(6)$ & $-0.2(1)$ & $0.18(7)$ & 43 & $-1.9(3)$ & $0.15(3)$ & $0.01(2)$ & $0.12(4)$ & 130 \\
SSP Z gr.            & $-1(1)$ & $0.02(9)$ & $-0.0(2)$ & $0.3(1)$ & 43 & $-1.3(5)$ & $0.09(4)$ & $0.01(4)$ & $0.05(5)$ & 131 \\
SSP $\alpha$         & $0.5(5)$ & $0.00(5)$ & $0.12(9)$ & $-0.06(5)$ & 43 & $-0.6(2)$ & $0.07(2)$ & $-0.01(2)$ & $0.03(3)$ & 130 \\
SSP $\alpha$ gr.     & $0(1)$ & $-0.02(9)$ & $0.1(2)$ & $-0.1(1)$ & 43 & $0.3(3)$ & $-0.03(3)$ & $-0.01(3)$ & $0.06(4)$ & 131 \\

%% file: main.bbl
\begin{thebibliography}{220}
\expandafter\ifx\csname natexlab\endcsname\relax\def\natexlab#1{#1}\fi

\bibitem[{{Afruni} {et~al.}(2019){Afruni}, {Fraternali}, \&
  {Pezzulli}}]{afruni19}
{Afruni}, A., {Fraternali}, F., \& {Pezzulli}, G. 2019, \aap, 625, A11

\bibitem[{{Algorry} {et~al.}(2014){Algorry}, {Navarro}, {Abadi}, {Sales},
  {Steinmetz}, \& {Piontek}}]{algorry14}
{Algorry}, D.~G., {Navarro}, J.~F., {Abadi}, M.~G., {et~al.} 2014, \mnras, 437,
  3596

\bibitem[{{Amorisco}(2015)}]{amorisco15}
{Amorisco}, N.~C. 2015, \mnras, 450, 575

\bibitem[{{Amorisco}(2017)}]{amorisco17}
{Amorisco}, N.~C. 2017, \mnras, 464, 2882

\bibitem[{{Ardila} {et~al.}(2018){Ardila}, {Alatalo}, {Lanz}, {Appleton},
  {Beaton}, {Bitsakis}, {Cales}, {Falc{\'o}n-Barroso}, {Kewley}, {Medling},
  {Mulchaey}, {Nyland}, {Rich}, \& {Urry}}]{ardila18}
{Ardila}, F., {Alatalo}, K., {Lanz}, L., {et~al.} 2018, \apj, 863, 28

\bibitem[{{Atkinson} {et~al.}(2013){Atkinson}, {Abraham}, \&
  {Ferguson}}]{atkinson13}
{Atkinson}, A.~M., {Abraham}, R.~G., \& {Ferguson}, A. M.~N. 2013, \apj, 765,
  28

\bibitem[{{Bah{\'e}} {et~al.}(2017){Bah{\'e}}, {Barnes}, {Dalla Vecchia},
  {Kay}, {White}, {McCarthy}, {Schaye}, {Bower}, {Crain}, {Theuns}, {Jenkins},
  {McGee}, {Schaller}, {Thomas}, \& {Trayford}}]{bahe17}
{Bah{\'e}}, Y.~M., {Barnes}, D.~J., {Dalla Vecchia}, C., {et~al.} 2017, \mnras,
  470, 4186

\bibitem[{{Beasley} {et~al.}(2018){Beasley}, {Trujillo}, {Leaman}, \&
  {Montes}}]{beasley18}
{Beasley}, M.~A., {Trujillo}, I., {Leaman}, R., \& {Montes}, M. 2018, \nat,
  555, 483

\bibitem[{{Bender} \& {Surma}(1992)}]{bender92}
{Bender}, R. \& {Surma}, P. 1992, \aap, 258, 250

\bibitem[{{Bennett} {et~al.}(2014){Bennett}, {Larson}, {Weiland}, \&
  {Hinshaw}}]{bennett14}
{Bennett}, C.~L., {Larson}, D., {Weiland}, J.~L., \& {Hinshaw}, G. 2014, \apj,
  794, 135

\bibitem[{{Bernardi} {et~al.}(2019){Bernardi}, {Dom{\'\i}nguez S{\'a}nchez},
  {Brownstein}, {Drory}, \& {Sheth}}]{bernardi19}
{Bernardi}, M., {Dom{\'\i}nguez S{\'a}nchez}, H., {Brownstein}, J.~R., {Drory},
  N., \& {Sheth}, R.~K. 2019, \mnras, 489, 5633

\bibitem[{{B{\'\i}lek}(2016)}]{bildiz}
{B{\'\i}lek}, M. 2016, arXiv e-prints, arXiv:1601.01240

\bibitem[{{B{\'\i}lek} {et~al.}(2020){B{\'\i}lek}, {Duc}, {Cuillandre}, {Gwyn},
  {Cappellari}, {Bekaert}, {Bonfini}, {Bitsakis}, {Paudel}, {Krajnovi{\'c}},
  {Durrell}, \& {Marleau}}]{bil20}
{B{\'\i}lek}, M., {Duc}, P.-A., {Cuillandre}, J.-C., {et~al.} 2020, \mnras,
  498, 2138

\bibitem[{{B{\'\i}lek} {et~al.}(2019){B{\'\i}lek}, {Samurovi{\'c}}, \&
  {Renaud}}]{bil19}
{B{\'\i}lek}, M., {Samurovi{\'c}}, S., \& {Renaud}, F. 2019, \aap, 625, A32

\bibitem[{{B{\'\i}lek} {et~al.}(2018){B{\'\i}lek}, {Thies}, {Kroupa}, \&
  {Famaey}}]{bil18}
{B{\'\i}lek}, M., {Thies}, I., {Kroupa}, P., \& {Famaey}, B. 2018, \aap, 614,
  A59

\bibitem[{B\'\i{}lek {et~al.}(2019)B\'\i{}lek, Thies, Kroupa, \&
  Famaey}]{bil19d}
B\'\i{}lek, M., Thies, I., Kroupa, P., \& Famaey, B. 2019, in {IAU Symposium
  355}: {The Realm of the Low Surface Brightness Universe}

\bibitem[{{Bitsakis} {et~al.}(2016){Bitsakis}, {Dultzin}, {Ciesla},
  {D{\'\i}az-Santos}, {Appleton}, {Charmandaris}, {Krongold}, {Guillard},
  {Alatalo}, {Zezas}, {Gonz{\'a}lez}, \& {Lanz}}]{bitsakis16}
{Bitsakis}, T., {Dultzin}, D., {Ciesla}, L., {et~al.} 2016, \mnras, 459, 957

\bibitem[{{Bitsakis} {et~al.}(2019){Bitsakis}, {S{\'a}nchez}, {Ciesla},
  {Bonfini}, {Charmandaris}, {Cervantes Sodi}, {Maragkoudakis}, {Diaz-Santos},
  \& {Zezas}}]{bitsakis19}
{Bitsakis}, T., {S{\'a}nchez}, S.~F., {Ciesla}, L., {et~al.} 2019, \mnras, 483,
  370

\bibitem[{{Blanton} \& {Moustakas}(2009)}]{blanton09}
{Blanton}, M.~R. \& {Moustakas}, J. 2009, \araa, 47, 159

\bibitem[{{Bournaud} {et~al.}(2005){Bournaud}, {Jog}, \& {Combes}}]{bournaud05}
{Bournaud}, F., {Jog}, C.~J., \& {Combes}, F. 2005, \aap, 437, 69

\bibitem[{{Bournaud} {et~al.}(2007){Bournaud}, {Jog}, \& {Combes}}]{bournaud07}
{Bournaud}, F., {Jog}, C.~J., \& {Combes}, F. 2007, \aap, 476, 1179

\bibitem[{{Bournaud} {et~al.}(2014){Bournaud}, {Perret}, {Renaud}, {Dekel},
  {Elmegreen}, {Elmegreen}, {Teyssier}, {Amram}, {Daddi}, {Duc}, {Elbaz},
  {Epinat}, {Gabor}, {Juneau}, {Kraljic}, \& {Le Floch'}}]{bournaud14}
{Bournaud}, F., {Perret}, V., {Renaud}, F., {et~al.} 2014, \apj, 780, 57

\bibitem[{{Brooks} \& {Christensen}(2016)}]{brooks16}
{Brooks}, A. \& {Christensen}, C. 2016, {Bulge Formation via Mergers in
  Cosmological Simulations}, ed. E.~{Laurikainen}, R.~{Peletier}, \&
  D.~{Gadotti}, Vol. 418, 317

\bibitem[{{Brough} {et~al.}(2017){Brough}, {van de Sande}, {Owers},
  {d'Eugenio}, {Sharp}, {Cortese}, {Scott}, {Croom}, {Bassett}, {Bekki},
  {Bland-Hawthorn}, {Bryant}, {Davies}, {Drinkwater}, {Driver}, {Foster},
  {Goldstein}, {L{\'o}pez-S{\'a}nchez}, {Medling}, {Sweet}, {Taranu}, {Tonini},
  {Yi}, {Goodwin}, {Lawrence}, \& {Richards}}]{brough17}
{Brough}, S., {van de Sande}, J., {Owers}, M.~S., {et~al.} 2017, \apj, 844, 59

\bibitem[{{Cappellari}(2008)}]{cappellari08}
{Cappellari}, M. 2008, \mnras, 390, 71

\bibitem[{{Cappellari}(2016)}]{cappellari16}
{Cappellari}, M. 2016, \araa, 54, 597

\bibitem[{{Cappellari} {et~al.}(2007){Cappellari}, {Emsellem}, {Bacon},
  {Bureau}, {Davies}, {de Zeeuw}, {Falc{\'o}n-Barroso}, {Krajnovi{\'c}},
  {Kuntschner}, {McDermid}, {Peletier}, {Sarzi}, {van den Bosch}, \& {van de
  Ven}}]{cappellari07}
{Cappellari}, M., {Emsellem}, E., {Bacon}, R., {et~al.} 2007, \mnras, 379, 418

\bibitem[{{Cappellari} {et~al.}(2011{\natexlab{a}}){Cappellari}, {Emsellem},
  {Krajnovi{\'c}}, {McDermid}, {Scott}, {Verdoes Kleijn}, {Young}, {Alatalo},
  {Bacon}, {Blitz}, {Bois}, {Bournaud}, {Bureau}, {Davies}, {Davis}, {de
  Zeeuw}, {Duc}, {Khochfar}, {Kuntschner}, {Lablanche}, {Morganti}, {Naab},
  {Oosterloo}, {Sarzi}, {Serra}, \& {Weijmans}}]{cappellari11a}
{Cappellari}, M., {Emsellem}, E., {Krajnovi{\'c}}, D., {et~al.}
  2011{\natexlab{a}}, \mnras, 413, 813

\bibitem[{{Cappellari} {et~al.}(2011{\natexlab{b}}){Cappellari}, {Emsellem},
  {Krajnovi{\'c}}, {McDermid}, {Serra}, {Alatalo}, {Blitz}, {Bois}, {Bournaud},
  {Bureau}, {Davies}, {Davis}, {de Zeeuw}, {Khochfar}, {Kuntschner},
  {Lablanche}, {Morganti}, {Naab}, {Oosterloo}, {Sarzi}, {Scott}, {Weijmans},
  \& {Young}}]{cappellari11b}
{Cappellari}, M., {Emsellem}, E., {Krajnovi{\'c}}, D., {et~al.}
  2011{\natexlab{b}}, \mnras, 416, 1680

\bibitem[{{Cappellari} {et~al.}(2013){Cappellari}, {Scott}, {Alatalo}, {Blitz},
  {Bois}, {Bournaud}, {Bureau}, {Crocker}, {Davies}, {Davis}, {de Zeeuw},
  {Duc}, {Emsellem}, {Khochfar}, {Krajnovi{\'c}}, {Kuntschner}, {McDermid},
  {Morganti}, {Naab}, {Oosterloo}, {Sarzi}, {Serra}, {Weijmans}, \&
  {Young}}]{cappellari13a}
{Cappellari}, M., {Scott}, N., {Alatalo}, K., {et~al.} 2013, \mnras, 432, 1709

\bibitem[{{Carleton} {et~al.}(2020){Carleton}, {Guo}, {Nayyeri}, {Cooper},
  {Rudnick}, \& {Whitaker}}]{carleton20}
{Carleton}, T., {Guo}, Y., {Nayyeri}, H., {et~al.} 2020, \mnras, 491, 2822

\bibitem[{{Carnall} {et~al.}(2022){Carnall}, {McLeod}, {McLure}, {Dunlop},
  {Begley}, {Cullen}, {Donnan}, {Hamadouche}, {Jewell}, {Jones}, {Pollock}, \&
  {Wild}}]{carnall22}
{Carnall}, A.~C., {McLeod}, D.~J., {McLure}, R.~J., {et~al.} 2022, arXiv
  e-prints, arXiv:2208.00986

\bibitem[{{Carnall} {et~al.}(2020){Carnall}, {Walker}, {McLure}, {Dunlop},
  {McLeod}, {Cullen}, {Wild}, {Amorin}, {Bolzonella}, {Castellano}, {Cimatti},
  {Cucciati}, {Fontana}, {Gargiulo}, {Garilli}, {Jarvis}, {Pentericci},
  {Pozzetti}, {Zamorani}, {Calabro}, {Hathi}, \& {Koekemoer}}]{carnall20}
{Carnall}, A.~C., {Walker}, S., {McLure}, R.~J., {et~al.} 2020, \mnras, 496,
  695

\bibitem[{{Cava} {et~al.}(2018){Cava}, {Schaerer}, {Richard},
  {P{\'e}rez-Gonz{\'a}lez}, {Dessauges-Zavadsky}, {Mayer}, \&
  {Tamburello}}]{cava18}
{Cava}, A., {Schaerer}, D., {Richard}, J., {et~al.} 2018, Nature Astronomy, 2,
  76

\bibitem[{{Chang} {et~al.}(2013){Chang}, {van der Wel}, {Rix}, {Wuyts},
  {Zibetti}, {Ramkumar}, \& {Holden}}]{chang13}
{Chang}, Y.-Y., {van der Wel}, A., {Rix}, H.-W., {et~al.} 2013, \apj, 762, 83

\bibitem[{{Chu} {et~al.}(2021){Chu}, {Durret}, \& {M{\'a}rquez}}]{chu21}
{Chu}, A., {Durret}, F., \& {M{\'a}rquez}, I. 2021, \aap, 649, A42

\bibitem[{{Chu} {et~al.}(2022){Chu}, {Sarron}, {Durret}, \&
  {M{\'a}rquez}}]{chu22}
{Chu}, A., {Sarron}, F., {Durret}, F., \& {M{\'a}rquez}, I. 2022, arXiv
  e-prints, arXiv:2206.14209

\bibitem[{{Clarke} {et~al.}(2019){Clarke}, {Debattista}, {Nidever}, {Loebman},
  {Simons}, {Kassin}, {Du}, {Ness}, {Fisher}, {Quinn}, {Wadsley}, {Freeman}, \&
  {Popescu}}]{clarke19}
{Clarke}, A.~J., {Debattista}, V.~P., {Nidever}, D.~L., {et~al.} 2019, \mnras,
  484, 3476

\bibitem[{{Cole} {et~al.}(2020){Cole}, {Bezanson}, {van der Wel}, {Bell},
  {D'Eugenio}, {Franx}, {Gallazzi}, {van Houdt}, {Muzzin}, {Pacifici}, {van de
  Sande}, {Sobral}, {Straatman}, \& {Wu}}]{cole20}
{Cole}, J., {Bezanson}, R., {van der Wel}, A., {et~al.} 2020, \apjl, 890, L25

\bibitem[{{Conselice} {et~al.}(2008){Conselice}, {Rajgor}, \&
  {Myers}}]{conselice08}
{Conselice}, C.~J., {Rajgor}, S., \& {Myers}, R. 2008, \mnras, 386, 909

\bibitem[{{Costantin} {et~al.}(2021){Costantin}, {P{\'e}rez-Gonz{\'a}lez},
  {M{\'e}ndez-Abreu}, {Huertas-Company}, {Dimauro}, {Alcalde-Pampliega},
  {Buitrago}, {Ceverino}, {Daddi}, {Dom{\'\i}nguez-S{\'a}nchez},
  {Espino-Briones}, {Hern{\'a}n-Caballero}, {Koekemoer}, \&
  {Rodighiero}}]{costantin21}
{Costantin}, L., {P{\'e}rez-Gonz{\'a}lez}, P.~G., {M{\'e}ndez-Abreu}, J.,
  {et~al.} 2021, \apj, 913, 125

\bibitem[{{C{\^o}t{\'e}} {et~al.}(1998){C{\^o}t{\'e}}, {Marzke}, \&
  {West}}]{cote98}
{C{\^o}t{\'e}}, P., {Marzke}, R.~O., \& {West}, M.~J. 1998, \apj, 501, 554

\bibitem[{{Daddi} {et~al.}(2005){Daddi}, {Renzini}, {Pirzkal}, {Cimatti},
  {Malhotra}, {Stiavelli}, {Xu}, {Pasquali}, {Rhoads}, {Brusa}, {di Serego
  Alighieri}, {Ferguson}, {Koekemoer}, {Moustakas}, {Panagia}, \&
  {Windhorst}}]{daddi05}
{Daddi}, E., {Renzini}, A., {Pirzkal}, N., {et~al.} 2005, \apj, 626, 680

\bibitem[{{Damjanov} {et~al.}(2009){Damjanov}, {McCarthy}, {Abraham},
  {Glazebrook}, {Yan}, {Mentuch}, {Le Borgne}, {Savaglio}, {Crampton},
  {Murowinski}, {Juneau}, {Carlberg}, {J{\o}rgensen}, {Roth}, {Chen}, \&
  {Marzke}}]{damjanov09}
{Damjanov}, I., {McCarthy}, P.~J., {Abraham}, R.~G., {et~al.} 2009, \apj, 695,
  101

\bibitem[{{de la Rosa} {et~al.}(2016){de la Rosa}, {La Barbera}, {Ferreras},
  {S{\'a}nchez Almeida}, {Dalla Vecchia}, {Mart{\'\i}nez-Valpuesta}, \&
  {Stringer}}]{delarosa19}
{de la Rosa}, I.~G., {La Barbera}, F., {Ferreras}, I., {et~al.} 2016, \mnras,
  457, 1916

\bibitem[{{De Lucia} \& {Blaizot}(2007)}]{delucia07}
{De Lucia}, G. \& {Blaizot}, J. 2007, \mnras, 375, 2

\bibitem[{{De Lucia} {et~al.}(2017){De Lucia}, {Fontanot}, \&
  {Hirschmann}}]{delucia17}
{De Lucia}, G., {Fontanot}, F., \& {Hirschmann}, M. 2017, \mnras, 466, L88

\bibitem[{{Dekel} \& {Birnboim}(2006)}]{dekel06}
{Dekel}, A. \& {Birnboim}, Y. 2006, \mnras, 368, 2

\bibitem[{{D'Eugenio} {et~al.}(2013){D'Eugenio}, {Houghton}, {Davies}, \&
  {Dalla Bont{\`a}}}]{deugenio13}
{D'Eugenio}, F., {Houghton}, R.~C.~W., {Davies}, R.~L., \& {Dalla Bont{\`a}},
  E. 2013, \mnras, 429, 1258

\bibitem[{{Di Matteo} {et~al.}(2009{\natexlab{a}}){Di Matteo}, {Jog},
  {Lehnert}, {Combes}, \& {Semelin}}]{dimatteo09b}
{Di Matteo}, P., {Jog}, C.~J., {Lehnert}, M.~D., {Combes}, F., \& {Semelin}, B.
  2009{\natexlab{a}}, \aap, 501, L9

\bibitem[{{Di Matteo} {et~al.}(2009{\natexlab{b}}){Di Matteo}, {Pipino},
  {Lehnert}, {Combes}, \& {Semelin}}]{dimatteo09}
{Di Matteo}, P., {Pipino}, A., {Lehnert}, M.~D., {Combes}, F., \& {Semelin}, B.
  2009{\natexlab{b}}, \aap, 499, 427

\bibitem[{{Dubinski}(1998)}]{dubinski98}
{Dubinski}, J. 1998, \apj, 502, 141

\bibitem[{{Duc} {et~al.}(2015){Duc}, {Cuillandre}, {Karabal}, {Cappellari},
  {Alatalo}, {Blitz}, {Bournaud}, {Bureau}, {Crocker}, {Davies}, {Davis}, {de
  Zeeuw}, {Emsellem}, {Khochfar}, {Krajnovi{\'c}}, {Kuntschner}, {McDermid},
  {Michel-Dansac}, {Morganti}, {Naab}, {Oosterloo}, {Paudel}, {Sarzi}, {Scott},
  {Serra}, {Weijmans}, \& {Young}}]{duc15}
{Duc}, P.-A., {Cuillandre}, J.-C., {Karabal}, E., {et~al.} 2015, \mnras, 446,
  120

\bibitem[{{Dudzevi{\v{c}}i{\={u}}t{\.{e}}}
  {et~al.}(2020){Dudzevi{\v{c}}i{\={u}}t{\.{e}}}, {Smail}, {Swinbank}, {Stach},
  {Almaini}, {da Cunha}, {An}, {Arumugam}, {Birkin}, {Blain}, {Chapman},
  {Chen}, {Conselice}, {Coppin}, {Dunlop}, {Farrah}, {Geach}, {Gullberg},
  {Hartley}, {Hodge}, {Ivison}, {Maltby}, {Scott}, {Simpson}, {Simpson},
  {Thomson}, {Walter}, {Wardlow}, {Weiss}, \& {van der Werf}}]{dudzeviciute20}
{Dudzevi{\v{c}}i{\={u}}t{\.{e}}}, U., {Smail}, I., {Swinbank}, A.~M., {et~al.}
  2020, \mnras, 494, 3828

\bibitem[{{Ebrov{\'a}} {et~al.}(2020){Ebrov{\'a}}, {B{\'\i}lek},
  {Y{\i}ld{\i}z}, \& {Eli{\'a}{\v{s}}ek}}]{ebrova20}
{Ebrov{\'a}}, I., {B{\'\i}lek}, M., {Y{\i}ld{\i}z}, M.~K., \&
  {Eli{\'a}{\v{s}}ek}, J. 2020, \aap, 634, A73

\bibitem[{{Ebrov{\'a}} {et~al.}(2021){Ebrov{\'a}}, {{\L}okas}, \&
  {Eli{\'a}{\v{s}}ek}}]{ebrova20b}
{Ebrov{\'a}}, I., {{\L}okas}, E.~L., \& {Eli{\'a}{\v{s}}ek}, J. 2021, \aap,
  647, A103

\bibitem[{{Elmegreen} {et~al.}(2008){Elmegreen}, {Bournaud}, \&
  {Elmegreen}}]{elmegreen08}
{Elmegreen}, B.~G., {Bournaud}, F., \& {Elmegreen}, D.~M. 2008, \apj, 688, 67

\bibitem[{{Elmegreen} \& {Elmegreen}(2014)}]{elmegreen14}
{Elmegreen}, D.~M. \& {Elmegreen}, B.~G. 2014, \apj, 781, 11

\bibitem[{{Elmegreen} {et~al.}(2009){Elmegreen}, {Elmegreen}, {Marcus},
  {Shahinyan}, {Yau}, \& {Petersen}}]{elmegreen09}
{Elmegreen}, D.~M., {Elmegreen}, B.~G., {Marcus}, M.~T., {et~al.} 2009, \apj,
  701, 306

\bibitem[{{Emsellem} {et~al.}(2011){Emsellem}, {Cappellari}, {Krajnovi{\'c}},
  {Alatalo}, {Blitz}, {Bois}, {Bournaud}, {Bureau}, {Davies}, {Davis}, {de
  Zeeuw}, {Khochfar}, {Kuntschner}, {Lablanche}, {McDermid}, {Morganti},
  {Naab}, {Oosterloo}, {Sarzi}, {Scott}, {Serra}, {van de Ven}, {Weijmans}, \&
  {Young}}]{emsellem11}
{Emsellem}, E., {Cappellari}, M., {Krajnovi{\'c}}, D., {et~al.} 2011, \mnras,
  414, 888

\bibitem[{{Emsellem} {et~al.}(2007){Emsellem}, {Cappellari}, {Krajnovi{\'c}},
  {van de Ven}, {Bacon}, {Bureau}, {Davies}, {de Zeeuw}, {Falc{\'o}n-Barroso},
  {Kuntschner}, {McDermid}, {Peletier}, \& {Sarzi}}]{emsellem07}
{Emsellem}, E., {Cappellari}, M., {Krajnovi{\'c}}, D., {et~al.} 2007, \mnras,
  379, 401

\bibitem[{{Ene} {et~al.}(2018){Ene}, {Ma}, {Veale}, {Greene}, {Thomas},
  {Blakeslee}, {Foster}, {Walsh}, {Ito}, \& {Goulding}}]{ene18}
{Ene}, I., {Ma}, C.-P., {Veale}, M., {et~al.} 2018, \mnras, 479, 2810

\bibitem[{{Estrada-Carpenter} {et~al.}(2020){Estrada-Carpenter}, {Papovich},
  {Momcheva}, {Brammer}, {Simons}, {Bridge}, {Cleri}, {Ferguson},
  {Finkelstein}, {Giavalisco}, {Jung}, {Matharu}, {Trump}, \&
  {Weiner}}]{estrada20}
{Estrada-Carpenter}, V., {Papovich}, C., {Momcheva}, I., {et~al.} 2020, \apj,
  898, 171

\bibitem[{{Faber} {et~al.}(1997){Faber}, {Tremaine}, {Ajhar}, {Byun},
  {Dressler}, {Gebhardt}, {Grillmair}, {Kormendy}, {Lauer}, \&
  {Richstone}}]{faber97}
{Faber}, S.~M., {Tremaine}, S., {Ajhar}, E.~A., {et~al.} 1997, \aj, 114, 1771

\bibitem[{{Faisst} {et~al.}(2019){Faisst}, {Capak}, {Emami}, {Tacchella}, \&
  {Larson}}]{faisst19}
{Faisst}, A.~L., {Capak}, P.~L., {Emami}, N., {Tacchella}, S., \& {Larson},
  K.~L. 2019, \apj, 884, 133

\bibitem[{{Fan} {et~al.}(2010){Fan}, {Lapi}, {Bressan}, {Bernardi}, {De Zotti},
  \& {Danese}}]{fan10}
{Fan}, L., {Lapi}, A., {Bressan}, A., {et~al.} 2010, \apj, 718, 1460

\bibitem[{{Fan} {et~al.}(2008){Fan}, {Lapi}, {De Zotti}, \& {Danese}}]{fan08}
{Fan}, L., {Lapi}, A., {De Zotti}, G., \& {Danese}, L. 2008, \apjl, 689, L101

\bibitem[{{Fensch} \& {Bournaud}(2021)}]{fensch20}
{Fensch}, J. \& {Bournaud}, F. 2021, \mnras, 505, 3579

\bibitem[{{Fisher} \& {Drory}(2011)}]{fisher11}
{Fisher}, D.~B. \& {Drory}, N. 2011, \apjl, 733, L47

\bibitem[{{Forrest} {et~al.}(2020){Forrest}, {Annunziatella}, {Wilson},
  {Marchesini}, {Muzzin}, {Cooper}, {Marsan}, {McConachie}, {Chan}, {Gomez},
  {Kado-Fong}, {L Barbera}, {Labb{\'e}}, {Lange-Vagle}, {Nantais}, {Nonino},
  {Pe{\~n}a}, {Saracco}, {Stefanon}, \& {van der Burg}}]{forrest19}
{Forrest}, B., {Annunziatella}, M., {Wilson}, G., {et~al.} 2020, \apjl, 890, L1

\bibitem[{{Foster} {et~al.}(2017){Foster}, {van de Sande}, {D'Eugenio},
  {Cortese}, {McDermid}, {Bland-Hawthorn}, {Brough}, {Bryant}, {Croom},
  {Goodwin}, {Konstantopoulos}, {Lawrence}, {L{\'o}pez-S{\'a}nchez}, {Medling},
  {Owers}, {Richards}, {Scott}, {Taranu}, {Tonini}, \& {Zafar}}]{foster17}
{Foster}, C., {van de Sande}, J., {D'Eugenio}, F., {et~al.} 2017, \mnras, 472,
  966

\bibitem[{{Ganda} {et~al.}(2007){Ganda}, {Peletier}, {McDermid},
  {Falc{\'o}n-Barroso}, {de Zeeuw}, {Bacon}, {Cappellari}, {Davies},
  {Emsellem}, {Krajnovi{\'c}}, {Kuntschner}, {Sarzi}, \& {van de
  Ven}}]{ganda07}
{Ganda}, K., {Peletier}, R.~F., {McDermid}, R.~M., {et~al.} 2007, \mnras, 380,
  506

\bibitem[{{Gavazzi} {et~al.}(2018){Gavazzi}, {Consolandi}, {Pedraglio},
  {Fossati}, {Fumagalli}, \& {Boselli}}]{gavazzi18}
{Gavazzi}, G., {Consolandi}, G., {Pedraglio}, S., {et~al.} 2018, \aap, 611, A28

\bibitem[{{Ghigna} {et~al.}(1998){Ghigna}, {Moore}, {Governato}, {Lake},
  {Quinn}, \& {Stadel}}]{ghigna98}
{Ghigna}, S., {Moore}, B., {Governato}, F., {et~al.} 1998, \mnras, 300, 146

\bibitem[{{Gonz{\'a}lez Delgado} {et~al.}(2017){Gonz{\'a}lez Delgado},
  {P{\'e}rez}, {Cid Fernand es}, {Garc{\'\i}a-Benito}, {L{\'o}pez
  Fern{\'a}ndez}, {Vale Asari}, {Cortijo-Ferrero}, {de Amorim}, {Lacerda},
  {S{\'a}nchez}, {Lehnert}, \& {Walcher}}]{gonzalez17}
{Gonz{\'a}lez Delgado}, R.~M., {P{\'e}rez}, E., {Cid Fernand es}, R., {et~al.}
  2017, \aap, 607, A128

\bibitem[{{Graham}(2013)}]{graham13}
{Graham}, A.~W. 2013, {Elliptical and Disk Galaxy Structure and Modern Scaling
  Laws}, ed. T.~D. {Oswalt} \& W.~C. {Keel}, Vol.~6, 91--140

\bibitem[{{Graham} {et~al.}(2019{\natexlab{a}}){Graham}, {Cappellari},
  {Bershady}, \& {Drory}}]{graham19b}
{Graham}, M.~T., {Cappellari}, M., {Bershady}, M.~A., \& {Drory}, N.
  2019{\natexlab{a}}, arXiv e-prints, arXiv:1911.06103

\bibitem[{{Graham} {et~al.}(2019{\natexlab{b}}){Graham}, {Cappellari},
  {Bershady}, \& {Drory}}]{graham19}
{Graham}, M.~T., {Cappellari}, M., {Bershady}, M.~A., \& {Drory}, N.
  2019{\natexlab{b}}, arXiv e-prints, arXiv:1910.05139

\bibitem[{{Graham} {et~al.}(2018){Graham}, {Cappellari}, {Li}, {Mao},
  {Bershady}, {Bizyaev}, {Brinkmann}, {Brownstein}, {Bundy}, {Drory}, {Law},
  {Pan}, {Thomas}, {Wake}, {Weijmans}, {Westfall}, \& {Yan}}]{graham18}
{Graham}, M.~T., {Cappellari}, M., {Li}, H., {et~al.} 2018, \mnras, 477, 4711

\bibitem[{{Greene} {et~al.}(2017){Greene}, {Leauthaud}, {Emsellem}, {Goddard},
  {Ge}, {Andrews}, {Brinkman}, {Brownstein}, {Greco}, {Law}, {Lin}, {Masters},
  {Merrifield}, {More}, {Okabe}, {Schneider}, {Thomas}, {Wake}, {Yan}, \&
  {Drory}}]{greene17}
{Greene}, J.~E., {Leauthaud}, A., {Emsellem}, E., {et~al.} 2017, \apjl, 851,
  L33

\bibitem[{{Gullberg} {et~al.}(2019){Gullberg}, {Smail}, {Swinbank},
  {Dudzevi{\v{c}}i{\={u}}t{\.{e}}}, {Stach}, {Thomson}, {Almaini}, {Chen},
  {Conselice}, {Cooke}, {Farrah}, {Ivison}, {Maltby}, {Micha{\l}owski},
  {Simpson}, {Scott}, {Wardlow}, \& {Weiss}}]{gullberg19}
{Gullberg}, B., {Smail}, I., {Swinbank}, A.~M., {et~al.} 2019, \mnras, 490,
  4956

\bibitem[{{Haan} {et~al.}(2013){Haan}, {Armus}, {Surace}, {Charmandaris},
  {Evans}, {Diaz-Santos}, {Melbourne}, {Mazzarella}, {Howell}, {Stierwalt},
  {Kim}, {Vavilkin}, {Sanders}, {Petric}, {Murphy}, {Braun}, {Bridge}, \&
  {Inami}}]{hahn13}
{Haan}, S., {Armus}, L., {Surace}, J.~A., {et~al.} 2013, \mnras, 434, 1264

\bibitem[{{Harris}(2001)}]{harrissaasfee}
{Harris}, W.~E. 2001, {Globular Cluster Systems}, Vol.~28, 223

\bibitem[{{Harrison}(2017)}]{harrison17}
{Harrison}, C.~M. 2017, Nature Astronomy, 1, 0165

\bibitem[{{Hau} \& {Thomson}(1994)}]{hau94}
{Hau}, G.~K.~T. \& {Thomson}, R.~C. 1994, \mnras, 270, L23

\bibitem[{{Hendel} \& {Johnston}(2015)}]{hendel15}
{Hendel}, D. \& {Johnston}, K.~V. 2015, \mnras, 454, 2472

\bibitem[{{Hill} {et~al.}(2017){Hill}, {Muzzin}, {Franx}, \&
  {Marchesini}}]{hill17}
{Hill}, A.~R., {Muzzin}, A., {Franx}, M., \& {Marchesini}, D. 2017, \apjl, 849,
  L26

\bibitem[{{Hill} {et~al.}(2019){Hill}, {van der Wel}, {Franx}, {Muzzin},
  {Skelton}, {Momcheva}, {van Dokkum}, \& {Whitaker}}]{hill19}
{Hill}, A.~R., {van der Wel}, A., {Franx}, M., {et~al.} 2019, \apj, 871, 76

\bibitem[{{Hilz} {et~al.}(2013){Hilz}, {Naab}, \& {Ostriker}}]{hilz13}
{Hilz}, M., {Naab}, T., \& {Ostriker}, J.~P. 2013, \mnras, 429, 2924

\bibitem[{{Hirschmann} {et~al.}(2015){Hirschmann}, {Naab}, {Ostriker},
  {Forbes}, {Duc}, {Dav{\'e}}, {Oser}, \& {Karabal}}]{hirschmann15}
{Hirschmann}, M., {Naab}, T., {Ostriker}, J.~P., {et~al.} 2015, \mnras, 449,
  528

\bibitem[{{Hodge} {et~al.}(2019){Hodge}, {Smail}, {Walter}, {da Cunha},
  {Swinbank}, {Rybak}, {Venemans}, {Brandt}, {Calistro Rivera}, {Chapman},
  {Chen}, {Cox}, {Dannerbauer}, {Decarli}, {Greve}, {Knudsen}, {Menten},
  {Schinnerer}, {Simpson}, {van der Werf}, {Wardlow}, \& {Weiss}}]{hodge19}
{Hodge}, J.~A., {Smail}, I., {Walter}, F., {et~al.} 2019, \apj, 876, 130

\bibitem[{{Hopkins} {et~al.}(2009{\natexlab{a}}){Hopkins}, {Bundy}, {Murray},
  {Quataert}, {Lauer}, \& {Ma}}]{hopkins09}
{Hopkins}, P.~F., {Bundy}, K., {Murray}, N., {et~al.} 2009{\natexlab{a}},
  \mnras, 398, 898

\bibitem[{{Hopkins} {et~al.}(2009{\natexlab{b}}){Hopkins}, {Cox}, {Dutta},
  {Hernquist}, {Kormendy}, \& {Lauer}}]{hopkins09b}
{Hopkins}, P.~F., {Cox}, T.~J., {Dutta}, S.~N., {et~al.} 2009{\natexlab{b}},
  \apjs, 181, 135

\bibitem[{{Hopkins} {et~al.}(2012){Hopkins}, {Kere{\v{s}}}, {Murray},
  {Quataert}, \& {Hernquist}}]{hopkins12}
{Hopkins}, P.~F., {Kere{\v{s}}}, D., {Murray}, N., {Quataert}, E., \&
  {Hernquist}, L. 2012, \mnras, 427, 968

\bibitem[{{Houghton} {et~al.}(2013){Houghton}, {Davies}, {D'Eugenio}, {Scott},
  {Thatte}, {Clarke}, {Tecza}, {Salter}, {Fogarty}, \& {Goodsall}}]{houghton13}
{Houghton}, R.~C.~W., {Davies}, R.~L., {D'Eugenio}, F., {et~al.} 2013, \mnras,
  436, 19

\bibitem[{{Huang} {et~al.}(2018){Huang}, {Leauthaud}, {Greene}, {Bundy}, {Lin},
  {Tanaka}, {Miyazaki}, \& {Komiyama}}]{huang18}
{Huang}, S., {Leauthaud}, A., {Greene}, J.~E., {et~al.} 2018, \mnras, 475, 3348

\bibitem[{{Huertas-Company} {et~al.}(2015){Huertas-Company},
  {P{\'e}rez-Gonz{\'a}lez}, {Mei}, {Shankar}, {Bernardi}, {Daddi}, {Barro},
  {Cabrera-Vives}, {Cattaneo}, {Dimauro}, \& {Gravet}}]{huertascompany15}
{Huertas-Company}, M., {P{\'e}rez-Gonz{\'a}lez}, P.~G., {Mei}, S., {et~al.}
  2015, \apj, 809, 95

\bibitem[{{Ishibashi} {et~al.}(2013){Ishibashi}, {Fabian}, \&
  {Canning}}]{ishibashi13}
{Ishibashi}, W., {Fabian}, A.~C., \& {Canning}, R.~E.~A. 2013, \mnras, 431,
  2350

\bibitem[{{Kado-Fong} {et~al.}(2018){Kado-Fong}, {Greene}, {Hendel},
  {Price-Whelan}, {Greco}, {Goulding}, {Huang}, {Johnston}, {Komiyama}, \&
  {Lee}}]{kadofong18}
{Kado-Fong}, E., {Greene}, J.~E., {Hendel}, D., {et~al.} 2018, \apj, 866, 103

\bibitem[{{Karademir} {et~al.}(2019){Karademir}, {Remus}, {Burkert}, {Dolag},
  {Hoffmann}, {Moster}, {Steinwandel}, \& {Zhang}}]{karademir19}
{Karademir}, G.~S., {Remus}, R.-S., {Burkert}, A., {et~al.} 2019, \mnras, 487,
  318

\bibitem[{{Kawinwanichakij} {et~al.}(2020){Kawinwanichakij}, {Papovich},
  {Ciardullo}, {Finkelstein}, {Stevans}, {Wold}, {Jogee}, {Sherman}, {Florez},
  \& {Gronwall}}]{kawinwanichakij20}
{Kawinwanichakij}, L., {Papovich}, C., {Ciardullo}, R., {et~al.} 2020, \apj,
  892, 7

\bibitem[{{Kelvin} {et~al.}(2014){Kelvin}, {Driver}, {Robotham}, {Taylor},
  {Graham}, {Alpaslan}, {Baldry}, {Bamford}, {Bauer}, {Bland-Hawthorn},
  {Brown}, {Colless}, {Conselice}, {Holwerda}, {Hopkins}, {Lara-L{\'o}pez},
  {Liske}, {L{\'o}pez-S{\'a}nchez}, {Loveday}, {Norberg}, {Phillipps},
  {Popescu}, {Prescott}, {Sansom}, \& {Tuffs}}]{kelvin14}
{Kelvin}, L.~S., {Driver}, S.~P., {Robotham}, A. S.~G., {et~al.} 2014, \mnras,
  444, 1647

\bibitem[{{Kere{\v{s}}} {et~al.}(2005){Kere{\v{s}}}, {Katz}, {Weinberg}, \&
  {Dav{\'e}}}]{keres05}
{Kere{\v{s}}}, D., {Katz}, N., {Weinberg}, D.~H., \& {Dav{\'e}}, R. 2005,
  \mnras, 363, 2

\bibitem[{{Kobayashi}(2004)}]{kobayashi04}
{Kobayashi}, C. 2004, \mnras, 347, 740

\bibitem[{{Kormendy} {et~al.}(2010){Kormendy}, {Drory}, {Bender}, \&
  {Cornell}}]{kormendy10}
{Kormendy}, J., {Drory}, N., {Bender}, R., \& {Cornell}, M.~E. 2010, \apj, 723,
  54

\bibitem[{{Krajnovi{\'c}} {et~al.}(2011){Krajnovi{\'c}}, {Emsellem},
  {Cappellari}, {Alatalo}, {Blitz}, {Bois}, {Bournaud}, {Bureau}, {Davies},
  {Davis}, {de Zeeuw}, {Khochfar}, {Kuntschner}, {Lablanche}, {McDermid},
  {Morganti}, {Naab}, {Oosterloo}, {Sarzi}, {Scott}, {Serra}, {Weijmans}, \&
  {Young}}]{krajnovic11}
{Krajnovi{\'c}}, D., {Emsellem}, E., {Cappellari}, M., {et~al.} 2011, \mnras,
  414, 2923

\bibitem[{{Krajnovi{\'c}} {et~al.}(2020){Krajnovi{\'c}}, {Ural}, {Kuntschner},
  {Goudfrooij}, {Wolfe}, {Cappellari}, {Davies}, {de Zeeuw}, {Duc}, {Emsellem},
  {Karick}, {McDermid}, {Mei}, \& {Naab}}]{krajnovic20}
{Krajnovi{\'c}}, D., {Ural}, U., {Kuntschner}, H., {et~al.} 2020, \aap, 635,
  A129

\bibitem[{{Kuntschner} {et~al.}(2010){Kuntschner}, {Emsellem}, {Bacon},
  {Cappellari}, {Davies}, {de Zeeuw}, {Falc{\'o}n-Barroso}, {Krajnovi{\'c}},
  {McDermid}, {Peletier}, {Sarzi}, {Shapiro}, {van den Bosch}, \& {van de
  Ven}}]{kuntschner10}
{Kuntschner}, H., {Emsellem}, E., {Bacon}, R., {et~al.} 2010, \mnras, 408, 97

\bibitem[{{Lagos} {et~al.}(2022){Lagos}, {Emsellem}, {van de Sande},
  {Harborne}, {Cortese}, {Davison}, {Foster}, \& {Wright}}]{lagos22}
{Lagos}, C. d.~P., {Emsellem}, E., {van de Sande}, J., {et~al.} 2022, \mnras,
  509, 4372

\bibitem[{{Lagos} {et~al.}(2018){Lagos}, {Schaye}, {Bah{\'e}}, {Van de Sande},
  {Kay}, {Barnes}, {Davis}, \& {Dalla Vecchia}}]{lagos18}
{Lagos}, C. d.~P., {Schaye}, J., {Bah{\'e}}, Y., {et~al.} 2018, \mnras, 476,
  4327

\bibitem[{{Larson} {et~al.}(1980){Larson}, {Tinsley}, \& {Caldwell}}]{larson80}
{Larson}, R.~B., {Tinsley}, B.~M., \& {Caldwell}, C.~N. 1980, \apj, 237, 692

\bibitem[{{Lauer} {et~al.}(1995){Lauer}, {Ajhar}, {Byun}, {Dressler}, {Faber},
  {Grillmair}, {Kormendy}, {Richstone}, \& {Tremaine}}]{lauer95}
{Lauer}, T.~R., {Ajhar}, E.~A., {Byun}, Y.~I., {et~al.} 1995, \aj, 110, 2622

\bibitem[{{Lee} {et~al.}(2013){Lee}, {Giavalisco}, {Williams}, {Guo}, {Lotz},
  {Van der Wel}, {Ferguson}, {Faber}, {Koekemoer}, {Grogin}, {Kocevski},
  {Conselice}, {Wuyts}, {Dekel}, {Kartaltepe}, \& {Bell}}]{lee13}
{Lee}, B., {Giavalisco}, M., {Williams}, C.~C., {et~al.} 2013, \apj, 774, 47

\bibitem[{{Lelli} {et~al.}(2017){Lelli}, {McGaugh}, {Schombert}, \&
  {Pawlowski}}]{lelli17}
{Lelli}, F., {McGaugh}, S.~S., {Schombert}, J.~M., \& {Pawlowski}, M.~S. 2017,
  \apj, 836, 152

\bibitem[{{Li} {et~al.}(2018{\natexlab{a}}){Li}, {Mao}, {Cappellari}, {Ge},
  {Long}, {Li}, {Mo}, {Li}, {Zheng}, {Bundy}, {Thomas}, {Brownstein}, {Roman
  Lopes}, {Law}, \& {Drory}}]{li18}
{Li}, H., {Mao}, S., {Cappellari}, M., {et~al.} 2018{\natexlab{a}}, \mnras,
  476, 1765

\bibitem[{{Li} {et~al.}(2018{\natexlab{b}}){Li}, {Mao}, {Cappellari}, {Graham},
  {Emsellem}, \& {Long}}]{li18b}
{Li}, H., {Mao}, S., {Cappellari}, M., {et~al.} 2018{\natexlab{b}}, \apjl, 863,
  L19

\bibitem[{{Li} {et~al.}(2017){Li}, {Bregman}, {Wang}, {Crain}, {Anderson}, \&
  {Zhang}}]{li17cgm}
{Li}, J.-T., {Bregman}, J.~N., {Wang}, Q.~D., {et~al.} 2017, \apjs, 233, 20

\bibitem[{{Li} {et~al.}(2014){Li}, {Crain}, \& {Wang}}]{li14cgm}
{Li}, J.-T., {Crain}, R.~A., \& {Wang}, Q.~D. 2014, \mnras, 440, 859

\bibitem[{{Liu} {et~al.}(2018){Liu}, {Daddi}, {Dickinson}, {Owen}, {Pannella},
  {Sargent}, {B{\'e}thermin}, {Magdis}, {Gao}, {Shu}, {Wang}, {Jin}, \&
  {Inami}}]{liu18}
{Liu}, D., {Daddi}, E., {Dickinson}, M., {et~al.} 2018, \apj, 853, 172

\bibitem[{{Lustig} {et~al.}(2021){Lustig}, {Strazzullo}, {D'Eugenio}, {Daddi},
  {Pannella}, {Renzini}, {Cimatti}, {Gobat}, {Jin}, {Mohr}, \&
  {Onodera}}]{lustig21}
{Lustig}, P., {Strazzullo}, V., {D'Eugenio}, C., {et~al.} 2021, \mnras, 501,
  2659

\bibitem[{{Madau} \& {Dickinson}(2014)}]{madau14}
{Madau}, P. \& {Dickinson}, M. 2014, \araa, 52, 415

\bibitem[{{Maiolino} \& {Mannucci}(2019)}]{maiolino19}
{Maiolino}, R. \& {Mannucci}, F. 2019, \aapr, 27, 3

\bibitem[{{Malin} \& {Carter}(1983)}]{MC83}
{Malin}, D.~F. \& {Carter}, D. 1983, \apj, 274, 534

\bibitem[{{Man} {et~al.}(2016){Man}, {Zirm}, \& {Toft}}]{man16}
{Man}, A. W.~S., {Zirm}, A.~W., \& {Toft}, S. 2016, \apj, 830, 89

\bibitem[{{Mancillas} {et~al.}(2019){Mancillas}, {Duc}, {Combes}, {Bournaud},
  {Emsellem}, {Martig}, \& {Michel-Dansac}}]{mancillas19}
{Mancillas}, B., {Duc}, P.-A., {Combes}, F., {et~al.} 2019, \aap, 632, A122

\bibitem[{{Mancini} {et~al.}(2019){Mancini}, {Daddi}, {Juneau}, {Renzini},
  {Rodighiero}, {Cappellari}, {Rodr{\'\i}guez-Mu{\~n}oz}, {Liu}, {Pannella},
  {Baronchelli}, {Franceschini}, {Bergamini}, {D'Eugenio}, \&
  {Puglisi}}]{mancini19}
{Mancini}, C., {Daddi}, E., {Juneau}, S., {et~al.} 2019, \mnras, 489, 1265

\bibitem[{{Martig} {et~al.}(2009){Martig}, {Bournaud}, {Teyssier}, \&
  {Dekel}}]{martig09}
{Martig}, M., {Bournaud}, F., {Teyssier}, R., \& {Dekel}, A. 2009, \apj, 707,
  250

\bibitem[{{Martig} {et~al.}(2013){Martig}, {Crocker}, {Bournaud}, {Emsellem},
  {Gabor}, {Alatalo}, {Blitz}, {Bois}, {Bureau}, {Cappellari}, {Davies},
  {Davis}, {Dekel}, {de Zeeuw}, {Duc}, {Falc{\'o}n-Barroso}, {Khochfar},
  {Krajnovi{\'c}}, {Kuntschner}, {Morganti}, {McDermid}, {Naab}, {Oosterloo},
  {Sarzi}, {Scott}, {Serra}, {Griffin}, {Teyssier}, {Weijmans}, \&
  {Young}}]{martig13}
{Martig}, M., {Crocker}, A.~F., {Bournaud}, F., {et~al.} 2013, \mnras, 432,
  1914

\bibitem[{{Mart{\'\i}n-Navarro} {et~al.}(2018){Mart{\'\i}n-Navarro},
  {Vazdekis}, {Falc{\'o}n-Barroso}, {La Barbera}, {Y{\i}ld{\i}r{\i}m}, \& {van
  de Ven}}]{martin18}
{Mart{\'\i}n-Navarro}, I., {Vazdekis}, A., {Falc{\'o}n-Barroso}, J., {et~al.}
  2018, \mnras, 475, 3700

\bibitem[{{Matteucci}(2014)}]{matteucci14}
{Matteucci}, F. 2014, Saas-Fee Advanced Course, 37, 145

\bibitem[{{McDermid} {et~al.}(2015){McDermid}, {Alatalo}, {Blitz}, {Bournaud},
  {Bureau}, {Cappellari}, {Crocker}, {Davies}, {Davis}, {de Zeeuw}, {Duc},
  {Emsellem}, {Khochfar}, {Krajnovi{\'c}}, {Kuntschner}, {Morganti}, {Naab},
  {Oosterloo}, {Sarzi}, {Scott}, {Serra}, {Weijmans}, \& {Young}}]{mcdermid15}
{McDermid}, R.~M., {Alatalo}, K., {Blitz}, L., {et~al.} 2015, \mnras, 448, 3484

\bibitem[{{Merlin} {et~al.}(2019){Merlin}, {Fortuni}, {Torelli}, {Santini},
  {Castellano}, {Fontana}, {Grazian}, {Pentericci}, {Pilo}, \&
  {Schmidt}}]{merlin19}
{Merlin}, E., {Fortuni}, F., {Torelli}, M., {et~al.} 2019, \mnras, 490, 3309

\bibitem[{{Merritt} {et~al.}(2020){Merritt}, {Pillepich}, {van Dokkum},
  {Nelson}, {Hernquist}, {Marinacci}, \& {Vogelsberger}}]{merritt20}
{Merritt}, A., {Pillepich}, A., {van Dokkum}, P., {et~al.} 2020, \mnras, 495,
  4570

\bibitem[{{Merritt} {et~al.}(2016){Merritt}, {van Dokkum}, {Abraham}, \&
  {Zhang}}]{merritt16}
{Merritt}, A., {van Dokkum}, P., {Abraham}, R., \& {Zhang}, J. 2016, \apj, 830,
  62

\bibitem[{{Mihos}(2003)}]{mihos03}
{Mihos}, C. 2003, arXiv e-prints, astro

\bibitem[{{Milgrom}(1983)}]{milg83}
{Milgrom}, M. 1983, \apj, 270, 365

\bibitem[{{Milosavljevi{\'c}} {et~al.}(2002){Milosavljevi{\'c}}, {Merritt},
  {Rest}, \& {van den Bosch}}]{miloslavljevic02}
{Milosavljevi{\'c}}, M., {Merritt}, D., {Rest}, A., \& {van den Bosch}, F.~C.
  2002, \mnras, 331, L51

\bibitem[{{Mo} {et~al.}(2010){Mo}, {van den Bosch}, \& {White}}]{mo10}
{Mo}, H., {van den Bosch}, F.~C., \& {White}, S. 2010, {Galaxy Formation and
  Evolution}

\bibitem[{{Montuori} {et~al.}(2010){Montuori}, {Di Matteo}, {Lehnert},
  {Combes}, \& {Semelin}}]{montuori10}
{Montuori}, M., {Di Matteo}, P., {Lehnert}, M.~D., {Combes}, F., \& {Semelin},
  B. 2010, \aap, 518, A56

\bibitem[{{Mortlock} {et~al.}(2013){Mortlock}, {Conselice}, {Hartley},
  {Ownsworth}, {Lani}, {Bluck}, {Almaini}, {Duncan}, {van der Wel},
  {Koekemoer}, {Dekel}, {Dav{\'e}}, {Ferguson}, {de Mello}, {Newman}, {Faber},
  {Grogin}, {Kocevski}, \& {Lai}}]{mortlock13}
{Mortlock}, A., {Conselice}, C.~J., {Hartley}, W.~G., {et~al.} 2013, \mnras,
  433, 1185

\bibitem[{{Mundy} {et~al.}(2017){Mundy}, {Conselice}, {Duncan}, {Almaini},
  {H{\"a}u{\ss}ler}, \& {Hartley}}]{mundy17}
{Mundy}, C.~J., {Conselice}, C.~J., {Duncan}, K.~J., {et~al.} 2017, \mnras,
  470, 3507

\bibitem[{{Naab} \& {Burkert}(2003)}]{naab03}
{Naab}, T. \& {Burkert}, A. 2003, \apj, 597, 893

\bibitem[{{Naab} {et~al.}(2009){Naab}, {Johansson}, \& {Ostriker}}]{naab09}
{Naab}, T., {Johansson}, P.~H., \& {Ostriker}, J.~P. 2009, \apjl, 699, L178

\bibitem[{{Naab} {et~al.}(2014){Naab}, {Oser}, {Emsellem}, {Cappellari},
  {Krajnovi{\'c}}, {McDermid}, {Alatalo}, {Bayet}, {Blitz}, {Bois}, {Bournaud},
  {Bureau}, {Crocker}, {Davies}, {Davis}, {de Zeeuw}, {Duc}, {Hirschmann},
  {Johansson}, {Khochfar}, {Kuntschner}, {Morganti}, {Oosterloo}, {Sarzi},
  {Scott}, {Serra}, {van de Ven}, {Weijmans}, \& {Young}}]{naab14}
{Naab}, T., {Oser}, L., {Emsellem}, E., {et~al.} 2014, \mnras, 444, 3357

\bibitem[{{Naab} \& {Ostriker}(2009)}]{naab09b}
{Naab}, T. \& {Ostriker}, J.~P. 2009, \apj, 690, 1452

\bibitem[{{Nagashima} {et~al.}(2005){Nagashima}, {Lacey}, {Okamoto}, {Baugh},
  {Frenk}, \& {Cole}}]{nagashima05}
{Nagashima}, M., {Lacey}, C.~G., {Okamoto}, T., {et~al.} 2005, \mnras, 363, L31

\bibitem[{{Navarro-Gonz{\'a}lez} {et~al.}(2013){Navarro-Gonz{\'a}lez},
  {Ricciardelli}, {Quilis}, \& {Vazdekis}}]{navarrogonzalez13}
{Navarro-Gonz{\'a}lez}, J., {Ricciardelli}, E., {Quilis}, V., \& {Vazdekis}, A.
  2013, \mnras, 436, 3507

\bibitem[{{Neistein} {et~al.}(2006){Neistein}, {van den Bosch}, \&
  {Dekel}}]{neistein06}
{Neistein}, E., {van den Bosch}, F.~C., \& {Dekel}, A. 2006, \mnras, 372, 933

\bibitem[{{Newman} {et~al.}(2018){Newman}, {Belli}, {Ellis}, \&
  {Patel}}]{newman18}
{Newman}, A.~B., {Belli}, S., {Ellis}, R.~S., \& {Patel}, S.~G. 2018, \apj,
  862, 126

\bibitem[{{Newman} {et~al.}(2012){Newman}, {Ellis}, {Bundy}, \&
  {Treu}}]{newman12}
{Newman}, A.~B., {Ellis}, R.~S., {Bundy}, K., \& {Treu}, T. 2012, \apj, 746,
  162

\bibitem[{{Nipoti} {et~al.}(2006){Nipoti}, {Londrillo}, \& {Ciotti}}]{nipoti06}
{Nipoti}, C., {Londrillo}, P., \& {Ciotti}, L. 2006, \mnras, 370, 681

\bibitem[{{Noguchi}(1999)}]{noguchi99}
{Noguchi}, M. 1999, \apj, 514, 77

\bibitem[{{Noirot} {et~al.}(2022){Noirot}, {Sawicki}, {Abraham},
  {Brada{\v{c}}}, {Iyer}, {Moutard}, {Pacifici}, {Ravindranath}, \&
  {Willott}}]{noirot22}
{Noirot}, G., {Sawicki}, M., {Abraham}, R., {et~al.} 2022, \mnras, 512, 3566

\bibitem[{{Okamoto} {et~al.}(2017){Okamoto}, {Nagashima}, {Lacey}, \&
  {Frenk}}]{okamoto17}
{Okamoto}, T., {Nagashima}, M., {Lacey}, C.~G., \& {Frenk}, C.~S. 2017, \mnras,
  464, 4866

\bibitem[{{Onodera} {et~al.}(2015){Onodera}, {Carollo}, {Renzini},
  {Cappellari}, {Mancini}, {Arimoto}, {Daddi}, {Gobat}, {Strazzullo},
  {Tacchella}, \& {Yamada}}]{onodera15}
{Onodera}, M., {Carollo}, C.~M., {Renzini}, A., {et~al.} 2015, \apj, 808, 161

\bibitem[{{Oser} {et~al.}(2010){Oser}, {Ostriker}, {Naab}, {Johansson}, \&
  {Burkert}}]{oser10}
{Oser}, L., {Ostriker}, J.~P., {Naab}, T., {Johansson}, P.~H., \& {Burkert}, A.
  2010, \apj, 725, 2312

\bibitem[{{Oyarz{\'u}n} {et~al.}(2019){Oyarz{\'u}n}, {Bundy}, {Westfall},
  {Belfiore}, {Thomas}, {Maraston}, {Lian}, {Arag{\'o}n-Salamanca}, {Zheng},
  {Gonzalez-Perez}, {Law}, {Drory}, \& {Andrews}}]{oyarzun19}
{Oyarz{\'u}n}, G.~A., {Bundy}, K., {Westfall}, K.~B., {et~al.} 2019, \apj, 880,
  111

\bibitem[{{Parikh} {et~al.}(2021){Parikh}, {Thomas}, {Maraston}, {Westfall},
  {Andrews}, {Boardman}, {Drory}, \& {Oyarzun}}]{parikh21}
{Parikh}, T., {Thomas}, D., {Maraston}, C., {et~al.} 2021, \mnras, 502, 5508

\bibitem[{{Patil} {et~al.}(2007){Patil}, {Pandey}, {Sahu}, \&
  {Kembhavi}}]{patil07}
{Patil}, M.~K., {Pandey}, S.~K., {Sahu}, D.~K., \& {Kembhavi}, A. 2007, \aap,
  461, 103

\bibitem[{{Peebles}(2020)}]{peebles20}
{Peebles}, P.~J.~E. 2020, \mnras, 498, 4386

\bibitem[{{Peebles} \& {Nusser}(2010)}]{peebles10}
{Peebles}, P.~J.~E. \& {Nusser}, A. 2010, \nat, 465, 565

\bibitem[{{Peletier} {et~al.}(2007){Peletier}, {Falc{\'o}n-Barroso}, {Bacon},
  {Cappellari}, {Davies}, {de Zeeuw}, {Emsellem}, {Ganda}, {Krajnovi{\'c}},
  {Kuntschner}, {McDermid}, {Sarzi}, \& {van de Ven}}]{peletier07}
{Peletier}, R.~F., {Falc{\'o}n-Barroso}, J., {Bacon}, R., {et~al.} 2007,
  \mnras, 379, 445

\bibitem[{{Peng} {et~al.}(2015){Peng}, {Maiolino}, \& {Cochrane}}]{peng15}
{Peng}, Y., {Maiolino}, R., \& {Cochrane}, R. 2015, \nat, 521, 192

\bibitem[{{Peng} {et~al.}(2010){Peng}, {Lilly}, {Kova{\v{c}}}, {Bolzonella},
  {Pozzetti}, {Renzini}, {Zamorani}, {Ilbert}, {Knobel}, {Iovino}, {Maier},
  {Cucciati}, {Tasca}, {Carollo}, {Silverman}, {Kampczyk}, {de Ravel},
  {Sanders}, {Scoville}, {Contini}, {Mainieri}, {Scodeggio}, {Kneib}, {Le
  F{\`e}vre}, {Bardelli}, {Bongiorno}, {Caputi}, {Coppa}, {de la Torre},
  {Franzetti}, {Garilli}, {Lamareille}, {Le Borgne}, {Le Brun}, {Mignoli},
  {Perez Montero}, {Pello}, {Ricciardelli}, {Tanaka}, {Tresse}, {Vergani},
  {Welikala}, {Zucca}, {Oesch}, {Abbas}, {Barnes}, {Bordoloi}, {Bottini},
  {Cappi}, {Cassata}, {Cimatti}, {Fumana}, {Hasinger}, {Koekemoer},
  {Leauthaud}, {Maccagni}, {Marinoni}, {McCracken}, {Memeo}, {Meneux}, {Nair},
  {Porciani}, {Presotto}, \& {Scaramella}}]{peng10}
{Peng}, Y.-j., {Lilly}, S.~J., {Kova{\v{c}}}, K., {et~al.} 2010, \apj, 721, 193

\bibitem[{{Penoyre} {et~al.}(2017){Penoyre}, {Moster}, {Sijacki}, \&
  {Genel}}]{penoyre17}
{Penoyre}, Z., {Moster}, B.~P., {Sijacki}, D., \& {Genel}, S. 2017, \mnras,
  468, 3883

\bibitem[{{Perez} {et~al.}(2011){Perez}, {Michel-Dansac}, \&
  {Tissera}}]{perez11}
{Perez}, J., {Michel-Dansac}, L., \& {Tissera}, P.~B. 2011, \mnras, 417, 580

\bibitem[{{Pierce} {et~al.}(2005){Pierce}, {Brodie}, {Forbes}, {Beasley},
  {Proctor}, \& {Strader}}]{pierce05}
{Pierce}, M., {Brodie}, J.~P., {Forbes}, D.~A., {et~al.} 2005, \mnras, 358, 419

\bibitem[{{Pipino} {et~al.}(2013){Pipino}, {Calura}, \& {Matteucci}}]{pipino13}
{Pipino}, A., {Calura}, F., \& {Matteucci}, F. 2013, \mnras, 432, 2541

\bibitem[{{Pipino} {et~al.}(2010){Pipino}, {D'Ercole}, {Chiappini}, \&
  {Matteucci}}]{pipino10}
{Pipino}, A., {D'Ercole}, A., {Chiappini}, C., \& {Matteucci}, F. 2010, \mnras,
  407, 1347

\bibitem[{{Pipino} {et~al.}(2009){Pipino}, {Devriendt}, {Thomas}, {Silk}, \&
  {Kaviraj}}]{pipino09}
{Pipino}, A., {Devriendt}, J.~E.~G., {Thomas}, D., {Silk}, J., \& {Kaviraj}, S.
  2009, \aap, 505, 1075

\bibitem[{{Pop} {et~al.}(2018){Pop}, {Pillepich}, {Amorisco}, \&
  {Hernquist}}]{pop18}
{Pop}, A.-R., {Pillepich}, A., {Amorisco}, N.~C., \& {Hernquist}, L. 2018,
  \mnras, 480, 1715

\bibitem[{{Proctor} \& {Sansom}(2002)}]{proctor02}
{Proctor}, R.~N. \& {Sansom}, A.~E. 2002, \mnras, 333, 517

\bibitem[{{Qu} {et~al.}(2010){Qu}, {Di Matteo}, {Lehnert}, {van Driel}, \&
  {Jog}}]{qu10}
{Qu}, Y., {Di Matteo}, P., {Lehnert}, M., {van Driel}, W., \& {Jog}, C.~J.
  2010, \aap, 515, A11

\bibitem[{{Read} \& {Trentham}(2005)}]{read05}
{Read}, J.~I. \& {Trentham}, N. 2005, Philosophical Transactions of the Royal
  Society of London Series A, 363, 2693

\bibitem[{{Rodriguez-Gomez} {et~al.}(2015){Rodriguez-Gomez}, {Genel},
  {Vogelsberger}, {Sijacki}, {Pillepich}, {Sales}, {Torrey}, {Snyder},
  {Nelson}, {Springel}, {Ma}, \& {Hernquist}}]{rodriguez15}
{Rodriguez-Gomez}, V., {Genel}, S., {Vogelsberger}, M., {et~al.} 2015, \mnras,
  449, 49

\bibitem[{{Rowlands} {et~al.}(2018){Rowlands}, {Heckman}, {Wild}, {Zakamska},
  {Rodriguez-Gomez}, {Barrera-Ballesteros}, {Lotz}, {Thilker}, {Andrews},
  {Boquien}, {Brinkmann}, {Brownstein}, {Hwang}, \& {Smethurst}}]{rowlands18}
{Rowlands}, K., {Heckman}, T., {Wild}, V., {et~al.} 2018, \mnras, 480, 2544

\bibitem[{{Rupke} {et~al.}(2010){Rupke}, {Kewley}, \& {Barnes}}]{rupke10}
{Rupke}, D. S.~N., {Kewley}, L.~J., \& {Barnes}, J.~E. 2010, \apjl, 710, L156

\bibitem[{{Sancisi} {et~al.}(2008){Sancisi}, {Fraternali}, {Oosterloo}, \& {van
  der Hulst}}]{sancisi08}
{Sancisi}, R., {Fraternali}, F., {Oosterloo}, T., \& {van der Hulst}, T. 2008,
  \aapr, 15, 189

\bibitem[{{Schaye} {et~al.}(2015){Schaye}, {Crain}, {Bower}, {Furlong},
  {Schaller}, {Theuns}, {Dalla Vecchia}, {Frenk}, {McCarthy}, {Helly},
  {Jenkins}, {Rosas-Guevara}, {White}, {Baes}, {Booth}, {Camps}, {Navarro},
  {Qu}, {Rahmati}, {Sawala}, {Thomas}, \& {Trayford}}]{schaye15}
{Schaye}, J., {Crain}, R.~A., {Bower}, R.~G., {et~al.} 2015, \mnras, 446, 521

\bibitem[{{Schreiber} {et~al.}(2018){Schreiber}, {Glazebrook}, {Nanayakkara},
  {Kacprzak}, {Labb{\'e}}, {Oesch}, {Yuan}, {Tran}, {Papovich}, {Spitler}, \&
  {Straatman}}]{schreiber18}
{Schreiber}, C., {Glazebrook}, K., {Nanayakkara}, T., {et~al.} 2018, \aap, 618,
  A85

\bibitem[{{Schulze} {et~al.}(2018){Schulze}, {Remus}, {Dolag}, {Burkert},
  {Emsellem}, \& {van de Ven}}]{schulze18}
{Schulze}, F., {Remus}, R.-S., {Dolag}, K., {et~al.} 2018, \mnras, 480, 4636

\bibitem[{{Schweizer}(1982)}]{schweizer82}
{Schweizer}, F. 1982, \apj, 252, 455

\bibitem[{{Scott} {et~al.}(2017){Scott}, {Brough}, {Croom}, {Davies}, {van de
  Sande}, {Allen}, {Bland-Hawthorn}, {Bryant}, {Cortese}, {D'Eugenio},
  {Federrath}, {Ferreras}, {Goodwin}, {Groves}, {Konstantopoulos}, {Lawrence},
  {Medling}, {Moffett}, {Owers}, {Richards}, {Robotham}, {Tonini}, \&
  {Yi}}]{scott17}
{Scott}, N., {Brough}, S., {Croom}, S.~M., {et~al.} 2017, \mnras, 472, 2833

\bibitem[{{Scott} {et~al.}(2013){Scott}, {Cappellari}, {Davies}, {Verdoes
  Kleijn}, {Bois}, {Alatalo}, {Blitz}, {Bournaud}, {Bureau}, {Crocker},
  {Davis}, {de Zeeuw}, {Duc}, {Emsellem}, {Khochfar}, {Krajnovi{\'c}},
  {Kuntschner}, {McDermid}, {Morganti}, {Naab}, {Oosterloo}, {Sarzi}, {Serra},
  {Weijmans}, \& {Young}}]{scott13}
{Scott}, N., {Cappellari}, M., {Davies}, R.~L., {et~al.} 2013, \mnras, 432,
  1894

\bibitem[{{Serra} {et~al.}(2014){Serra}, {Oser}, {Krajnovi{\'c}}, {Naab},
  {Oosterloo}, {Morganti}, {Cappellari}, {Emsellem}, {Young}, {Blitz}, {Davis},
  {Duc}, {Hirschmann}, {Weijmans}, {Alatalo}, {Bayet}, {Bois}, {Bournaud},
  {Bureau}, {Crocker}, {Davies}, {de Zeeuw}, {Khochfar}, {Kuntschner},
  {Lablanche}, {McDermid}, {Sarzi}, \& {Scott}}]{serra14}
{Serra}, P., {Oser}, L., {Krajnovi{\'c}}, D., {et~al.} 2014, \mnras, 444, 3388

\bibitem[{{Serra} \& {Trager}(2007)}]{serra07}
{Serra}, P. \& {Trager}, S.~C. 2007, \mnras, 374, 769

\bibitem[{{Simons} {et~al.}(2017){Simons}, {Kassin}, {Weiner}, {Faber},
  {Trump}, {Heckman}, {Koo}, {Pacifici}, {Primack}, {Snyder}, \& {de la
  Vega}}]{simons17}
{Simons}, R.~C., {Kassin}, S.~A., {Weiner}, B.~J., {et~al.} 2017, \apj, 843, 46

\bibitem[{{Sola} {et~al.}(2022){Sola}, {Duc}, {Richards}, {Paiement}, {Urbano},
  {Klehammer}, {B{\'\i}lek}, {Cuillandre}, {Gwyn}, \& {McConnachie}}]{sola22}
{Sola}, E., {Duc}, P.-A., {Richards}, F., {et~al.} 2022, \aap, 662, A124

\bibitem[{{Spolaor} {et~al.}(2009){Spolaor}, {Proctor}, {Forbes}, \&
  {Couch}}]{spolaor09}
{Spolaor}, M., {Proctor}, R.~N., {Forbes}, D.~A., \& {Couch}, W.~J. 2009,
  \apjl, 691, L138

\bibitem[{{Statler}(1991)}]{statler91}
{Statler}, T.~S. 1991, \apjl, 382, L11

\bibitem[{{Stevans} {et~al.}(2021){Stevans}, {Finkelstein}, {Kawinwanichakij},
  {Wold}, {Papovich}, {Somerville}, {Yung}, {Sherman}, {Ciardullo}, {Dav{\'e}},
  {Florez}, {Gronwall}, \& {Jogee}}]{stevans21}
{Stevans}, M.~L., {Finkelstein}, S.~L., {Kawinwanichakij}, L., {et~al.} 2021,
  \apj, 921, 58

\bibitem[{{Tacchella} {et~al.}(2022){Tacchella}, {Conroy}, {Faber}, {Johnson},
  {Leja}, {Barro}, {Cunningham}, {Deason}, {Guhathakurta}, {Guo}, {Hernquist},
  {Koo}, {McKinnon}, {Rockosi}, {Speagle}, {van Dokkum}, \&
  {Yesuf}}]{tacchella22}
{Tacchella}, S., {Conroy}, C., {Faber}, S.~M., {et~al.} 2022, \apj, 926, 134

\bibitem[{{Tadaki} {et~al.}(2018){Tadaki}, {Iono}, {Yun}, {Aretxaga},
  {Hatsukade}, {Hughes}, {Ikarashi}, {Izumi}, {Kawabe}, {Kohno}, {Lee},
  {Matsuda}, {Nakanishi}, {Saito}, {Tamura}, {Ueda}, {Umehata}, {Wilson},
  {Michiyama}, {Ando}, \& {Kamieneski}}]{tadaki18}
{Tadaki}, K., {Iono}, D., {Yun}, M.~S., {et~al.} 2018, \nat, 560, 613

\bibitem[{{Taylor} \& {Kobayashi}(2017)}]{taylor17}
{Taylor}, P. \& {Kobayashi}, C. 2017, \mnras, 471, 3856

\bibitem[{{Thomas} {et~al.}(2003){Thomas}, {Bender}, {Hopp}, {Maraston}, \&
  {Greggio}}]{thomas03}
{Thomas}, D., {Bender}, R., {Hopp}, U., {Maraston}, C., \& {Greggio}, L. 2003,
  \apss, 284, 599

\bibitem[{{Thomas} {et~al.}(1999){Thomas}, {Greggio}, \& {Bender}}]{thomas99}
{Thomas}, D., {Greggio}, L., \& {Bender}, R. 1999, \mnras, 302, 537

\bibitem[{{Thomas} {et~al.}(2002){Thomas}, {Maraston}, \& {Bender}}]{thomas02}
{Thomas}, D., {Maraston}, C., \& {Bender}, R. 2002, \apss, 281, 371

\bibitem[{{Thomas} {et~al.}(2010){Thomas}, {Maraston}, {Schawinski}, {Sarzi},
  \& {Silk}}]{thomas10}
{Thomas}, D., {Maraston}, C., {Schawinski}, K., {Sarzi}, M., \& {Silk}, J.
  2010, \mnras, 404, 1775

\bibitem[{{Toft} {et~al.}(2007){Toft}, {van Dokkum}, {Franx}, {Labbe},
  {F{\"o}rster Schreiber}, {Wuyts}, {Webb}, {Rudnick}, {Zirm}, {Kriek}, {van
  der Werf}, {Blakeslee}, {Illingworth}, {Rix}, {Papovich}, \&
  {Moorwood}}]{toft07}
{Toft}, S., {van Dokkum}, P., {Franx}, M., {et~al.} 2007, \apj, 671, 285

\bibitem[{{Toomre}(1977)}]{toomre77}
{Toomre}, A. 1977, in Evolution of Galaxies and Stellar Populations, ed. B.~M.
  {Tinsley} \& D.~C. {Larson}, Richard B.~Gehret, 401

\bibitem[{{Trujillo} {et~al.}(2011){Trujillo}, {Ferreras}, \& {de La
  Rosa}}]{trujillo11}
{Trujillo}, I., {Ferreras}, I., \& {de La Rosa}, I.~G. 2011, \mnras, 415, 3903

\bibitem[{{Trujillo} {et~al.}(2006){Trujillo}, {F{\"o}rster Schreiber},
  {Rudnick}, {Barden}, {Franx}, {Rix}, {Caldwell}, {McIntosh}, {Toft},
  {H{\"a}ussler}, {Zirm}, {van Dokkum}, {Labb{\'e}}, {Moorwood},
  {R{\"o}ttgering}, {van der Wel}, {van der Werf}, \& {van
  Starkenburg}}]{trujillo06}
{Trujillo}, I., {F{\"o}rster Schreiber}, N.~M., {Rudnick}, G., {et~al.} 2006,
  \apj, 650, 18

\bibitem[{{Trussler} {et~al.}(2020){Trussler}, {Maiolino}, {Maraston}, {Peng},
  {Thomas}, {Goddard}, \& {Lian}}]{trussler20}
{Trussler}, J., {Maiolino}, R., {Maraston}, C., {et~al.} 2020, \mnras, 491,
  5406

\bibitem[{{van de Sande} {et~al.}(2021{\natexlab{a}}){van de Sande}, {Croom},
  {Bland-Hawthorn}, {Cortese}, {Scott}, {Lagos}, {D'Eugenio}, {Bryant},
  {Brough}, {Catinella}, {Foster}, {Groves}, {Harborne},
  {L{\'o}pez-S{\'a}nchez}, {McDermid}, {Medling}, {Owers}, {Richards}, {Sweet},
  \& {Vaughan}}]{vandesande21b}
{van de Sande}, J., {Croom}, S.~M., {Bland-Hawthorn}, J., {et~al.}
  2021{\natexlab{a}}, \mnras, 508, 2307

\bibitem[{{van de Sande} {et~al.}(2021{\natexlab{b}}){van de Sande}, {Vaughan},
  {Cortese}, {Scott}, {Bland-Hawthorn}, {Croom}, {Lagos}, {Brough}, {Bryant},
  {Devriendt}, {Dubois}, {D'Eugenio}, {Foster}, {Fraser-McKelvie}, {Harborne},
  {Lawrence}, {Oh}, {Owers}, {Poci}, {Remus}, {Richards}, {Schulze}, {Sweet},
  {Varidel}, \& {Welker}}]{vandesande21}
{van de Sande}, J., {Vaughan}, S.~P., {Cortese}, L., {et~al.}
  2021{\natexlab{b}}, \mnras, 505, 3078

\bibitem[{{van der Vlugt} \& {Costa}(2019)}]{vandervlugt19}
{van der Vlugt}, D. \& {Costa}, T. 2019, \mnras, 490, 4918

\bibitem[{{van Dokkum} {et~al.}(2009){van Dokkum}, {Kriek}, \&
  {Franx}}]{vandokkum09}
{van Dokkum}, P.~G., {Kriek}, M., \& {Franx}, M. 2009, \nat, 460, 717

\bibitem[{{Veale} {et~al.}(2017){Veale}, {Ma}, {Greene}, {Thomas}, {Blakeslee},
  {McConnell}, {Walsh}, \& {Ito}}]{veale17}
{Veale}, M., {Ma}, C.-P., {Greene}, J.~E., {et~al.} 2017, \mnras, 471, 1428

\bibitem[{{Ventou} {et~al.}(2017){Ventou}, {Contini}, {Bouch{\'e}}, {Epinat},
  {Brinchmann}, {Bacon}, {Inami}, {Lam}, {Drake}, {Garel}, {Michel-Dansac},
  {Pello}, {Steinmetz}, {Weilbacher}, {Wisotzki}, \& {Carollo}}]{ventou17}
{Ventou}, E., {Contini}, T., {Bouch{\'e}}, N., {et~al.} 2017, \aap, 608, A9

\bibitem[{{Ventou} {et~al.}(2019){Ventou}, {Contini}, {Bouch{\'e}}, {Epinat},
  {Brinchmann}, {Inami}, {Richard}, {Schroetter}, {Soucail}, {Steinmetz}, \&
  {Weilbacher}}]{ventou19}
{Ventou}, E., {Contini}, T., {Bouch{\'e}}, N., {et~al.} 2019, \aap, 631, A87

\bibitem[{{Vincenzo} {et~al.}(2018){Vincenzo}, {Kobayashi}, \&
  {Taylor}}]{vincenzo18}
{Vincenzo}, F., {Kobayashi}, C., \& {Taylor}, P. 2018, \mnras, 480, L38

\bibitem[{{Wang} {et~al.}(2012){Wang}, {Hammer}, {Athanassoula}, {Puech},
  {Yang}, \& {Flores}}]{wang12}
{Wang}, J., {Hammer}, F., {Athanassoula}, E., {et~al.} 2012, \aap, 538, A121

\bibitem[{{Weijmans} {et~al.}(2014){Weijmans}, {de Zeeuw}, {Emsellem},
  {Krajnovi{\'c}}, {Lablanche}, {Alatalo}, {Blitz}, {Bois}, {Bournaud},
  {Bureau}, {Cappellari}, {Crocker}, {Davies}, {Davis}, {Duc}, {Khochfar},
  {Kuntschner}, {McDermid}, {Morganti}, {Naab}, {Oosterloo}, {Sarzi}, {Scott},
  {Serra}, {Verdoes Kleijn}, \& {Young}}]{weijmans14}
{Weijmans}, A.-M., {de Zeeuw}, P.~T., {Emsellem}, E., {et~al.} 2014, \mnras,
  444, 3340

\bibitem[{{Wilkins} {et~al.}(2019){Wilkins}, {Lovell}, \&
  {Stanway}}]{wilkins19}
{Wilkins}, S.~M., {Lovell}, C.~C., \& {Stanway}, E.~R. 2019, \mnras, 490, 5359

\bibitem[{{Wright}(2006)}]{wright06}
{Wright}, E.~L. 2006, \pasp, 118, 1711

\bibitem[{{Y{\i}ld{\i}r{\i}m} {et~al.}(2017){Y{\i}ld{\i}r{\i}m}, {van den
  Bosch}, {van de Ven}, {Mart{\'\i}n-Navarro}, {Walsh}, {Husemann},
  {G{\"u}ltekin}, \& {Gebhardt}}]{yildirim17}
{Y{\i}ld{\i}r{\i}m}, A., {van den Bosch}, R. C.~E., {van de Ven}, G., {et~al.}
  2017, \mnras, 468, 4216

\bibitem[{{Y{\i}ld{\i}z} {et~al.}(2020){Y{\i}ld{\i}z}, {Peletier}, {Duc}, \&
  {Serra}}]{yildiz20}
{Y{\i}ld{\i}z}, M.~K., {Peletier}, R.~F., {Duc}, P.~A., \& {Serra}, P. 2020,
  \aap, 636, A8

\bibitem[{{Young} {et~al.}(2011){Young}, {Bureau}, {Davis}, {Combes},
  {McDermid}, {Alatalo}, {Blitz}, {Bois}, {Bournaud}, {Cappellari}, {Davies},
  {de Zeeuw}, {Emsellem}, {Khochfar}, {Krajnovi{\'c}}, {Kuntschner},
  {Lablanche}, {Morganti}, {Naab}, {Oosterloo}, {Sarzi}, {Scott}, {Serra}, \&
  {Weijmans}}]{young11}
{Young}, L.~M., {Bureau}, M., {Davis}, T.~A., {et~al.} 2011, \mnras, 414, 940

\bibitem[{{Young} {et~al.}(2020){Young}, {Krajnovi{\'c}}, {Duc}, \&
  {Serra}}]{young20}
{Young}, L.~M., {Krajnovi{\'c}}, D., {Duc}, P.-A., \& {Serra}, P. 2020, \mnras,
  495, 1433

\bibitem[{{Zibetti} {et~al.}(2020){Zibetti}, {Gallazzi}, {Hirschmann},
  {Consolandi}, {Falc{\'o}n-Barroso}, {van de Ven}, \& {Lyubenova}}]{zibetti19}
{Zibetti}, S., {Gallazzi}, A.~R., {Hirschmann}, M., {et~al.} 2020, \mnras, 491,
  3562

\end{thebibliography}
